%% file: main.tex
\documentclass{biometrika}
\usepackage{amsmath,mathtools,mathrsfs}
\usepackage{newtxtext}
\usepackage[subscriptcorrection]{newtxmath}
\usepackage{dsfont,bm}
\usepackage{graphicx,booktabs,multirow,array,makecell,threeparttable}
\usepackage{placeins}
\makeatletter
\renewcommand{\arabic}[1]{\expandafter\@arabic\csname c@#1\endcsname}
\makeatother
\usepackage{float,algorithm,algpseudocode}

\usepackage[table]{xcolor}
\usepackage{marginnote,xurl}
\usepackage[colorlinks=true,linkcolor=blue,urlcolor=blue,citecolor=blue]{hyperref}
\providecommand{\captionsetup}[1]{}
\floatstyle{plain}
\restylefloat{algorithm}
\theoremstyle{plain}
\theoremnumbering{arabic}
\theoremsymbol{}
\theoremheaderfont{\hskip\parindent\normalfont\itshape}
\theorembodyfont{\normalfont}
\newtheorem{tensorassumption}{Assumption}

\let\hat\widehat

\def\argmax{{\mathop{\text{\rm argmax}}}}

\def\tr{{\mathop{\text{\rm Tr}}}}

\def\E{\mathbb E}
\def\RR{\mathbb R}
\def\tr{\operatorname{tr}}

\def\one{\mathbf1_p}
\def\Uh{\widehat{\mathbf U}_r}

\def\ff#1{\left\|#1\right\|_F}
\def\op#1{\left\|#1\right\|_2}

\newcounter{mcomcount}

\makeatletter
\providecommand{\paragraph}{\@startsection{paragraph}{4}{\z@}{1ex plus .2ex}{-1em}{\normalfont\itshape}}
\makeatother

\input{manuscript-macros.tex}

\begin{document}
\jname{Biometrika}
\markboth{}{Matrix Elliptical Sliced Inverse Regression}
\title{Matrix elliptical sliced inverse regression: structured dimension reduction for heavy-tailed data}
\author{Dingzi Guo}
\affil{Department of Statistics and Data Science,\\ National University of Singapore,  Singapore.}

\author{Wenhui Li}
\affil{School of Mathematics, Shandong University, China.}

\author{Yalin Wang\textsuperscript{*}\thanks{\textsuperscript{*} Corresponding author}}
\affil{School of Mathematics, Shandong University, China.
\email{wangyalin@mail.sdu.edu.cn}}

\author{\and Wang Zhou}
\affil{Department of Statistics and Data Science,\\ National University of Singapore,  Singapore.}
\maketitle
 \begin{abstract}

Sliced inverse regression is widely used for sufficient dimension reduction, but existing matrix extensions that rely on covariance-based statistics can be unstable under heavy-tailed distributions. We propose matrix elliptical sliced inverse regression (MESIR) for estimating response-relevant row and column reduction subspaces under matrix elliptical distributions. 
We establish a mode-wise linearity relation linking the inverse regression curve to these subspaces through the corresponding scatter matrices. MESIR combines inverse-regression Kendall's tau matrices with eigenvalue-calibrated marginal Kendall's tau matrices. The calibration recovers the row and column scatter matrices needed to identify the reduction subspaces.
Theoretically, we derive convergence rates for the calibrated marginal and inverse-regression estimators and establish subspace consistency under weak moment conditions, allowing the matrix dimensions and number of slices to diverge with the sample size. We further develop a consistent eigenvalue-ratio procedure for structural dimension selection and extend the method and theory to higher-order tensor predictors. Thorough simulation studies are conducted to show competitive performance under light tails and increasingly pronounced gains as the tails become heavier. Applications to COVID-19 CT images and radar target recognition demonstrate favorable empirical performance.
\end{abstract}


\section{Introduction}

In the era of big data, advances in data acquisition technologies have made high-dimensional matrix-valued data increasingly prevalent in biomedical imaging, longitudinal studies, and spatial applications \citep{zhou2014regularized,pfeiffer2021least,li2018tucker}. While these rich measurements offer opportunities to study complex relationships with a response of interest, their high dimensionality poses challenges for statistical estimation and interpretation, particularly when the sample size is limited. Dimension reduction thus plays an important role in extracting informative representations of matrix-valued predictors. In supervised learning problems, the central goal of this reduction is to retain information relevant to the response while preserving the underlying matrix structure.
Sufficient dimension reduction (SDR) provides a general framework for reducing predictor dimension while retaining all information about the conditional distribution of the response. For a vector-valued predictor $\mathbf{x}\in\mathbb{R}^{p}$ and a response variable $Y$, SDR seeks a full-column-rank matrix \(\mathbf{B}\in\mathbb{R}^{p\times d}\), with \(d\le p\), such that
\[
Y \perp\!\!\!\perp \mathbf{x}
\mid
\mathbf{B}^{\top}\mathbf{x},
\]
where $\perp\!\!\!\perp$ denotes independence, and \(\mathbf{B}^{\mathsf T}\mathbf{x}\in\mathbb{R}^{d}\) is a reduced predictor. The column space $\operatorname{Span}(\mathbf B)$ is a dimension reduction subspace. Under mild conditions\citep{cook1996graphics}, the intersection of all such subspaces defines the \emph{central subspace} in SDR. A variety of methods have been developed to estimate the central subspace \citep{yin2003estimating,cook2005sufficient,li2007directional,
Xia2007constructive,lin2018consistency,zhang2019robust,
zeng2024subspace}.
Among these methods, sliced inverse regression (SIR) \citep{li1991sliced} is particularly attractive for its simplicity and flexibility. It identifies reduction directions through variation in the inverse regression curve $\mathbb E(\mathbf x\mid Y=y)$ as $y$ varies. 
Projecting the original vector-valued predictor onto these reduction directions yields a few linear combinations that capture information relevant to the response. This inverse-regression approach motivates the development of dimension reduction methods for matrix-valued predictors that identify response-relevant directions along the row and column modes while preserving the matrix structure.

For a matrix-valued predictor $\mathbf{X}\in\mathbb{R}^{p\times q}$, a straightforward approach is to vectorize $\mathbf{X}$ and apply classical SDR, seeking a matrix $\mathbf{B}\in\mathbb{R}^{pq\times r}$, with $r\le pq$, such that
\begin{equation}
Y \perp\!\!\!\perp \operatorname{vec}(\mathbf{X})
\mid
\mathbf{B}^{\top}\operatorname{vec}(\mathbf{X}).
\end{equation}
However, this approach does not exploit the row--column structure
of $\mathbf{X}$ and requires estimating an unrestricted
$pq\times r$ reduction matrix, which involves a large number
of parameters when both $p$ and $q$ are large. 
To preserve the matrix structure while reducing the number of
parameters, we consider a bilinear reduction
$\mathbf B_1^\top \mathbf X \mathbf B_2$.
Specifically, we seek matrices
$\mathbf B_1\in\mathbb R^{p\times d_1}$ and
$\mathbf B_2\in\mathbb R^{q\times d_2}$ such that
\begin{equation}
Y \perp\!\!\!\perp \mathbf X
\mid
\mathbf B_1^\top \mathbf X \mathbf B_2,
\label{MESIR_core}
\end{equation}
where $d_1\le p$ and $d_2\le q$.
The spaces $\operatorname{Span}(\mathbf B_1)$ and
$\operatorname{Span}(\mathbf B_2)$ define the row and column
reduction subspaces, respectively.
The resulting $d_1\times d_2$ predictor retains all information
in $\mathbf X$ relevant to $Y$ while preserving its matrix
structure. 
Together, these subspaces determine the dimension folding subspace $\operatorname{Span}(\mathbf B_2\otimes\mathbf B_1)$,
where $\otimes$ denotes the Kronecker product.
When $d_1$ and $d_2$ are the smallest dimensions for which
\eqref{MESIR_core} holds, this subspace is called the
\emph{central tensor subspace} (CTS)
\citep{Li2010dimensionfolding,ding2015tensor},
denoted by $\mathcal S_{Y\mid\mathbf X}$.

Some existing studies have considered the estimation of row and column reduction subspaces. \citet{Li2010dimensionfolding} introduced dimension folding and developed moment-based SDR methods that reduce both dimensions while preserving the matrix structure. Building on this framework, \citet{ding2015tensor} proposed tensor sliced inverse regression (TSIR), which, in the matrix case, exploits inverse regression along the row and column modes to estimate the corresponding reduction subspaces. More recently, \citet{lee2023sufficient} proposed the principal support matrix machine (PSMM) for matrix-valued predictors, which estimates the reduction subspaces through a sequence of classification problems formed by slicing the response.

Matrix-valued predictors may exhibit heavy-tailed behavior and occasional extreme observations in practical applications \citep{he2025new,barigozzi2026tail}. Although dimension reduction methods have been developed for matrix-valued predictors, their robustness to heavy-tailed distributions remains comparatively underexplored. Existing research on robust SDR has focused primarily on vector-valued predictors. For example, 
\citet{chen2022high} proposed elliptical sliced inverse regression (ESIR), which uses multivariate Kendall's tau to accommodate heavy-tailed elliptical predictors, and \citet{li2026high} extended this approach to high-dimensional settings through sliced inverse Kendall's tau estimation (SIKE), which avoids estimating mean and requires
no moment conditions on the predictors.
\citet{zeng2026robust} proposed robust sliced inverse regression
for heavy-tailed data (ROSE) for high-dimensional predictors. 
Other approaches employ dependence measures, such as ball covariance \citep{zhang2019robust} and $\alpha$-distance covariance \citep{huang2025robust}, to mitigate sensitivity to outliers. However, these methods are not specifically designed for matrix- or tensor-valued predictors and do not directly address the estimation of their structured dimension reduction subspaces.

To address this gap, we propose matrix elliptical sliced inverse regression (MESIR), a structured SDR approach for matrix-valued predictors under matrix elliptical distributions. MESIR incorporates the matrix Kendall's tau estimator proposed by \cite{he2025new}, which is based on normalized pairwise matrix differences and mitigates the influence of large-magnitude observations while preserving the intrinsic row/column structure of the predictors.

Our contributions are fourfold. 
First, we establish a mode-wise linearity relation linking the inverse regression curve to the row and column reduction subspaces through their respective scatter matrices. We construct inverse-regression Kendall's tau matrices to capture response-relevant directions transformed by the row and column scatter matrices. Since marginal Kendall's tau matrices share the eigenspaces of the corresponding scatter matrices but generally have different relative eigenvalues, we develop an eigenvalue calibration procedure to recover the scatter matrices. Combining the eigenvalue-calibrated marginal Kendall's tau matrices with the inverse-regression Kendall's tau matrices yields estimators of the row and column reduction subspaces. Second, to our knowledge, we provide the first convergence theory for sufficient dimension reduction with heavy-tailed matrix-valued predictors and extend the theory to higher-order tensor-valued predictors. We derive convergence rates for both the calibrated marginal and inverse-regression Kendall's tau estimators and establish consistency of the estimated row, column, and resulting central tensor subspaces. The theory allows the matrix dimensions \(p\) and \(q\) and the number of slices \(H\) to diverge with the sample size \(n\). Notably, the theory requires only finite entrywise \(m\)th moments for some \(m\ge2\), a weaker moment requirement than those commonly imposed in classical SIR and matrix SDR theory \citep{zhu1995asymptotics,zhu2006sliced,ding2015tensor}, allowing the analysis to accommodate heavier-tailed predictors. Third, we develop an eigenvalue-ratio procedure for selecting the row and column structural dimensions and establish the consistency of the resulting dimension estimators. Fourth, we extend MESIR to higher-order tensor-valued predictors, resulting in tensor elliptical sliced inverse regression (TESIR). We develop the corresponding estimation procedure, establish convergence and subspace consistency results, and evaluate its performance through tensor simulation studies. Extensive simulation studies show competitive performance under light tails and increasingly pronounced gains as the tails become heavier. Applications to COVID-19 CT images and MSTAR radar images further demonstrate the method's empirical effectiveness.

\subsection{Notation}

Throughout this paper, bold symbols denote vectors or matrices, and calligraphic uppercase letters denote tensors or subspaces, as specified in context. For a vector \(\mathbf v\), \(\|\mathbf v\|_2\) denotes its Euclidean norm. For a matrix \(\mathbf A\), \(\mathbf A^\top\), \(\operatorname{tr}(\mathbf A)\), \(\operatorname{rank}(\mathbf A)\), and \(\operatorname{vec}(\mathbf A)\)  denote its transpose, trace, rank, and column-wise vectorization. The norms \(\|\mathbf A\|_2\) and \(\|\mathbf A\|_F\) denote its spectral and Frobenius norms, respectively. For a symmetric matrix \(\mathbf A\in\mathbb R^{p\times p}\), \(\lambda_{\max}(\mathbf A)\) and \(\lambda_{\min}(\mathbf A)\) denote its largest and smallest eigenvalues, respectively. \(\operatorname{Span}(\mathbf A)\) denotes the column space of \(\mathbf A\). For a full-column-rank matrix \(\mathbf B\), let \(\mathbf P_{\mathbf B}=\mathbf B(\mathbf B^\top\mathbf B)^{-1}\mathbf B^\top\) denote the orthogonal projection onto \(\operatorname{Span}(\mathbf B)\), and let \(\mathbf P^\top_{\mathbf B(\mathbf A)}=\mathbf A\mathbf B(\mathbf B^\top\mathbf A\mathbf B)^{-1}\mathbf B^\top\) denote the projection onto \(\operatorname{Span}(\mathbf B)\) with respect to matrix \(\mathbf A\), where \(\mathbf B\in\mathbb R^{p\times d}\) and \(\mathbf A\in\mathbb R^{p\times p}\). More generally, \(\mathbf P_{\mathcal S}\) denotes the orthogonal projector onto a subspace \(\mathcal S\). We follow the tensor notation conventions of \citet{kolda2009tensor}.
For a $K$-mode tensor
$\mathcal X\in\mathbb R^{p_1\times\cdots\times p_K}$,
$\mathcal X_{(k)}\in\mathbb R^{p_k\times p_{-k}}$
denotes its mode-$k$ unfolding, where
$p_{-k}=\prod_{j\ne k}p_j$.
The mode-$k$ product of $\mathcal X$ with a conformable matrix
$\mathbf A$ is denoted by $\mathcal X\times_k\mathbf A$,
and $\|\mathcal X\|_F$ denotes its Frobenius norm. The symbol \(\otimes\) denotes the Kronecker product. For positive sequences \(a_n\) and \(b_n\), \(a_n\lesssim b_n\) means \(a_n\le Cb_n\) for some constant \(C>0\), and \(a_n\asymp b_n\) means \(a_n\lesssim b_n\) and \(b_n\lesssim a_n\).

\section{Methodology}
\label{MESIR_method}

\subsection{Preliminaries}
\label{subsec:preliminaries}
For matrix-valued predictors, the row and column structures provide important information for dimension reduction. In practice, such predictors may also exhibit heavy-tailed behavior. The matrix elliptical distribution provides a natural framework for modeling such data, with row and column scatter matrices describing the dependence structure and a flexible radial component accommodating heavy-tailed behavior \citep{hoff2016limitations,gupta2018matrix,Tomarchio2020107050,WANG2025105467}. Specifically, a random matrix 
$\mathbf{X}\in\mathbb{R}^{p\times q}$ follows a matrix elliptical distribution if
\[
\mathbf{X}\overset{d}{=}
\mathbf{M}+\xi\mathbf{A}\mathbf{U}\mathbf{B}^{\top},
\]
where $\mathbf{M}\in\mathbb{R}^{p\times q}$ is a deterministic location matrix, $\mathbf{A}\in\mathbb{R}^{p\times p}$ and $\mathbf{B}\in\mathbb{R}^{q\times q}$ are nonsingular, $\operatorname{vec}(\mathbf{U})$ is uniformly distributed on the unit sphere in $\mathbb{R}^{pq}$, and $\xi$ is a nonnegative random variable 
independent of $\mathbf{U}$. Let
\(
\boldsymbol{\Sigma}=\mathbf{A}\mathbf{A}^{\top}, 
\boldsymbol{\Omega}=\mathbf{B}\mathbf{B}^{\top}.
\)
We write
\(
\mathbf{X}\sim
\mathcal{MED}_{p\times q}
\bigl(
\mathbf{M}, 
\boldsymbol{\Sigma}\otimes\boldsymbol{\Omega},
\varphi
\bigr),
\)
where $\boldsymbol{\Sigma}$ and $\boldsymbol{\Omega}$ describe the row and column scatter structures, respectively, and $\varphi$ characterizes the radial distribution.
Without loss of generality, we normalize $\operatorname{tr}(\boldsymbol{\Sigma})=p$ and $\operatorname{tr}(\boldsymbol{\Omega})=q$, absorbing their scales into the radial variable $\xi$. Different choices of \(\varphi\) accommodate different tail behaviors and give rise to familiar matrix-valued distributions, including the matrix normal and \(t\) distributions. Further details on matrix elliptical distributions can be found in \citet{gupta2018matrix}.

% Heavy-tailed distributions can induce substantial variation in observation magnitudes, making covariance-based quantities unstable or even ill-defined. The separation between scatter structure and radial variation in the matrix elliptical distribution motivates the use of normalized pairwise differences to characterize the row and column scatter structures. To this end,

Following \citet{he2025new}, let \(\mathbf X'\) be an independent copy of \(\mathbf X\), and define the marginal row and column Kendall's tau matrices by
\begin{equation}
\mathbf K_r
=\mathbb E\left[
\frac{(\mathbf X-\mathbf X')(\mathbf X-\mathbf X')^\top}
{\|\mathbf X-\mathbf X'\|_F^2/p}
\right],\ \ 
\mathbf K_c=\mathbb E\left[
\frac{(\mathbf X-\mathbf X')^\top(\mathbf X-\mathbf X')}
{\|\mathbf X-\mathbf X'\|_F^2/q}
\right].
\label{Kr-Kc}
\end{equation}
% The normalization by \(\|\mathbf{X}-\mathbf{X}'\|_F^2\) removes the overall magnitude of each pairwise difference while retaining its row- and column-wise directional information.
Both $\mathbf{K}_r$ and $\mathbf{K}_c$ are positive semidefinite. Without the scaling factors \(p\) and \(q\), the two Kendall's tau matrices have unit trace and therefore have average eigenvalues \(1/p\) and \(1/q\), respectively. We include the factors \(p\) and \(q\) to normalize the average eigenvalues to one \citep{chen2016target,chen2022high}. This rescaling changes only the eigenvalue magnitudes and leaves the corresponding eigenspaces unchanged.
Under the matrix elliptical model, \(\mathbf K_r\) and \(\mathbf K_c\) share the eigenspaces of \(\boldsymbol\Sigma\) and \(\boldsymbol\Omega\), respectively \citep{he2025new}. 

% {\color{red}Their relative eigenvalues, however, generally differ from those of the corresponding scatter matrices. Thus, the marginal Kendall's tau matrices provide the scatter eigendirections, while eigenvalue calibration is needed to recover the scatter shapes used for subspace estimation.}

\subsection{Matrix elliptical sliced inverse regression}
\label{subsec:mesir}

In classical sliced inverse regression, the covariance matrix of the inverse conditional mean and the marginal covariance matrix jointly identify the reduction subspace under suitable conditions \citep{li1991sliced,chen2010aos}. Building on this framework, we combine inverse-regression Kendall's tau matrices with eigenvalue-calibrated marginal Kendall's tau matrices to estimate the row and column reduction subspaces under matrix elliptical distributions.

A key condition for establishing such characterizations is the linearity condition, which links the inverse regression curve \(y\mapsto \mathbb E(\mathbf X\mid Y=y)\) to the underlying reduction subspace \citep{li2018sufficient,jin2025unified}. For matrix-valued predictors, the following proposition shows that this mode-wise linearity condition holds under the matrix elliptical model.

\begin{proposition}[Mode-wise linearity condition]
\label{prop:matrix_linearity}
Let $\mathbf A\in\mathbb R^{p\times d_1}$ $(d_1\le p)$ and
$\mathbf B\in\mathbb R^{q\times d_2}$ $(d_2\le q)$ be two full-column-rank matrices.
Suppose that
\(\mathbf X\sim\mathcal{MED}_{p\times q}
(\mathbf M,\boldsymbol\Sigma\otimes\boldsymbol\Omega,\varphi)\), we have
\[
\begin{aligned}
\mathbb E(\mathbf X\mid{\mathbf A}^\top\mathbf X)
&=
\mathbf M+
\boldsymbol\Sigma{\mathbf A}
({\mathbf A}^\top\boldsymbol\Sigma{\mathbf A})^{-1}
{\mathbf A}^\top(\mathbf X-\mathbf M),\\
\mathbb E(\mathbf X\mid\mathbf X{\mathbf B})
&=
\mathbf M+
(\mathbf X-\mathbf M){\mathbf B}
({\mathbf B}^\top\boldsymbol\Omega{\mathbf B})^{-1}
{\mathbf B}^\top\boldsymbol\Omega.
\end{aligned}
\]
\end{proposition}

Applying Proposition~\ref{prop:matrix_linearity} to the row and column reduction bases \(\mathbf B_1\) and \(\mathbf B_2\) in \eqref{MESIR_core}, we obtain
\begin{align*}
\mathbb{E}(\mathbf{X} - \mathbf M \mid Y) 
&=
\mathbb{E}\bigl(
\mathbb{E}(\mathbf{X}-\mathbf M \mid (Y,\mathbf{B}_1^\top \mathbf{X}))
\mid Y
\bigr) =
\mathbb{E}\bigl(
\mathbb{E}(\mathbf{X}-\mathbf M \mid \mathbf{B}_1^\top \mathbf{X})
\mid Y
\bigr) \\
&=
\boldsymbol{\Sigma}\mathbf{B}_1
(\mathbf{B}_1^\top\boldsymbol{\Sigma}\mathbf{B}_1)^{-1} \mathbf{B}_1^\top
\mathbb{E}(\mathbf{X}-\mathbf M \mid Y)
=
\mathbf{P}^\top_{\mathbf{B}_1(\boldsymbol{\Sigma})}
\mathbb{E}(\mathbf{X} -\mathbf M \mid Y).
\end{align*}
Similarly, we observe \(\mathbb{E}(\mathbf{X} - \mathbf M \mid Y) = \mathbb{E}(\mathbf{X} - \mathbf M \mid Y)\mathbf{P}_{\mathbf{B}_2(\boldsymbol{\Omega})} .\)
Therefore, the row and column inverse regression curves satisfy
\begin{equation}
\label{eq:inverse_regression_spaces}
\begin{aligned}
\operatorname{Span}\{
\mathbb E(\mathbf X\mid Y)-\mathbb E(\mathbf X)
\}
\subseteq
\operatorname{Span}(\boldsymbol\Sigma\mathbf B_1),\quad
\operatorname{Span}\{
\mathbb E(\mathbf X^\top\mid Y)-\mathbb E(\mathbf X^\top)
\}
\subseteq
\operatorname{Span}(\boldsymbol\Omega\mathbf B_2).
\end{aligned}
\end{equation}
For a mean-zero random matrix \(\mathbf A\), we use \(\operatorname{Cov}(\mathbf A)=\mathbb E(\mathbf A\mathbf A^\top)\) to denote its column covariance matrix. When the corresponding second moments exist, the above results imply that
\begin{equation}
\label{eq:cov-sir}
\begin{aligned}
\operatorname{Span}\!\left[
\operatorname{Cov}\{\mathbb E(\mathbf X\mid Y)\}
\right]
\subseteq
\operatorname{Span}(\boldsymbol\Sigma\mathbf B_1),\quad
\operatorname{Span}\!\left[
\operatorname{Cov}\{\mathbb E(\mathbf X^\top\mid Y)\}
\right]
\subseteq
\operatorname{Span}(\boldsymbol\Omega\mathbf B_2).
\end{aligned}
\end{equation}
These inclusions suggest extracting the row and column
directions contained in the inverse regression curve.
% {\color{red}Building on the Kendall constructions for vector- and
% matrix-valued predictors \citep{Han02012018,he2025new},
% we apply normalized pairwise differences to the inverse
% conditional means to capture these directions while reducing
% sensitivity to the magnitudes of the differences.}
Let \((\widetilde{\mathbf X},\widetilde Y)\) be an independent
copy of \((\mathbf X,Y)\), and define the row and column
inverse-regression Kendall's tau matrices by

\begin{equation}
\label{KrIR-KcIR}
\begin{aligned}
\mathbf K^{\mathrm{IR}}_{r}
&=
\mathbb E\left[
\frac{
(\mathbb E(\mathbf X\mid Y)
-\mathbb E(\widetilde{\mathbf X}\mid\widetilde Y))
(\mathbb E(\mathbf X\mid Y)
-\mathbb E(\widetilde{\mathbf X}\mid\widetilde Y))^\top
}{
\|\mathbb E(\mathbf X\mid Y)
-\mathbb E(\widetilde{\mathbf X}\mid\widetilde Y)\|_F^2/p
}
\right], \\[1ex]
\mathbf K^{\mathrm{IR}}_{c}
&=
\mathbb E\left[
\frac{
(\mathbb E(\mathbf X\mid Y)
-\mathbb E(\widetilde{\mathbf X}\mid\widetilde Y))^\top
(\mathbb E(\mathbf X\mid Y)
-\mathbb E(\widetilde{\mathbf X}\mid\widetilde Y))
}{
\|\mathbb E(\mathbf X\mid Y)
-\mathbb E(\widetilde{\mathbf X}\mid\widetilde Y)\|_F^2/q
}
\right].
\end{aligned}
\end{equation}
The inverse-regression Kendall's tau matrices defined in \eqref{KrIR-KcIR} share the same column spaces as their
covariance counterparts in \eqref{eq:cov-sir}: 
$$
\operatorname{Span}(\mathbf K^{\mathrm{IR}}_{r})
=
\operatorname{Span}\!\left[\operatorname{Cov}\{\mathbb E(\mathbf X\mid Y\}\right],\quad \operatorname{Span}(\mathbf K^{\mathrm{IR}}_{c})
=
\operatorname{Span}\!\left[\operatorname{Cov}\{\mathbb E(\mathbf X^\top\mid Y)\}\right].
$$
Since inverse-regression directions are determined by the
corresponding eigenspaces rather than the eigenvalue magnitudes,
these Kendall's tau matrices can replace their covariance
counterparts for extracting those directions.

It follows from \eqref{eq:cov-sir} that
\[
\operatorname{Span}(\mathbf K_r^{\mathrm{IR}})
\subseteq\operatorname{Span}(\boldsymbol\Sigma\mathbf B_1),
\qquad
\operatorname{Span}(\mathbf K_c^{\mathrm{IR}})
\subseteq\operatorname{Span}(\boldsymbol\Omega\mathbf B_2).
\]
Thus, the inverse-regression Kendall's tau matrices $\mathbf K_r^{\mathrm{IR}}$ and $\mathbf K_c^{\mathrm{IR}}$ are degenerate in directions orthogonal to $\operatorname{Span}(\boldsymbol\Sigma\mathbf B_1)$ and $\operatorname{Span}(\boldsymbol\Omega\mathbf B_2)$, respectively. We therefore consider the matrices of the leading $d_1$ and $d_2$ orthonormal eigenvectors of $\mathbf K_r^{\mathrm{IR}}$ and $\mathbf K_c^{\mathrm{IR}}$, denoted by $\boldsymbol\Gamma_1$ and $\boldsymbol\Gamma_2$, respectively. Transforming back by $\boldsymbol\Sigma^{-1}$ and $\boldsymbol\Omega^{-1}$ yields reduction directions in the corresponding row and column reduction subspaces. The Kronecker products of these row and column reduction directions further yield directions in the central tensor subspace $\mathcal S_{Y\mid\mathbf X}$.

The recovery of the row and column reduction subspaces thus depends on the scatter matrices $\boldsymbol\Sigma$ and $\boldsymbol\Omega$. Although \(\mathbf K_r\) and \(\mathbf K_c\) share eigenvectors with \(\boldsymbol\Sigma\) and \(\boldsymbol\Omega\), respectively, the corresponding eigenvalues are generally not proportional. Consequently, replacing $\boldsymbol\Sigma^{-1}$ and $\boldsymbol\Omega^{-1}$ with $\mathbf K_r^{-1}$ and $\mathbf K_c^{-1}$ need not preserve the target reduction subspaces. We therefore retain the marginal Kendall's tau eigenvectors and calibrate their eigenvalues to recover the row and column scatter matrices. 

For the following calibration, we impose the normalization \(\operatorname{tr}(\boldsymbol\Sigma)=p\) to fix the scale of the row scatter matrix. Under the matrix elliptical model, let $\mathbf W=\boldsymbol\Sigma^{-1/2}(\mathbf X-\widetilde{\mathbf X})$. Row orthogonal invariance implies
$\mathbb E\{\mathbf W\mathbf W^\top/
\operatorname{tr}(\mathbf W\mathbf W^\top)\}=\mathbf I_p/p$.
Multiplying both sides by $\boldsymbol\Sigma^{1/2}$ gives
\begin{equation}
\label{eq:population_scatter_calibration}
\boldsymbol\Sigma
=\mathbb E\left\{
\frac{(\mathbf X-\widetilde{\mathbf X})(\mathbf X-\widetilde{\mathbf X})^\top}
{\operatorname{tr}((\mathbf X-\widetilde{\mathbf X})^\top\boldsymbol\Sigma^{-1}(\mathbf X-\widetilde{\mathbf X}))/p}
\right\}.
\end{equation}
Since $\mathbf K_r$ and $\boldsymbol\Sigma$ share the same eigenvectors, the diagonal entries of \eqref{eq:population_scatter_calibration} yield equations for the eigenvalues of $\boldsymbol\Sigma$.  Replacing the population expectations with their sample counterparts while holding the estimated eigenvectors fixed gives the empirical calibration equations in \eqref{eq:mesir_empirical_calibration}. The column-mode equations follow analogously by considering $\mathbf X^\top$.

The sample MESIR procedure combines eigenvalue-calibrated marginal Kendall's tau
matrices with inverse-regression Kendall's tau matrices constructed
from slice means. Given i.i.d. observations
$\{(\mathbf X_i,Y_i)\}_{i=1}^n$, the sample marginal Kendall's tau matrices are
\begin{equation}\label{eq:mesir_sample_marginal}
\widehat{\mathbf K}_r
=\frac{2}{n(n-1)}\sum_{i<j}
\frac{(\mathbf X_i-\mathbf X_j)(\mathbf X_i-\mathbf X_j)^\top}{\|(\mathbf X_i-\mathbf X_j)\|_F^2/p},
\qquad
\widehat{\mathbf K}_c
=\frac{2}{n(n-1)}\sum_{i<j}
\frac{(\mathbf X_i-\mathbf X_j)^\top(\mathbf X_i-\mathbf X_j)}{\|(\mathbf X_i-\mathbf X_j)\|_F^2/q}.
\end{equation}
Let $\widehat{\mathbf U}_r=(\widehat{\mathbf u}_{r,1},\ldots,
\widehat{\mathbf u}_{r,p})$ be an orthogonal matrix whose columns are eigenvectors of $\widehat{\mathbf K}_r$.
Holding these eigenvectors fixed, we obtain the following empirical calibration equations from the diagonal entries of  \eqref{eq:population_scatter_calibration}:
\begin{equation}\label{eq:mesir_empirical_calibration}
\widehat\lambda_{r,a}
=\frac{2p}{n(n-1)}\sum_{i<j}
\frac{\|\widehat{\mathbf u}_{r,a}^\top(\mathbf X_i-\mathbf X_j)\|_2^2}{\sum_{b=1}^p \|\widehat{\mathbf u}_{r,b}^\top(\mathbf X_i-\mathbf X_j)\|_2^2/\widehat\lambda_{r,b}},
\quad \text{s.t.} \quad\sum_{a=1}^p\widehat\lambda_{r,a}=p, \quad a=1,\ldots,p.
\end{equation}
A positive solution to \eqref{eq:mesir_empirical_calibration} defines the eigenvalue-calibrated marginal Kendall's tau matrix
\begin{equation}\label{eq:mesir_sample_shapes}
\widehat{\mathbf K}_r^{\mathrm{cal}}
=\widehat{\mathbf U}_r
\operatorname{diag}(\widehat\lambda_{r,1},\ldots,\widehat\lambda_{r,p})
\widehat{\mathbf U}_r^\top,
\end{equation}
which estimates the row scatter matrix $\boldsymbol\Sigma$. Under the conditions of Proposition~\ref{cr:thm:Cr},
the sample calibration equations in
\eqref{eq:mesir_empirical_calibration} admit a unique positive
solution with probability tending to one. The column estimator $\widehat{\mathbf K}_c^{\mathrm{cal}}$
is obtained from $\mathbf X_i^\top$, with eigenvalues summing
to $q$. We solve \eqref{eq:mesir_empirical_calibration} numerically by adapting the recursive updates of \citet{hettmansperger2002practical} to a fixed eigenbasis,
as in \citet{li2026high}; implementation details are given in Algorithm~\ref{alg:mesir_calibration}.

To estimate the inverse-regression Kendall's tau matrices \(\mathbf K_r^{\mathrm{IR}}\) and \(\mathbf K_c^{\mathrm{IR}}\), partition the responses
into $H$ slices $J_1,\ldots,J_H$ of approximately equal size.
In slice $J_s$, estimate $\mathbb E(\mathbf X\mid Y\in J_s)$ by
$\widehat{\mathbf M}_s=n_s^{-1}\sum_{i:Y_i\in J_s}\mathbf X_i$,
where $n_s$ is the slice size. Substituting $\{\widehat{\mathbf M}_s\}_{s=1}^{H}$ into
\eqref{KrIR-KcIR} gives
\begin{equation}
\label{eq:mesir_sample_ir}
\begin{aligned}
\widehat{\mathbf K}_{r}^{\mathrm{IR}}
&=\frac{2}{H(H-1)}\sum_{1\le s<s'\le H}
\frac{(\widehat{\mathbf M}_s-\widehat{\mathbf M}_{s'})
(\widehat{\mathbf M}_s-\widehat{\mathbf M}_{s'})^\top}
{\|\widehat{\mathbf M}_s-\widehat{\mathbf M}_{s'}\|_F^2/p},\\
\widehat{\mathbf K}_{c}^{\mathrm{IR}}
&=\frac{2}{H(H-1)}\sum_{1\le s<s'\le H}
\frac{(\widehat{\mathbf M}_s-\widehat{\mathbf M}_{s'})^\top
(\widehat{\mathbf M}_s-\widehat{\mathbf M}_{s'})}
{\|\widehat{\mathbf M}_s-\widehat{\mathbf M}_{s'}\|_F^2/q}.
\end{aligned}
\end{equation}
Let $\widehat{\boldsymbol\Gamma}_1$ and
$\widehat{\boldsymbol\Gamma}_2$ contain the leading $d_1$ and $d_2$ orthonormal eigenvectors of
$\widehat{\mathbf K}_r^{\mathrm{IR}}$ and
$\widehat{\mathbf K}_c^{\mathrm{IR}}$, respectively.
Applying the inverse transformations with the eigenvalue-calibrated marginal Kendall's tau matrices gives the estimated reduction bases
\begin{equation}\label{eq:mesir_sample_bases}
\widehat{\mathbf B}_1
=(\widehat{\mathbf K}_r^{\mathrm{cal}})^{-1}
\widehat{\boldsymbol\Gamma}_1,
\qquad
\widehat{\mathbf B}_2
=(\widehat{\mathbf K}_c^{\mathrm{cal}})^{-1}
\widehat{\boldsymbol\Gamma}_2.
\end{equation}
The column spaces of $\widehat{\mathbf B}_1$ and $\widehat{\mathbf B}_2$ estimate the row and column reduction subspaces, respectively, while $\operatorname{Span}(\widehat{\mathbf B}_2)\otimes\operatorname{Span}(\widehat{\mathbf B}_1)$ estimates the central tensor subspace $\mathcal S_{Y\mid\mathbf X}$. Algorithm~\ref{alg:mesir} summarizes the estimation procedure.

\begin{algorithm}
\caption{MESIR for row and column reduction subspace estimation}
\label{alg:mesir}
\begin{algorithmic}[1]
\Require Observations $\{(\mathbf X_i,Y_i)\}_{i=1}^n$,
number of slices $H$ and structural dimensions $(d_1,d_2)$.
\State Partition the responses into $H$ slices and compute
the slice means $\widehat{\mathbf M}_s$.
\State Compute the sample marginal Kendall's tau matrices using
\eqref{eq:mesir_sample_marginal} and the inverse-regression
Kendall's tau matrices using \eqref{eq:mesir_sample_ir}.
Take $\widehat{\mathbf K}_r$ and $\widehat{\mathbf K}_c$
as inputs to Algorithm~\ref{alg:mesir_calibration}
to obtain the eigenvalue-calibrated marginal Kendall's tau matrices
$\widehat{\mathbf K}_r^{\mathrm{cal}}$ and
$\widehat{\mathbf K}_c^{\mathrm{cal}}$.
\State Form $\widehat{\boldsymbol\Gamma}_1$ and
$\widehat{\boldsymbol\Gamma}_2$ from the leading $d_1$ and $d_2$
eigenvectors of $\widehat{\mathbf K}_r^{\mathrm{IR}}$ and
$\widehat{\mathbf K}_c^{\mathrm{IR}}$, respectively.
\State Compute
$\widehat{\mathbf B}_1=(\widehat{\mathbf K}_r^{\mathrm{cal}})^{-1}
\widehat{\boldsymbol\Gamma}_1$ and
$\widehat{\mathbf B}_2=(\widehat{\mathbf K}_c^{\mathrm{cal}})^{-1}
\widehat{\boldsymbol\Gamma}_2$.
\Ensure $\widehat{\mathbf B}_1,\widehat{\mathbf B}_2$
and $\widehat{\mathcal S}_{Y\mid\mathbf X}
=\operatorname{Span}(\widehat{\mathbf B}_2)
\otimes\operatorname{Span}(\widehat{\mathbf B}_1)$.
\end{algorithmic}
\end{algorithm}

\begin{algorithm}[!htbp]
\caption{Iterative eigenvalue calibration of marginal Kendall's tau matrices}
\label{alg:mesir_calibration}
\begin{algorithmic}[1]
\Require Observations $\{\mathbf X_i\}_{i=1}^n$,
marginal Kendall's tau matrices
$\widehat{\mathbf K}_r$ and $\widehat{\mathbf K}_c$,
iterative tolerance $\varepsilon>0$,
and maximum iterations $m_{\max}$.

\State Obtain the orthogonal eigenvector matrices
$\widehat{\mathbf U}_r
=(\widehat{\mathbf u}_{r,1},\ldots,\widehat{\mathbf u}_{r,p})$
and
$\widehat{\mathbf U}_c
=(\widehat{\mathbf u}_{c,1},\ldots,\widehat{\mathbf u}_{c,q})$ of $\widehat{\mathbf K}_r$ and
$\widehat{\mathbf K}_c$,
respectively.

\State Initialize $\lambda_{r,a}^{(0)}>0$,
$a=1,\ldots,p$, using the eigenvalues of
$\widehat{\mathbf K}_r$, with sufficiently small eigenvalues replaced by a positive constant.

\State For $m=0,\ldots,m_{\max}-1$, compute
\[
\widetilde\lambda_{r,a}^{(m+1)}
=\frac{2p}{n(n-1)}\sum_{i<j}
\frac{\|\widehat{\mathbf u}_{r,a}^{\top}
(\mathbf X_i-\mathbf X_j)\|_2^2}
{\sum_{b=1}^p
\|\widehat{\mathbf u}_{r,b}^{\top}
(\mathbf X_i-\mathbf X_j)\|_2^2/\lambda_{r,b}^{(m)}},
\quad a=1,\ldots,p.
\]

\State Stop if
$\sqrt{\sum_{a=1}^p
(
\widetilde\lambda_{r,a}^{(m+1)}/
\lambda_{r,a}^{(m)}-1
)^2}
\leq\varepsilon$;
otherwise, set
$\lambda_{r,a}^{(m+1)}
= p\,\widetilde\lambda_{r,a}^{(m+1)}
/\sum_{b=1}^p\widetilde\lambda_{r,b}^{(m+1)}$,
$a=1,\ldots,p$, and repeat.

\State Set $\widehat\lambda_{r,a}=\lambda_{r,a}^{(m)}$
upon convergence and form
$\widehat{\mathbf K}_r^{\mathrm{cal}}
=\widehat{\mathbf U}_r
\operatorname{diag}(\widehat\lambda_{r,1},\ldots,
\widehat\lambda_{r,p})
\widehat{\mathbf U}_r^\top$.

\State Repeat for the column mode, replacing $p$ with $q$ and
$\widehat{\mathbf u}_{r,a}^{\top}(\mathbf X_i-\mathbf X_j)$ with
$(\mathbf X_i-\mathbf X_j)\widehat{\mathbf u}_{c,a}$.

\Ensure The row and column eigenvalue-calibrated marginal Kendall's tau matrices
$\widehat{\mathbf K}_r^{\mathrm{cal}}$, $\widehat{\mathbf K}_c^{\mathrm{cal}}$.
\end{algorithmic}
\end{algorithm}

\subsection{Structural dimension selection}
\label{subsec:mesir_model_selection}

In the preceding sections, we have assumed that the structural dimensions \((d_1,d_2)\) are known. In practice, both \(d_1\) and \(d_2\) are unknown and must be estimated from the data. We therefore introduce an eigenvalue-ratio criterion for selecting the row and column structural dimensions.

As established in Section~\ref{subsec:mesir}, the leading eigenspaces of the inverse-regression Kendall's tau matrices \(\mathbf K_r^{\mathrm{IR}}\) and \(\mathbf K_c^{\mathrm{IR}}\) recover \(\operatorname{Span}(\boldsymbol{\Sigma}\mathbf B_1)\) and \(\operatorname{Span}(\boldsymbol{\Omega}\mathbf B_2)\), respectively. Accordingly, \(d_1\) and \(d_2\) are the numbers of nonzero eigenvalues of these two population matrices, while the remaining eigenvalues are zero. Let \(\widehat\lambda^{(r)}_1\ge\cdots\ge\widehat\lambda^{(r)}_p\) and \(\widehat\lambda^{(c)}_1\ge\cdots\ge\widehat\lambda^{(c)}_q\) denote the ordered eigenvalues of the sample inverse-regression Kendall's tau matrices \(\widehat{\mathbf K}_{r}^{\mathrm{IR}}\) and \(\widehat{\mathbf K}_{c}^{\mathrm{IR}}\), respectively. Motivated by existing eigenvalue-ratio methods for dimension determination \citep{ahn2013eigenvalue,he2025new}, we estimate \(d_1\) and \(d_2\) from adjacent eigenvalue ratios. Since adjacent sample eigenvalues beyond the true structural dimensions may both approach zero, their ratios can be unstable. We therefore introduce ridge terms to stabilize these ratios \citep{YU2019104543,ZHU2020106910}. Specifically, define
$$
a_{n,r}=p^{3/2}q^{1/2}H^{-1}+pH^{1/2}n^{-1/2}, \qquad a_{n,c}=q^{3/2}p^{1/2}H^{-1}+qH^{1/2}n^{-1/2},
$$
and let \(\tau_{n,r}=a_{n,r}^{\gamma}\) and \(\tau_{n,c}=a_{n,c}^{\gamma}\) for some \(0<\gamma<1\). For prespecified maximum row and column dimensions \(d_{1,\max}<p\) and \(d_{2,\max}<q\), respectively, we define
\begin{equation}
\widehat d_1
=
\argmax_{1\le j\le d_{1,\max}}
\frac{\widehat\lambda^{(r)}_j+\tau_{n,r}}
{\widehat\lambda^{(r)}_{j+1}+\tau_{n,r}},
\qquad
\widehat d_2
=
\argmax_{1\le j\le d_{2,\max}}
\frac{\widehat\lambda^{(c)}_j+\tau_{n,c}}
{\widehat\lambda^{(c)}_{j+1}+\tau_{n,c}}.
\label{eq:mesir_dimension_ratio}
\end{equation}
The quantities $a_{n,r}$ and $a_{n,c}$ denote the operator-norm error rates for the row and column inverse-regression Kendall's tau matrices, respectively, as established in Section~\ref{section-MESIR Teory}. When $a_{n,r},a_{n,c}\to0$, the corresponding ridge terms \(\tau_{n,r}\) and \(\tau_{n,c}\) also vanish, but at slower rates than these error rates. They therefore stabilize ratios associated with zero population eigenvalues while remaining asymptotically negligible relative to the nonzero eigenvalues, allowing the ratio at the boundary between the nonzero and zero spectra to be distinguished from the remaining ratios.

Given the estimated dimensions \((\widehat d_1,\widehat d_2)\), we apply Algorithm~\ref{alg:mesir} to estimate the reduction subspaces. Specifically, we take the leading \(\widehat d_1\) and \(\widehat d_2\) eigenvectors of \(\widehat{\mathbf K}_{r}^{\mathrm{IR}}\) and \(\widehat{\mathbf K}_{c}^{\mathrm{IR}}\) to form \(\widehat{\boldsymbol\Gamma}_1\) and \(\widehat{\boldsymbol\Gamma}_2\), respectively, and then estimate the basis matrices for the row and column reduction subspaces \(\widehat{\mathbf B}_1=(\widehat{\mathbf K}_r^{\mathrm{cal}})^{-1}\widehat{\boldsymbol\Gamma}_1\) and \(\widehat{\mathbf B}_2=(\widehat{\mathbf K}_c^{\mathrm{cal}})^{-1}\widehat{\boldsymbol\Gamma}_2\).

\paragraph{Extension to tensor-valued predictors.}

The proposed framework also extends to tensor-valued predictors by exploiting their mode-wise structure. Specifically, the row--column formulation for matrix-valued predictors is generalized to multiple modes through tensor unfoldings, yielding a reduction subspace for each mode and a corresponding tensor-product representation of the CTS. The resulting tensor elliptical sliced inverse regression (TESIR) procedure, together with its theoretical properties, including convergence rates and subspace consistency, is presented in Appendix~\ref{subsec:tensor_mesir}.

\section{Theoretical Analysis}  
\label{section-MESIR Teory}
In this section, we establish the asymptotic properties of MESIR while allowing the predictor dimensions to diverge. The analysis concerns the sample inverse-regression
and eigenvalue-calibrated marginal Kendall's tau matrices, the recovery of the row and column reduction subspaces, and the consistency of structural dimension selection. For theoretical convenience, we assume that \(n=Hl\) and that each slice contains exactly \(l\) observations, with both \(H\) and \(l\) allowed to diverge with \(n\). We present the analysis for the row mode; the corresponding results for the column mode follow analogously.

Denote the inverse regression curve by $\mathbf{m}(Y)
=
\mathbb{E}(\mathbf{X}\mid Y)
\in
\mathbb{R}^{p\times q}$. We write
\(
\mathbf{X}
=
\mathbf{m}(Y)+\boldsymbol{\varepsilon},
\)
where \(\boldsymbol\varepsilon\) is the residual matrix. For the sample \(\{(\mathbf X_i,Y_i)\}_{i=1}^n\), let
\[
\mathbf{X}_i
=
\mathbf{m}(Y_i)+\boldsymbol{\varepsilon}_i
=: \mathbf{m}_i+\boldsymbol{\varepsilon}_i,
\qquad
i=1,\ldots,n.
\]

We first establish convergence rates for the sample inverse-regression
and eigenvalue-calibrated marginal Kendall's tau matrices in
Propositions~\ref{prop:Kxs_error_main} and~\ref{cr:thm:Cr}, respectively.
Combining these results, we derive the convergence rate in
Theorem~\ref{thm:mesir_consistency_main} and establish consistency
of the estimated reduction subspaces in
Corollary~\ref{cor:mesir_consistency_row}. Finally, Theorem~\ref{thm:mesir_dimension_selection_consistency} establishes consistency of the structural dimension estimators. The following regularity conditions are imposed throughout the analysis.

\begin{assumption}
\label{assump1}
Let
\(
\mathbf{M}_s
=
\mathbb{E}(\mathbf{X}\mid Y\in J_s),
s=1,\ldots,H,
\)
and for observations $\mathbf{X}_{sj}$ in slice $J_s$, write
\(
\mathbf{m}_{sj}
=
\mathbb{E}(\mathbf{X}\mid Y=Y_{sj}).
\)
Assume that the inverse regression curve satisfies the following
slice-separation and local regularity conditions:
\[
\max_{s\neq s'}
\frac{1}{\|\mathbf{M}_s-\mathbf{M}_{s'}\|_F}
<\infty,
\qquad
\max_{s\neq s'}
\max_{1\le j,k\le l}
\frac{
\|\mathbf{m}_{sj}-\mathbf{m}_{s'k}\|_F
}{
\|\mathbf{M}_s-\mathbf{M}_{s'}\|_F
}
<\infty,
\]
and
\[
\max_s
\max_{1\le j,k\le l}
|Y_{sk}-Y_{sj}|
=
O_p(H^{-1}).
\]
\end{assumption}

\begin{assumption}
\label{assump2}
Let \(\mathbf m(y)=\mathbb E(\mathbf X\mid Y=y)=\bigl(m_{ab}(y)\bigr)_{1\le a\le p,\;1\le b\le q}\). Assume that each \(m_{ab}(y)\) is continuously differentiable in \(y\), with
\[
\sup_{1\le a\le p,\;1\le b\le q}
\sup_y
\left|
\frac{d}{dy}m_{ab}(y)
\right|
=
O(1).
\]
\end{assumption}

\begin{assumption}
\label{assump:bounded_K}
Let $\lambda_j^{(r)}=\lambda_j(\mathbf K_r^{\mathrm{IR}})$
and $\lambda_j^{(c)}=\lambda_j(\mathbf K_c^{\mathrm{IR}})$
denote the ordered eigenvalues.
Assume that there exist constants
$0<c_r<C_r<\infty$ and $0<c_c<C_c<\infty$ such that
\(c_r \le \lambda_{d_1}^{(r)}
\le \cdots \le \lambda_1^{(r)} \le C_r
\) and 
\(
c_c \le \lambda_{d_2}^{(c)}
\le \cdots \le \lambda_1^{(c)} \le C_c.\)
\end{assumption}

\begin{assumption}
\label{assump3}
Assume that \(\mathbf X\sim\mathcal{MED}_{p\times q}(\mathbf M,\boldsymbol{\Sigma}\otimes\boldsymbol{\Omega},\varphi)\) and that, for some constant \(m\ge2\),
$$
\sup_{1\le a\le p,\;1\le b\le q}\mathbb E|X_{ab}|^m<\infty.
$$
\end{assumption}

\begin{assumption}
\label{assump4}
Assume that there exist constants
$0<c_{\Sigma}\le C_{\Sigma}<\infty$ and
$0<c_{\Omega}\le C_{\Omega}<\infty$
such that
\(
c_{\Sigma}\le
\lambda_p(\boldsymbol{\Sigma})
\le\lambda_1(\boldsymbol{\Sigma})
\le C_{\Sigma},
\)
and
\(
c_{\Omega}\le
\lambda_q(\boldsymbol{\Omega})
\le\lambda_1(\boldsymbol{\Omega})
\le C_{\Omega}.
\)
\end{assumption}

Assumptions~\ref{assump1}--\ref{assump2} control the separation between slices and the local variation of the inverse regression curve, in the spirit of regularity conditions commonly used in sliced inverse regression \citep{zhu2006sliced,ding2015tensor,chen2022high}. Assumption~\ref{assump:bounded_K} provides sufficient signal strength for identifying the informative eigenspace and consistently selecting the structural dimension \citep{yu2015useful}. Assumption~\ref{assump3} requires only finite entrywise \(m\)th moments for some \(m\ge2\), in line with elliptical SIR \citep{chen2022high}. Classical SIR theory typically requires fourth or higher-order moments \citep{hsing1992asymptotic,zhu1995asymptotics,zhu2006sliced}, while \citet{ding2015tensor} imposes a higher-order moment condition on the vectorized predictor in matrix sufficient dimension reduction. This weaker requirement is facilitated by the elliptical structure and the normalized Kendall's tau construction. Finally, Assumption~\ref{assump4} provides the spectral regularity needed for stable recovery of the row reduction subspace \citep{Han02012018}.

We first quantify the estimation error of the sample row inverse-regression Kendall's tau matrix \(\widehat{\mathbf K}_{r}^{\mathrm{IR}}\), constructed from the slice means \(\{\widehat{\mathbf M}_s\}_{s=1}^H\), relative to its population counterpart \(\mathbf K_r^{\mathrm{IR}}\).

\begin{proposition}
\label{prop:Kxs_error_main}
Under Assumptions~\ref{assump1}--\ref{assump3}, if \(p=o(n^{1/2})\), then
\[
\left\|\widehat{\mathbf K}_{r}^{\mathrm{IR}}
-
\mathbf K_{r}^{\mathrm{IR}}\right\|_2
=
O_p(p^{3/2}q^{1/2}H^{-1}+pH^{1/2}n^{-1/2}).
\]
\end{proposition}

Proposition~\ref{prop:Kxs_error_main} reveals two sources of estimation error. The term \(p^{3/2}q^{1/2}H^{-1}\) arises from replacing the inverse regression curve \(\mathbf m(Y)\) by the population slice means \(\{\mathbf M_s\}_{s=1}^H\), whereas \(pH^{1/2}n^{-1/2}=p/\sqrt l\) arises from estimating \(\mathbf M_s\) by the sample slice means \(\widehat{\mathbf M}_s\). Increasing \(H\) reduces the slicing approximation error at the cost of larger sampling error, leading to a trade-off between the two terms \citep{chen2022high,ding2015tensor}. When the column dimension \(q=1\), Proposition~\ref{prop:Kxs_error_main} reduces to the vector-valued setting considered by \citet{chen2022high}, yielding the convergence rate $O_p\!\left(p^{3/2}H^{-1}+pH^{1/2}n^{-1/2}\right)$.

We next establish the convergence rate of the calibrated marginal
Kendall's tau matrix $\widehat{\mathbf K}_r^{\mathrm{cal}}$ to the row
scatter shape $\boldsymbol\Sigma$.
\begin{proposition}\label{cr:thm:Cr}
Suppose that
\(\mathbf X\sim\mathcal{MED}_{p\times q}
(\mathbf M,\boldsymbol{\Sigma}\otimes\boldsymbol{\Omega},\varphi)\),
Assumption~\ref{assump4} holds, and \(p=o(n^{1/2})\).
Then, with probability tending to one,
\eqref{eq:mesir_empirical_calibration} admits a unique positive
solution. The resulting eigenvalue-calibrated marginal Kendall's tau matrix
$\widehat{\mathbf K}_r^{\mathrm{cal}}$ satisfies
\[
 \op{\widehat{\mathbf K}_r^{\mathrm{cal}}-\boldsymbol{\Sigma}}
 =O_p(pn^{-1/2}).
\]
\end{proposition}

We know that the row reduction subspace is identified through \(\boldsymbol{\Sigma}^{-1}\mathbf K_r^{\mathrm{IR}}\). Accordingly, \((\widehat{\mathbf K}_r^{\mathrm{cal}})^{-1}\widehat{\mathbf K}_r^{\mathrm{IR}}\) serves as its sample counterpart. The following theorem establishes the convergence rate of this estimator.

\begin{theorem}
\label{thm:mesir_consistency_main}
Under the conditions of Proposition~\ref{prop:Kxs_error_main} and ~\ref{cr:thm:Cr}, we have
\[
\left\|
(\widehat{\mathbf K}_r^{\mathrm{cal}})^{-1}\widehat{\mathbf K}_{r}^{\mathrm{IR}}
-
\boldsymbol{\Sigma}^{-1}\mathbf K_{r}^{\mathrm{IR}}
\right\|_2
=
O_p(p^{3/2}q^{1/2}H^{-1})
+
O_p\!\left(
pH^{1/2}n^{-1/2}
\right).
\]
\end{theorem}
The calibration error contributes an additional
$O_p(pn^{-1/2})$ term, which is absorbed into
$O_p(pH^{1/2}n^{-1/2})$ in the displayed rate.

For \(j=1,2\), let \(\mathbf P_{\mathbf B_j}\) and \(\mathbf P_{\widehat{\mathbf B}_j}\) denote the orthogonal projections onto the true and estimated reduction subspaces, respectively. Combining Theorem~\ref{thm:mesir_consistency_main} with a variant of the Davis--Kahan theorem \citep{vu2013minimax} yields the following corollary.

\begin{corollary}
\label{cor:mesir_consistency_row}
Under the conditions of Proposition~\ref{prop:Kxs_error_main} and ~\ref{cr:thm:Cr}, we have  
\[
\left\|
\mathbf P_{\widehat{\mathbf B}_1}
-
\mathbf P_{\mathbf B_1}
\right\|_2
=
O_p\left(
p^{3/2}q^{1/2}H^{-1}
+
pH^{1/2}n^{-1/2}
\right).
\]
The column-mode result follows analogously by interchanging \(p\) and \(q\).
\end{corollary}

Since the projection matrix onto the CTS can be written as \(\mathbf P_{\mathcal S_{Y|X}}
=
\mathbf P_{\mathbf B_2}\otimes \mathbf P_{\mathbf B_1},\) Corollary~\ref{cor:mesir_consistency_row} and its column-mode counterpart, together with the continuity of the Kronecker product, imply that
\[
\begin{aligned}
\left\|
\mathbf P_{\widehat{\mathcal S}_{Y\mid\mathbf X}}
-\mathbf P_{\mathcal S_{Y\mid\mathbf X}}\right\|_2
\xrightarrow{p}0.
\end{aligned}
\]

The preceding results establish consistency of the estimated CTS when \(d_1\) and \(d_2\) are known. We next establish consistency of the eigenvalue-ratio estimators in Section~\ref{subsec:mesir_model_selection} when the structural dimensions are unknown.

\begin{theorem}
\label{thm:mesir_dimension_selection_consistency}
Suppose that Assumptions~\ref{assump1}--\ref{assump3} hold and that
\(a_{n,r}=o(1)\) and \(a_{n,c}=o(1)\).
Let \(\widehat d_1\) and \(\widehat d_2\) be defined as in
\eqref{eq:mesir_dimension_ratio}. Then
\[
\mathbb P(\widehat d_1=d_1)\to1,
\qquad
\mathbb P(\widehat d_2=d_2)\to1.
\]
\end{theorem}

Here \(a_{n,r}\) and \(a_{n,c}\), defined in Section~\ref{subsec:mesir_model_selection}, quantify the operator-norm estimation errors of the row and column inverse-regression Kendall's tau matrices. Since the ridge terms dominate these estimation errors while vanishing asymptotically, the eigenvalue ratio at the boundary between the nonzero and zero spectra can be consistently distinguished from the remaining ratios.

\section{Simulation Studies}
\label{sec:MESIR_simulation}
The simulation study evaluates the performance of MESIR under different tail behaviours, focusing on the recovery of the row and column dimension reduction subspaces and the CTS, as well as the impact of estimating the structural dimensions. We consider the following methods: (1) \emph{MESIR}, implemented with the true structural dimensions; (2) \emph{MESIR-Auto}, which estimates \(d_1\) and \(d_2\) using the procedure introduced in Section~\ref{subsec:mesir_model_selection}; (3) tensor sliced inverse regression (\emph{TSIR}) of \citet{ding2015tensor}, implemented with the true structural dimensions; and (4) elliptical sliced inverse regression (\emph{Vector ESIR}) of \citet{chen2022high}, applied to the vectorized predictor with the reduction dimension set to \(d_1d_2\). MESIR, MESIR-Auto, and TSIR directly estimate the row and column reduction subspaces by exploiting the matrix structure, whereas Vector ESIR is evaluated only for recovery of the CTS.

We assess subspace recovery using two metrics: the squared multiple correlation coefficient \(R_k^2\) \citep{li1991sliced,chen2022high} and the projection distance \(D_k\) \citep{zeng2024subspace}. Let \(k=1,2\) index the row and column modes, respectively, with \((p_1,p_2)=(p,q)\) and \((\boldsymbol\Sigma_1,\boldsymbol\Sigma_2)=(\boldsymbol\Sigma,\boldsymbol\Omega)\).
% For notational convenience, both metrics are defined in a unified \(K\)-mode framework, with \(K=2\) for matrix predictors and \(K=3\) for third-order tensor predictors. 
For mode \(k\), let \(\mathcal B_k\subseteq\mathbb R^{p_k}\) denote the true reduction subspace with \(\dim(\mathcal B_k)=d_k\), and let \(\widehat{\mathcal B}_k=\operatorname{Span}(\widehat{\mathbf B}_k)=\operatorname{Span}\{\widehat{\boldsymbol\gamma}_k^{(1)},\ldots,\widehat{\boldsymbol\gamma}_k^{(\widehat d_k)}\}\) denote its estimate, where \(\widehat{\boldsymbol\gamma}_k^{(j)}\in\mathbb R^{p_k}\). For each estimated direction \(\widehat{\boldsymbol\gamma}_k^{(j)}\), the squared multiple correlation \(R^2\{\widehat{\boldsymbol\gamma}_k^{(j)}\}\) measures its alignment with the true subspace \(\mathcal B_k\). We summarize the recovery of the mode-\(k\) reduction subspace by averaging these directional correlations over the \(d_k\) directions, yielding
$$
R_k^2
=
\frac{1}{d_k}
\sum_{j=1}^{d_k}
R^2\left(\hat{\boldsymbol\gamma}_k^{(j)}\right),
\quad R^2\left(\hat{\boldsymbol\gamma}_k^{(j)}\right)
=
\max_{\boldsymbol\beta\in
\mathcal B_k\setminus\{\boldsymbol 0\}}
\frac{
\left[
\left(\hat{\boldsymbol\gamma}_k^{(j)}\right)^\top
{\boldsymbol{\Sigma}}_k\boldsymbol\beta
\right]^2
}{
\left[
\left(\hat{\boldsymbol\gamma}_k^{(j)}\right)^\top
\boldsymbol{\Sigma}_k\hat{\boldsymbol\gamma}_k^{(j)}
\right]
\left(\boldsymbol\beta^\top{\boldsymbol{\Sigma}}_k\boldsymbol\beta\right)
}.
$$

The projection distance \(D_k\) measures the discrepancy between the estimated subspace \(\widehat{\mathcal B}_k\) and the true subspace \(\mathcal B_k\) through their orthogonal projection matrices, and is defined as
\begin{equation*}
D_k
=
\frac{
\|
{\mathbf P}_{\mathcal B_k}-{\mathbf P}_{\widehat{\mathcal B}_k}
\|_F
}{
\sqrt{2d_k}
},
\qquad k=1,\ldots,K,
\end{equation*}
where \(\mathbf P_{\mathcal B_k}\) and \(\mathbf P_{\widehat{\mathcal B}_k}\) denote the orthogonal projections onto \(\mathcal B_k\) and \(\widehat{\mathcal B}_k\), respectively. Larger \(R_k^2\) and smaller \(D_k\) indicate more accurate subspace recovery. When \(\widehat d_k=d_k\), both metrices lie in \([0,1]\).
% {\color{red}For matrix predictors, \(K=2\), with \((p_1,p_2)=(p,q)\) and \((\boldsymbol\Sigma_1,\boldsymbol\Sigma_2)=(\boldsymbol\Sigma,\boldsymbol\Omega)\) corresponding to the row and column modes.} For tensor predictors, the same notation is used for each mode. 
In addition to mode-wise subspace recovery, we evaluate recovery of the CTS. Let
\(
\mathcal B
=
\operatorname{Span}(\mathbf B_K\otimes\cdots\otimes \mathbf B_1)
\)
and
\(
\widehat{\mathcal B}
=
\operatorname{Span}(\widehat{\mathbf B}_K\otimes\cdots\otimes\widehat{\mathbf B}_1)
\) denote the true and estimated CTSs, respectively. The \(R^2\) and \(D\) are defined analogously using \(\mathcal B\) and \(\widehat{\mathcal B}\).

\paragraph{Simulation settings.} We generate the predictor matrices independently according to
\[
{\mathbf X}=\xi^{1/2}{\mathbf U},\qquad
\operatorname{vec}({\mathbf U})\sim N_{pq}({\mathbf0},{\mathbf I}_{pq}),
\]
where \(\xi\) is independent of \(\mathbf U\). The row and column scatter matrices are set to $(\boldsymbol\Sigma,\boldsymbol\Omega)=(\mathbf I_p,\mathbf I_q)$. Setting \(\xi=1\) yields the matrix normal distribution, while \(\xi=\nu/W\) with \(W\sim\chi_\nu^2\) yields a matrix \(t\) distribution with \(\nu\) degrees of freedom. To examine performance across different tail behaviors, we consider the normal, $t_3$,
$t_2$, and Cauchy ($\nu=1$) distributions.

Throughout the matrix experiments, we set \(p=q=20\) and consider sample sizes \(n\in\{500,1000\}\). In all models, the response error is \(\sigma\varepsilon\), where \(\varepsilon\sim N(0,1)\) is independent of \(\mathbf X\) and \(\sigma=0.05\). We use \(H=5\) response slices of approximately equal size. We construct nonlinear response models using selected entries of the matrix predictor. For Models A1--A3, let \(\mathbf B_1=\mathbf B_2=\mathbf e_1\) and \(t=\mathbf B_1^\top\mathbf X\mathbf B_2=X_{11}\), where $X_{11}$ denotes the $(1,1)$ entry of $\mathbf X$.
We consider
\begin{align*}
\text{A1:}\quad
Y&=\frac{t}{1+\exp\{8(|t|-9.5)\}}+\sigma\varepsilon,
\\
\text{A2:}\quad
Y&=t\Phi\left(\frac{9-|t|}{0.5}\right)+\sigma\varepsilon,
\\
\text{A3:}\quad
Y&=\operatorname{sgn}(t)|t|^{3/4}
\left[\frac12+\frac{\arctan\{10(9.5-|t|)\}}{\pi}\right]
+\sigma\varepsilon.
\end{align*}
Here \(\Phi\) denotes the standard normal distribution function, and \(\mathbf e_j\) denotes the \(j\)th standard basis vector. The three models share the same one-dimensional reduction structure but differ in the nonlinear relationship between \(Y\) and \(t\). The true row and column reduction subspaces are both \(\operatorname{Span}(\mathbf e_1)\), so that \(d_1=d_2=1\).

For Models B1--B3, let \(\mathbf B_1=\mathbf B_2=(\mathbf e_1,\mathbf e_2)\) and \(\mathbf T=\mathbf B_1^\top\mathbf X\mathbf B_2\). Define
\[
r=\|{\mathbf T}\|_F,\quad
a_1=T_{11}-T_{12}+T_{21}+T_{22},\quad
a_2=T_{11}-T_{12}-T_{21}-T_{22},\quad
u=\frac{|a_1|}{|a_1|+|a_2|}.
\]
We consider
\begin{align*}
\text{B1:}\quad
Y&=r\operatorname{sgn}(a_2)
\left[1+40\left\{\frac12+
\frac{\arctan\{80(u-0.70)\}}{\pi}\right\}\right]
+\sigma\varepsilon,
\\
\text{B2:}\quad
Y&=r\operatorname{sgn}(a_2)(1+80u^8)+\sigma\varepsilon,
\\
\text{B3:}\quad
Y&=r\operatorname{sgn}(a_2)\exp(5u^4)+\sigma\varepsilon.
\end{align*}
The three models share the same two-dimensional reduction structure but differ in the nonlinear relationship between \(Y\) and \(\mathbf T\). The true row and column reduction subspaces are both \(\operatorname{Span}(\mathbf e_1,\mathbf e_2)\), so that \(d_1=d_2=2\). 
% MESIR and TSIR are implemented using the true structural dimensions, whereas MESIR-Auto uses the dimensions selected from the data. For Vector ESIR, we retain \(d_1d_2\) directions for the CTS.

\begin{table}[!htbp]
\centering
\caption{Row subspace recovery for Models A1--A3.}
\label{tab:A1_A3_row_compare}
\footnotesize
\setlength{\tabcolsep}{5pt}
\renewcommand{\arraystretch}{1}
\begin{tabular}{lc|ccc|ccc}
\toprule
Model & Metric & \multicolumn{3}{c|}{$n=500$} & \multicolumn{3}{c}{$n=1000$} \\
& & TSIR & MESIR & MESIR-Auto & TSIR & MESIR & MESIR-Auto \\
\midrule
\multicolumn{2}{c}{} & \multicolumn{6}{c}{\textbf{Normal Distribution}} \\
\midrule
\multirow{2}{*}{A1} & $R_1^2$ & \textbf{0.958 (0.014)} & 0.949 (0.018) & 0.949 (0.018) & \textbf{0.978 (0.007)} & 0.976 (0.009) & 0.976 (0.009) \\
 & $D_1$ & \textbf{0.203 (0.033)} & 0.221 (0.040) & 0.221 (0.040) & \textbf{0.145 (0.025)} & 0.151 (0.028) & 0.151 (0.028) \\
\multirow{2}{*}{A2} & $R_1^2$ & \textbf{0.957 (0.012)} & 0.949 (0.015) & 0.949 (0.015) & \textbf{0.979 (0.006)} & 0.976 (0.008) & 0.976 (0.008) \\
 & $D_1$ & \textbf{0.205 (0.029)} & 0.223 (0.033) & 0.223 (0.033) & \textbf{0.142 (0.021)} & 0.153 (0.026) & 0.153 (0.026) \\
\multirow{2}{*}{A3} & $R_1^2$ & \textbf{0.958 (0.013)} & 0.949 (0.016) & 0.949 (0.016) & \textbf{0.980 (0.006)} & 0.977 (0.007) & 0.977 (0.007) \\
 & $D_1$ & \textbf{0.202 (0.032)} & 0.224 (0.035) & 0.224 (0.035) & \textbf{0.141 (0.021)} & 0.150 (0.022) & 0.150 (0.022) \\
\midrule
\multicolumn{2}{c}{} & \multicolumn{6}{c}{\textbf{$t_3$ Distribution}} \\
\midrule
\multirow{2}{*}{A1} & $R_1^2$ & 0.877 (0.152) & \textbf{0.905 (0.030)} & \textbf{0.905 (0.030)} & 0.938 (0.098) & \textbf{0.954 (0.016)} & \textbf{0.954 (0.016)} \\
 & $D_1$ & 0.325 (0.131) & \textbf{0.304 (0.049)} & \textbf{0.304 (0.049)} & 0.231 (0.092) & \textbf{0.211 (0.037)} & \textbf{0.211 (0.037)} \\
\multirow{2}{*}{A2} & $R_1^2$ & 0.895 (0.087) & \textbf{0.904 (0.033)} & \textbf{0.904 (0.033)} & 0.949 (0.027) & \textbf{0.957 (0.015)} & \textbf{0.957 (0.015)} \\
 & $D_1$ & 0.313 (0.083) & \textbf{0.306 (0.051)} & \textbf{0.306 (0.051)} & 0.219 (0.052) & \textbf{0.206 (0.035)} & \textbf{0.206 (0.035)} \\
\multirow{2}{*}{A3} & $R_1^2$ & 0.903 (0.094) & \textbf{0.908 (0.032)} & \textbf{0.908 (0.032)} & 0.950 (0.019) & \textbf{0.956 (0.012)} & \textbf{0.956 (0.012)} \\
 & $D_1$ & \textbf{0.298 (0.088)} & 0.299 (0.052) & 0.299 (0.052) & 0.221 (0.040) & \textbf{0.207 (0.029)} & \textbf{0.207 (0.029)} \\
\midrule
\multicolumn{2}{c}{} & \multicolumn{6}{c}{\textbf{$t_2$ Distribution}} \\
\midrule
\multirow{2}{*}{A1} & $R_1^2$ & 0.652 (0.331) & \textbf{0.830 (0.066)} & 0.811 (0.117) & 0.774 (0.297) & \textbf{0.920 (0.032)} & \textbf{0.920 (0.032)} \\
 & $D_1$ & 0.536 (0.249) & \textbf{0.405 (0.077)} & 0.425 (0.129) & 0.413 (0.236) & \textbf{0.277 (0.053)} & \textbf{0.277 (0.053)} \\
\multirow{2}{*}{A2} & $R_1^2$ & 0.678 (0.323) & \textbf{0.835 (0.062)} & 0.816 (0.112) & 0.793 (0.280) & \textbf{0.925 (0.025)} & \textbf{0.925 (0.025)} \\
 & $D_1$ & 0.513 (0.244) & \textbf{0.400 (0.073)} & 0.419 (0.119) & 0.395 (0.226) & \textbf{0.271 (0.045)} & \textbf{0.271 (0.045)} \\
\multirow{2}{*}{A3} & $R_1^2$ & 0.643 (0.336) & \textbf{0.829 (0.069)} & 0.811 (0.125) & 0.750 (0.308) & \textbf{0.926 (0.027)} & \textbf{0.926 (0.027)} \\
 & $D_1$ & 0.543 (0.251) & \textbf{0.406 (0.080)} & 0.429 (0.155) & 0.435 (0.247) & \textbf{0.268 (0.051)} & \textbf{0.268 (0.051)} \\
\midrule
\multicolumn{2}{c}{} & \multicolumn{6}{c}{\textbf{Cauchy Distribution}} \\
\midrule
\multirow{2}{*}{A1} & $R_1^2$ & 0.043 (0.049) & \textbf{0.270 (0.214)} & 0.237 (0.170) & 0.056 (0.069) & \textbf{0.574 (0.224)} & 0.468 (0.250) \\
 & $D_1$ & 0.978 (0.025) & \textbf{0.844 (0.135)} & 1.072 (0.302) & 0.971 (0.037) & \textbf{0.633 (0.162)} & 0.806 (0.305) \\
\multirow{2}{*}{A2} & $R_1^2$ & 0.048 (0.057) & \textbf{0.334 (0.225)} & 0.254 (0.186) & 0.044 (0.055) & \textbf{0.565 (0.259)} & 0.527 (0.267) \\
 & $D_1$ & 0.975 (0.030) & \textbf{0.803 (0.145)} & 1.058 (0.291) & 0.977 (0.029) & \textbf{0.632 (0.189)} & 0.754 (0.338) \\
\multirow{2}{*}{A3} & $R_1^2$ & 0.039 (0.055) & \textbf{0.318 (0.244)} & 0.276 (0.203) & 0.055 (0.072) & \textbf{0.569 (0.237)} & 0.503 (0.245) \\
 & $D_1$ & 0.980 (0.029) & \textbf{0.811 (0.155)} & 1.019 (0.296) & 0.971 (0.039) & \textbf{0.634 (0.172)} & 0.780 (0.326) \\
\bottomrule
\end{tabular}
\end{table}

\begin{table}[!htbp]
\centering
\caption{Row subspace recovery for Models B1--B3.}
\label{tab:B1_B3_row_compare}
\footnotesize
\setlength{\tabcolsep}{5pt}
\renewcommand{\arraystretch}{1}
\begin{tabular}{lc|ccc|ccc}
\toprule
Model & Metric & \multicolumn{3}{c|}{$n=500$} & \multicolumn{3}{c}{$n=1000$} \\
& & TSIR & MESIR & MESIR-Auto & TSIR & MESIR & MESIR-Auto \\
\midrule
\multicolumn{2}{c}{} & \multicolumn{6}{c}{\textbf{Normal Distribution}} \\
\midrule
\multirow{2}{*}{B1} & $R_1^2$ & \textbf{0.878 (0.028)} & 0.815 (0.048) & 0.813 (0.088) & \textbf{0.940 (0.013)} & 0.920 (0.016) & 0.920 (0.016) \\
 & $D_1$ & \textbf{0.347 (0.041)} & 0.426 (0.055) & 0.468 (0.099) & \textbf{0.244 (0.027)} & 0.282 (0.029) & 0.282 (0.029) \\
\multirow{2}{*}{B2} & $R_1^2$ & \textbf{0.878 (0.028)} & 0.824 (0.048) & 0.823 (0.073) & \textbf{0.943 (0.014)} & 0.925 (0.018) & 0.925 (0.018) \\
 & $D_1$ & \textbf{0.347 (0.040)} & 0.416 (0.056) & 0.443 (0.090) & \textbf{0.238 (0.029)} & 0.271 (0.032) & 0.271 (0.032) \\
\multirow{2}{*}{B3} & $R_1^2$ & \textbf{0.878 (0.031)} & 0.827 (0.046) & 0.821 (0.076) & \textbf{0.940 (0.012)} & 0.922 (0.018) & 0.922 (0.018) \\
 & $D_1$ & \textbf{0.346 (0.043)} & 0.412 (0.053) & 0.446 (0.090) & \textbf{0.243 (0.025)} & 0.278 (0.031) & 0.278 (0.031) \\
\midrule
\multicolumn{2}{c}{} & \multicolumn{6}{c}{\textbf{$t_3$ Distribution}} \\
\midrule
\multirow{2}{*}{B1} & $R_1^2$ & 0.668 (0.188) & \textbf{0.720 (0.106)} & 0.701 (0.145) & 0.878 (0.096) & \textbf{0.889 (0.027)} & 0.883 (0.048) \\
 & $D_1$ & 0.557 (0.148) & \textbf{0.521 (0.091)} & 0.574 (0.131) & 0.336 (0.094) & \textbf{0.331 (0.040)} & 0.341 (0.061) \\
\multirow{2}{*}{B2} & $R_1^2$ & 0.694 (0.168) & \textbf{0.733 (0.085)} & 0.714 (0.147) & \textbf{0.893 (0.037)} & 0.888 (0.025) & 0.889 (0.025) \\
 & $D_1$ & 0.534 (0.143) & \textbf{0.511 (0.078)} & 0.563 (0.137) & \textbf{0.323 (0.051)} & 0.333 (0.038) & 0.339 (0.052) \\
\multirow{2}{*}{B3} & $R_1^2$ & 0.628 (0.163) & \textbf{0.711 (0.095)} & 0.710 (0.122) & 0.870 (0.051) & 0.878 (0.035) & \textbf{0.879 (0.034)} \\
 & $D_1$ & 0.595 (0.133) & \textbf{0.531 (0.084)} & 0.563 (0.107) & 0.356 (0.064) & \textbf{0.346 (0.050)} & 0.351 (0.060) \\
\midrule
\multicolumn{2}{c}{} & \multicolumn{6}{c}{\textbf{$t_2$ Distribution}} \\
\midrule
\multirow{2}{*}{B1} & $R_1^2$ & 0.346 (0.188) & 0.606 (0.137) & \textbf{0.608 (0.181)} & 0.571 (0.213) & \textbf{0.827 (0.067)} & 0.812 (0.109) \\
 & $D_1$ & 0.799 (0.126) & \textbf{0.618 (0.108)} & 0.664 (0.143) & 0.633 (0.170) & \textbf{0.409 (0.072)} & 0.447 (0.119) \\
\multirow{2}{*}{B2} & $R_1^2$ & 0.349 (0.141) & \textbf{0.618 (0.111)} & 0.607 (0.161) & 0.625 (0.206) & \textbf{0.829 (0.050)} & 0.827 (0.059) \\
 & $D_1$ & 0.802 (0.093) & \textbf{0.612 (0.088)} & 0.670 (0.126) & 0.590 (0.165) & \textbf{0.409 (0.060)} & 0.436 (0.084) \\
\multirow{2}{*}{B3} & $R_1^2$ & 0.345 (0.146) & 0.590 (0.136) & \textbf{0.591 (0.166)} & 0.578 (0.215) & \textbf{0.820 (0.068)} & 0.819 (0.080) \\
 & $D_1$ & 0.804 (0.094) & \textbf{0.632 (0.104)} & 0.678 (0.130) & 0.627 (0.169) & \textbf{0.418 (0.076)} & 0.439 (0.097) \\
\midrule
\multicolumn{2}{c}{} & \multicolumn{6}{c}{\textbf{Cauchy Distribution}} \\
\midrule
\multirow{2}{*}{B1} & $R_1^2$ & 0.145 (0.095) & \textbf{0.342 (0.126)} & \textbf{0.342 (0.139)} & 0.175 (0.104) & \textbf{0.578 (0.136)} & 0.551 (0.173) \\
 & $D_1$ & 0.923 (0.053) & \textbf{0.807 (0.081)} & 0.855 (0.137) & 0.907 (0.058) & \textbf{0.642 (0.104)} & 0.710 (0.140) \\
\multirow{2}{*}{B2} & $R_1^2$ & 0.138 (0.091) & \textbf{0.376 (0.133)} & 0.371 (0.154) & 0.170 (0.099) & \textbf{0.582 (0.138)} & 0.565 (0.164) \\
 & $D_1$ & 0.927 (0.050) & \textbf{0.786 (0.084)} & 0.823 (0.131) & 0.910 (0.056) & \textbf{0.638 (0.107)} & 0.703 (0.141) \\
\multirow{2}{*}{B3} & $R_1^2$ & 0.126 (0.070) & \textbf{0.355 (0.108)} & 0.352 (0.153) & 0.158 (0.098) & \textbf{0.577 (0.130)} & 0.570 (0.162) \\
 & $D_1$ & 0.934 (0.038) & \textbf{0.800 (0.068)} & 0.852 (0.136) & 0.916 (0.055) & \textbf{0.643 (0.099)} & 0.694 (0.133) \\
\bottomrule
\end{tabular}
\end{table}

% Source: A raw simulation CSV files, n=500 and n=1000; 100 successful repetitions per cell.
% Entries are mean (sample standard deviation). Bold: best displayed mean within n; ties at 3 decimals are all bold.
\begin{table}[!htbp]
\centering
\caption{CTS recovery for Models A1--A3.}
\label{tab:A1_A3_kron_compare}
\footnotesize
\setlength{\tabcolsep}{5pt}
\renewcommand{\arraystretch}{1}
\begin{tabular}{lc|ccc|ccc}
\toprule
Model & Metric & \multicolumn{3}{c|}{$n=500$} & \multicolumn{3}{c}{$n=1000$} \\
& & TSIR & MESIR & MESIR-Auto & TSIR & MESIR & MESIR-Auto \\
\midrule
\multicolumn{2}{c}{} & \multicolumn{6}{c}{\textbf{Normal Distribution}} \\
\midrule
\multirow{2}{*}{A1} & $R^2$ & \textbf{0.918 (0.017)} & 0.901 (0.022) & 0.901 (0.022) & \textbf{0.958 (0.010)} & 0.953 (0.011) & 0.953 (0.011) \\
 & $D$ & \textbf{0.286 (0.029)} & 0.312 (0.035) & 0.312 (0.035) & \textbf{0.204 (0.024)} & 0.216 (0.025) & 0.216 (0.025) \\
\multirow{2}{*}{A2} & $R^2$ & \textbf{0.917 (0.017)} & 0.902 (0.020) & 0.902 (0.020) & \textbf{0.959 (0.009)} & 0.953 (0.011) & 0.953 (0.011) \\
 & $D$ & \textbf{0.287 (0.029)} & 0.311 (0.032) & 0.311 (0.032) & \textbf{0.202 (0.023)} & 0.215 (0.025) & 0.215 (0.025) \\
\multirow{2}{*}{A3} & $R^2$ & \textbf{0.919 (0.017)} & 0.902 (0.021) & 0.902 (0.021) & \textbf{0.960 (0.009)} & 0.955 (0.011) & 0.955 (0.011) \\
 & $D$ & \textbf{0.283 (0.031)} & 0.310 (0.035) & 0.310 (0.035) & \textbf{0.199 (0.023)} & 0.212 (0.024) & 0.212 (0.024) \\
\midrule
\multicolumn{2}{c}{} & \multicolumn{6}{c}{\textbf{$t_3$ Distribution}} \\
\midrule
\multirow{2}{*}{A1} & $R^2$ & 0.789 (0.166) & \textbf{0.820 (0.044)} & \textbf{0.820 (0.044)} & 0.889 (0.101) & \textbf{0.911 (0.023)} & \textbf{0.911 (0.023)} \\
 & $D$ & 0.442 (0.126) & \textbf{0.421 (0.051)} & \textbf{0.421 (0.051)} & 0.321 (0.092) & \textbf{0.297 (0.038)} & \textbf{0.297 (0.038)} \\
\multirow{2}{*}{A2} & $R^2$ & 0.816 (0.104) & \textbf{0.826 (0.043)} & \textbf{0.826 (0.043)} & 0.901 (0.050) & \textbf{0.917 (0.019)} & \textbf{0.917 (0.019)} \\
 & $D$ & 0.420 (0.088) & \textbf{0.414 (0.051)} & \textbf{0.414 (0.051)} & 0.307 (0.067) & \textbf{0.286 (0.033)} & \textbf{0.286 (0.033)} \\
\multirow{2}{*}{A3} & $R^2$ & 0.812 (0.096) & \textbf{0.817 (0.046)} & \textbf{0.817 (0.046)} & 0.899 (0.035) & \textbf{0.912 (0.018)} & \textbf{0.912 (0.018)} \\
 & $D$ & 0.426 (0.082) & \textbf{0.424 (0.052)} & \textbf{0.424 (0.052)} & 0.313 (0.051) & \textbf{0.295 (0.031)} & \textbf{0.295 (0.031)} \\
\midrule
\multicolumn{2}{c}{} & \multicolumn{6}{c}{\textbf{$t_2$ Distribution}} \\
\midrule
\multirow{2}{*}{A1} & $R^2$ & 0.522 (0.305) & \textbf{0.691 (0.088)} & 0.667 (0.137) & 0.687 (0.282) & \textbf{0.847 (0.048)} & \textbf{0.847 (0.048)} \\
 & $D$ & 0.661 (0.205) & \textbf{0.551 (0.077)} & 0.581 (0.154) & 0.520 (0.207) & \textbf{0.387 (0.057)} & \textbf{0.387 (0.057)} \\
\multirow{2}{*}{A2} & $R^2$ & 0.560 (0.291) & \textbf{0.693 (0.090)} & 0.666 (0.145) & 0.706 (0.272) & \textbf{0.855 (0.042)} & \textbf{0.855 (0.042)} \\
 & $D$ & 0.634 (0.197) & \textbf{0.549 (0.078)} & 0.581 (0.152) & 0.504 (0.200) & \textbf{0.377 (0.054)} & \textbf{0.377 (0.054)} \\
\multirow{2}{*}{A3} & $R^2$ & 0.523 (0.307) & \textbf{0.688 (0.091)} & 0.663 (0.150) & 0.654 (0.304) & \textbf{0.852 (0.040)} & \textbf{0.852 (0.040)} \\
 & $D$ & 0.660 (0.204) & \textbf{0.553 (0.078)} & 0.593 (0.206) & 0.546 (0.220) & \textbf{0.381 (0.052)} & \textbf{0.381 (0.052)} \\
\midrule
\multicolumn{2}{c}{} & \multicolumn{6}{c}{\textbf{Cauchy Distribution}} \\
\midrule
\multirow{2}{*}{A1} & $R^2$ & 0.003 (0.005) & \textbf{0.103 (0.132)} & 0.071 (0.096) & 0.003 (0.009) & \textbf{0.361 (0.205)} & 0.271 (0.217) \\
 & $D$ & 0.999 (0.003) & \textbf{0.944 (0.076)} & 1.617 (0.678) & 0.998 (0.004) & \textbf{0.789 (0.130)} & 1.152 (0.517) \\
\multirow{2}{*}{A2} & $R^2$ & 0.002 (0.004) & \textbf{0.120 (0.142)} & 0.081 (0.112) & 0.002 (0.006) & \textbf{0.359 (0.228)} & 0.291 (0.226) \\
 & $D$ & 0.999 (0.002) & \textbf{0.935 (0.081)} & 1.587 (0.643) & 0.999 (0.003) & \textbf{0.787 (0.145)} & 1.151 (0.595) \\
\multirow{2}{*}{A3} & $R^2$ & 0.002 (0.005) & \textbf{0.121 (0.144)} & 0.083 (0.103) & 0.003 (0.006) & \textbf{0.352 (0.214)} & 0.280 (0.209) \\
 & $D$ & 0.999 (0.003) & \textbf{0.934 (0.082)} & 1.552 (0.598) & 0.999 (0.003) & \textbf{0.794 (0.133)} & 1.124 (0.519) \\
\bottomrule
\end{tabular}
\end{table}

\begin{table}[!htbp]
\centering
\caption{CTS recovery for Models B1--B3.}
\label{tab:B1_B3_kron_compare}
\footnotesize
\setlength{\tabcolsep}{5pt}
\renewcommand{\arraystretch}{1}
\begin{tabular}{lc|ccc|ccc}
\toprule
Model & Metric & \multicolumn{3}{c|}{$n=500$} & \multicolumn{3}{c}{$n=1000$} \\
& & TSIR & MESIR & MESIR-Auto & TSIR & MESIR & MESIR-Auto \\
\midrule
\multicolumn{2}{c}{} & \multicolumn{6}{c}{\textbf{Normal Distribution}} \\
\midrule
\multirow{2}{*}{B1} & $R^2$ & \textbf{0.775 (0.041)} & 0.676 (0.065) & 0.670 (0.097) & \textbf{0.887 (0.018)} & 0.854 (0.022) & 0.854 (0.022) \\
 & $D$ & \textbf{0.473 (0.043)} & 0.567 (0.054) & 0.602 (0.089) & \textbf{0.335 (0.027)} & 0.381 (0.028) & 0.381 (0.028) \\
\multirow{2}{*}{B2} & $R^2$ & \textbf{0.775 (0.038)} & 0.678 (0.055) & 0.670 (0.087) & \textbf{0.889 (0.018)} & 0.857 (0.024) & 0.857 (0.024) \\
 & $D$ & \textbf{0.473 (0.040)} & 0.565 (0.047) & 0.591 (0.083) & \textbf{0.332 (0.027)} & 0.376 (0.032) & 0.376 (0.032) \\
\multirow{2}{*}{B3} & $R^2$ & \textbf{0.771 (0.038)} & 0.680 (0.059) & 0.667 (0.098) & \textbf{0.884 (0.016)} & 0.851 (0.023) & 0.851 (0.023) \\
 & $D$ & \textbf{0.477 (0.040)} & 0.564 (0.051) & 0.596 (0.087) & \textbf{0.340 (0.024)} & 0.385 (0.030) & 0.385 (0.030) \\
\midrule
\multicolumn{2}{c}{} & \multicolumn{6}{c}{\textbf{$t_3$ Distribution}} \\
\midrule
\multirow{2}{*}{B1} & $R^2$ & 0.489 (0.201) & \textbf{0.533 (0.111)} & 0.514 (0.146) & 0.781 (0.116) & \textbf{0.790 (0.040)} & 0.785 (0.056) \\
 & $D$ & 0.702 (0.134) & \textbf{0.679 (0.076)} & 0.732 (0.153) & 0.458 (0.097) & \textbf{0.456 (0.043)} & 0.466 (0.062) \\
\multirow{2}{*}{B2} & $R^2$ & 0.501 (0.204) & \textbf{0.539 (0.097)} & 0.515 (0.146) & \textbf{0.797 (0.062)} & 0.789 (0.036) & 0.790 (0.036) \\
 & $D$ & 0.693 (0.138) & \textbf{0.676 (0.070)} & 0.730 (0.155) & \textbf{0.446 (0.064)} & 0.457 (0.039) & 0.462 (0.048) \\
\multirow{2}{*}{B3} & $R^2$ & 0.413 (0.188) & \textbf{0.510 (0.102)} & 0.509 (0.138) & 0.758 (0.084) & \textbf{0.771 (0.046)} & 0.768 (0.059) \\
 & $D$ & 0.756 (0.124) & \textbf{0.696 (0.073)} & 0.725 (0.111) & 0.486 (0.077) & \textbf{0.476 (0.049)} & 0.484 (0.063) \\
\midrule
\multicolumn{2}{c}{} & \multicolumn{6}{c}{\textbf{$t_2$ Distribution}} \\
\midrule
\multirow{2}{*}{B1} & $R^2$ & 0.151 (0.145) & 0.374 (0.135) & \textbf{0.388 (0.170)} & 0.369 (0.231) & \textbf{0.686 (0.086)} & 0.666 (0.131) \\
 & $D$ & 0.918 (0.085) & \textbf{0.787 (0.086)} & 0.820 (0.177) & 0.779 (0.154) & \textbf{0.556 (0.072)} & 0.602 (0.132) \\
\multirow{2}{*}{B2} & $R^2$ & 0.135 (0.106) & 0.375 (0.120) & \textbf{0.384 (0.162)} & 0.432 (0.229) & \textbf{0.686 (0.068)} & 0.678 (0.093) \\
 & $D$ & 0.928 (0.062) & \textbf{0.787 (0.077)} & 0.832 (0.184) & 0.738 (0.154) & \textbf{0.557 (0.059)} & 0.586 (0.089) \\
\multirow{2}{*}{B3} & $R^2$ & 0.135 (0.110) & \textbf{0.358 (0.131)} & 0.355 (0.159) & 0.370 (0.232) & \textbf{0.677 (0.093)} & 0.663 (0.119) \\
 & $D$ & 0.928 (0.062) & \textbf{0.797 (0.081)} & 0.853 (0.217) & 0.779 (0.151) & \textbf{0.563 (0.078)} & 0.589 (0.100) \\
\midrule
\multicolumn{2}{c}{} & \multicolumn{6}{c}{\textbf{Cauchy Distribution}} \\
\midrule
\multirow{2}{*}{B1} & $R^2$ & 0.023 (0.023) & 0.127 (0.070) & \textbf{0.128 (0.087)} & 0.037 (0.041) & \textbf{0.343 (0.121)} & 0.311 (0.148) \\
 & $D$ & 0.988 (0.012) & \textbf{0.934 (0.039)} & 1.034 (0.249) & 0.981 (0.021) & \textbf{0.807 (0.075)} & 0.902 (0.199) \\
\multirow{2}{*}{B2} & $R^2$ & 0.023 (0.025) & 0.135 (0.083) & \textbf{0.146 (0.100)} & 0.034 (0.038) & \textbf{0.343 (0.129)} & 0.312 (0.139) \\
 & $D $ & 0.988 (0.013) & \textbf{0.929 (0.046)} & 0.987 (0.259) & 0.983 (0.020) & \textbf{0.806 (0.082)} & 0.894 (0.226) \\
\multirow{2}{*}{B3} & $R^2$ & 0.020 (0.017) & \textbf{0.136 (0.073)} & 0.132 (0.084) & 0.031 (0.037) & \textbf{0.340 (0.122)} & 0.321 (0.145) \\
 & $D$ & 0.990 (0.009) & \textbf{0.929 (0.041)} & 1.028 (0.252) & 0.984 (0.019) & \textbf{0.809 (0.077)} & 0.883 (0.201) \\
\bottomrule
\end{tabular}
\end{table}

\paragraph{Simulation results.}
Tables~\ref{tab:A1_A3_row_compare}--\ref{tab:B1_B3_kron_compare} report the recovery performance of the row reduction subspaces and the CTSs for Models A1--A3 and B1--B3. Each entry is averaged over \(100\) replications, with standard deviations reported in parentheses. Since the column reduction subspace results exhibit patterns similar to the row counterparts, they are reported in Appendix~\ref{simulation results}.

When the structural dimensions are known, TSIR generally provides better subspace recovery under normally distributed predictors. For \(t_3\) predictors, MESIR and TSIR perform comparably across Models A1--A3 and B1--B3, with MESIR showing a modest advantage in most settings. This advantage is more pronounced at smaller sample sizes. As the tails become heavier, the advantage of MESIR becomes more apparent. Under \(t_2\) and Cauchy predictors, MESIR generally achieves larger \(R^2\) values and smaller projection distances than TSIR for both the row reduction subspace and the CTS, while Vector ESIR tends to be less accurate in recovering the CTS. Increasing the sample size from \(n=500\) to \(n=1000\) generally improves subspace recovery across all methods. Although recovery of the CTS remains difficult in the Cauchy setting, MESIR retains a clear relative advantage over the competing methods.

MESIR-Auto estimates the structural dimensions from the data, whereas MESIR uses the true dimensions. Across most settings, MESIR-Auto achieves subspace recovery comparable to MESIR in terms of both \(R^2\) and \(D\), indicating that estimating the structural dimensions incurs little loss in recovery accuracy. Some deterioration is observed under the Cauchy distribution, where the extremely heavy tails make dimension selection more challenging. Overall, MESIR remains competitive under light-tailed distributions, while its relative advantage over the competing methods becomes more pronounced as the tails grow heavier, consistent with findings for vector-valued elliptical SIR \citep{chen2022high}. Notably, although the \(t_2\) and Cauchy distributions do not satisfy the moment condition in Assumption~\ref{assump3}, MESIR still performs well relative to the competing methods in these settings.

\paragraph{Extension to tensor-valued predictors.}
We further evaluate the tensor extension, TESIR, using third-order tensor predictors. Detailed simulation settings, model specifications, and results for both mode-wise and CTS recovery are provided in Appendix~\ref{subsec:tensor_simulation}.

\section{Real Data Analysis}
\label{sec:real_data}

\subsection{COVID-19 
recognition}
\label{sec:COVID-19 
recognition}
Coronavirus disease 2019 (COVID-19), caused by severe acute respiratory syndrome coronavirus 2 (SARS-CoV-2), spread rapidly worldwide and posed a major public health challenge. Chest computed tomography (CT) provides detailed imaging information on lung tissues and can serve as a complementary tool for identifying COVID-19-related abnormalities. We analyze the publicly available COVID-CT dataset, which has also been used in previous studies \citep{zhang2023low,yang2020covid}.\footnote{\url{https://github.com/desaisrkr/https-github.com-UCSD-AI4H-COVID-CT}} The dataset contains 349 COVID-19-positive and 397 COVID-19-negative CT images, mainly in PNG and JPEG formats.

\begin{figure}[!htbp]
\centering
\includegraphics[width=0.7\linewidth]{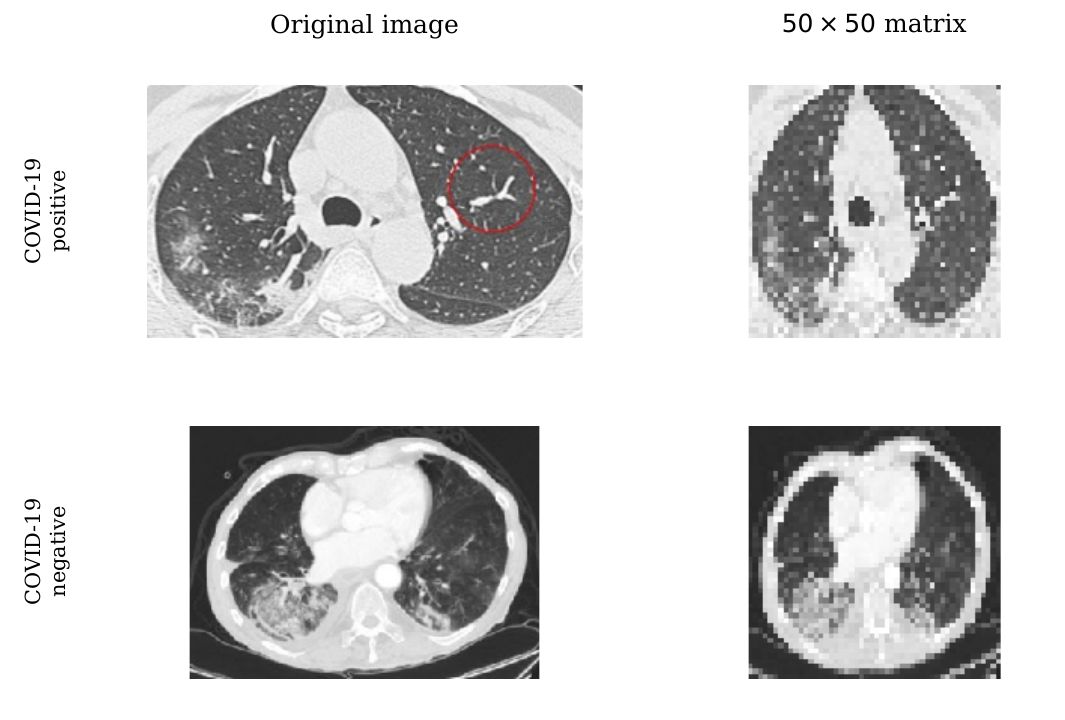}
\caption{Representative COVID-19-positive (top) and COVID-19-negative (bottom) CT images before (left) and after (right) resizing to \(50\times50\).}
\label{fig:ct_comparison}
\end{figure}

\begin{figure}[!htbp]
\centering
\includegraphics[width=0.9\linewidth]{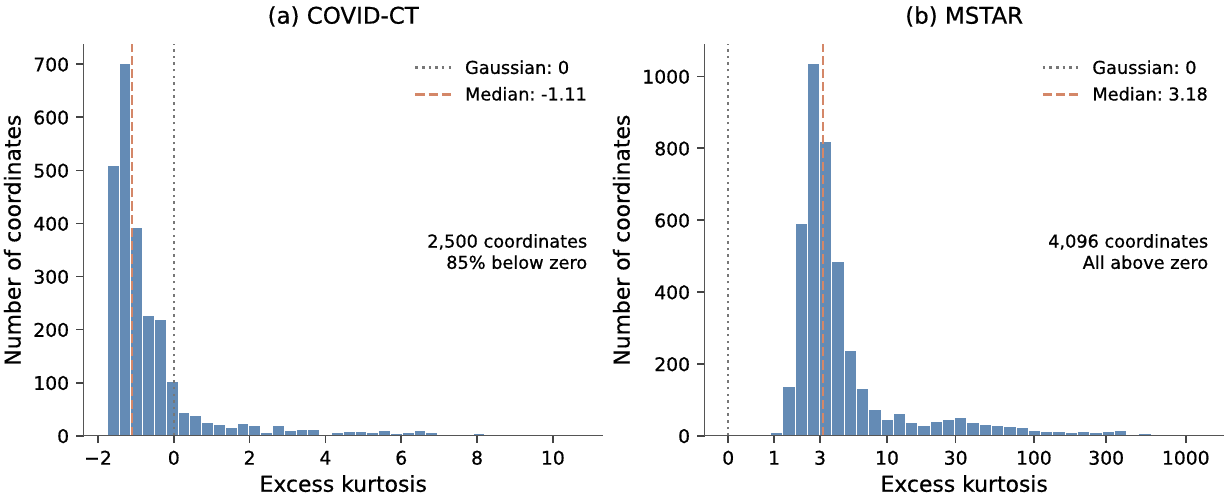}
\caption{Coordinate-wise excess kurtosis for (a) the COVID-CT dataset in Section~\ref{sec:COVID-19 recognition} and (b) the MSTAR dataset in Section~\ref{subsec:mstar}.}
\label{fig:ct_mstar_tail}
\end{figure}

We preprocess the images by converting them to grayscale, scaling the pixel intensities to \([0,1]\), and resizing each image to \(50\times50\) using the R package \texttt{EBImage}. Each observation is therefore represented as a matrix \(\mathbf X_i\in\mathbb R^{50\times50}\), with a binary response indicating COVID-19 status. Representative images from the two groups are shown in Figure~\ref{fig:ct_comparison}. Figure~\ref{fig:ct_mstar_tail}(a) summarizes the tail behavior of the preprocessed CT images using coordinate-wise excess kurtosis. Among the \(2{,}500\) pixel coordinates, approximately \(85\%\) have negative excess kurtosis, with a median of \(-1.11\), suggesting predominantly light-tailed marginal behavior.

We compare the classification performance of MESIR, TSIR, and Vector ESIR.  We assess the estimated reduction subspaces through downstream classification performance. For MESIR and TSIR, we set \(d_1=d_2=3\), yielding nine-dimensional reduced features and allowing the two matrix-based methods to be compared under the same dimension. For Vector ESIR, the two response classes are treated as slices, so one direction is retained. We randomly split the data into training and test sets in a stratified \(70/30\) ratio.
For each split, we first estimate the reduction subspace using the training data. The resulting low-dimensional features are then standardized and used to train three classifiers: random forest (RF), \(k\)-nearest neighbors (\(k\)NN), and quadratic discriminant analysis (QDA). We repeat this procedure \(50\) times using the same split for all methods within each replication, and report the means and standard deviations of precision, recall, F1 score, and AUC in Table~\ref{tab:ct_original}.

\begin{figure}[!htbp]
\centering
\includegraphics[width=0.9\linewidth]{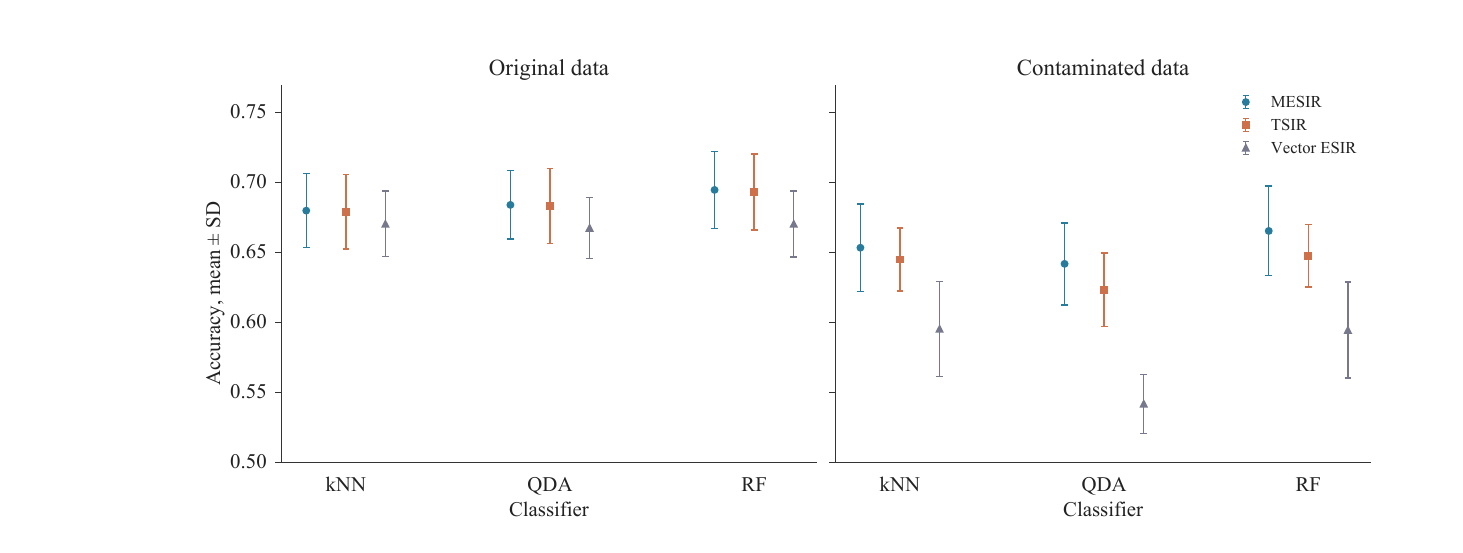}
\caption{Classification accuracy on the original and contaminated COVID-CT data.
Error bars indicate one standard deviation across 50 splits.}
\label{fig:ct_accuracy}
\end{figure}

Because the original CT images exhibit predominantly light-tailed marginal behavior, we further examine the robustness of the competing methods to substantial image-level scale variation. {Specifically, we randomly select \(10\%\) of the images and multiply each selected matrix by an independent scale factor \(1+\min\{|T_i|,15\}\), where \(T_i\sim t_2\). }The truncation prevents a few extreme realizations from dominating the experiment while retaining pronounced scale variation. The contaminated dataset is generated once before the training-test splits, and the same evaluation procedure is then applied. The results are reported in Table~\ref{tab:ct_contaminated} and Figure~\ref{fig:ct_accuracy}.

\begin{table}[!htbp]
\centering
\caption{Classification performance on the original COVID-CT dataset.}
\label{tab:ct_original}
\resizebox{0.85\linewidth}{!}{%
\begin{tabular}{llcccc}
\toprule
Classifier & Method & Precision & Recall & F1 & AUC \\
\midrule
RF & MESIR & \textbf{0.678 (0.034)} & 0.667 (0.048) & \textbf{0.671 (0.032)} & \textbf{0.765 (0.027)} \\
 & TSIR & 0.675 (0.032) & \textbf{0.670 (0.051)} & 0.671 (0.033) & 0.765 (0.026) \\
 & Vector ESIR & 0.652 (0.028) & 0.639 (0.042) & 0.645 (0.027) & 0.674 (0.023) \\
\midrule
kNN & MESIR & 0.690 (0.037) & 0.579 (0.047) & 0.629 (0.035) & 0.739 (0.032) \\
 & TSIR & \textbf{0.691 (0.040)} & 0.575 (0.048) & 0.626 (0.034) & \textbf{0.742 (0.033)} \\
 & Vector ESIR & 0.654 (0.028) & \textbf{0.632 (0.041)} & \textbf{0.642 (0.028)} & 0.674 (0.023) \\
\midrule
QDA & MESIR & 0.682 (0.031) & \textbf{0.615 (0.055)} & \textbf{0.645 (0.035)} & 0.746 (0.021) \\
 & TSIR & 0.681 (0.033) & 0.614 (0.057) & 0.644 (0.037) & \textbf{0.748 (0.021)} \\
 & Vector ESIR & \textbf{0.689 (0.033)} & 0.532 (0.047) & 0.599 (0.033) & 0.734 (0.028) \\
\bottomrule
\end{tabular}%
}
\end{table}

\begin{table}[!htbp]
\centering
\caption{Classification performance on the contaminated COVID-CT dataset.}
\label{tab:ct_contaminated}
\resizebox{0.85\linewidth}{!}{%
\begin{tabular}{llcccc}
\toprule
Classifier & Method & Precision & Recall & F1 & AUC \\
\midrule
RF & MESIR & \textbf{0.654 (0.031)} & \textbf{0.642 (0.048)} & \textbf{0.647 (0.028)} & \textbf{0.735 (0.026)} \\
 & TSIR & 0.642 (0.038) & 0.594 (0.063) & 0.616 (0.045) & 0.706 (0.033) \\
 & Vector ESIR & 0.583 (0.041) & 0.580 (0.076) & 0.580 (0.055) & 0.609 (0.040) \\
\midrule
kNN & MESIR & \textbf{0.650 (0.040)} & 0.569 (0.049) & \textbf{0.605 (0.035)} & \textbf{0.712 (0.030)} \\
 & TSIR & 0.646 (0.039) & 0.533 (0.053) & 0.583 (0.041) & 0.690 (0.029) \\
 & Vector ESIR & 0.584 (0.041) & \textbf{0.577 (0.076)} & 0.579 (0.055) & 0.609 (0.040) \\
\midrule
QDA & MESIR & \textbf{0.661 (0.041)} & 0.528 (0.092) & \textbf{0.582 (0.058)} & \textbf{0.692 (0.029)} \\
 & TSIR & 0.643 (0.040) & 0.454 (0.070) & 0.530 (0.053) & 0.671 (0.028) \\
 & Vector ESIR & 0.563 (0.142) & \textbf{0.723 (0.379)} & 0.524 (0.175) & 0.631 (0.055) \\
\bottomrule
\end{tabular}%
}
\end{table}

{On the original data, MESIR achieves mean classification accuracies broadly comparable to those of TSIR across the three classifiers. This similarity is consistent with the relatively light-tailed nature of the CT data.} After introducing scale contamination, the performance of all methods deteriorates, but MESIR achieves the highest mean accuracy and shows a more pronounced advantage over both competing methods. Figure~\ref{fig:ct_accuracy} further shows that MESIR experiences the smallest decrease in mean accuracy across the three classifiers, indicating greater robustness to heavy-tailed scale contamination.

\subsection{Radar target recognition}
\label{subsec:mstar}

We further investigate radar target recognition using the Moving and Stationary Target Acquisition and Recognition (MSTAR) dataset, a widely used benchmark for synthetic aperture radar (SAR) target recognition \citep{chen2016target,kechagias2021automatic}.\footnote{\url{https://github.com/omae11/MSTAR-dataset}} The subset used in our analysis contains 5,172 SAR images from ten target classes: 2S1, BMP2, BRDM2, BTR60, BTR70, D7, T62, T72, ZIL131, and ZSU234. These classes correspond to different types of vehicles and equipment. For preprocessing, we extract the magnitude image, crop the central \(128\times128\) region, and downsample it to \(64\times64\) by averaging non-overlapping \(2\times2\) pixel blocks. Each observation is therefore represented by a matrix predictor \(\mathbf X_i\in\mathbb{R}^{64\times64}\), with \(Y_i\in\{1,\ldots,10\}\) indicating its target class.

Figure~\ref{fig:ct_mstar_tail}(b) shows substantially heavier marginal tails for the MSTAR data than for the COVID-CT data. All \(4{,}096\) pixel coordinates have positive excess kurtosis, with a median of \(3.18\), whereas the excess kurtosis values for COVID-CT are predominantly negative. This pronounced difference in marginal tail behavior makes MSTAR a useful dataset for evaluating the performance of MESIR under heavier-tailed data. We pool the original subsets and apply the same data-splitting, training, and evaluation procedure as in the COVID-CT analysis. As before, we set \(d_1=d_2=3\) for MESIR and TSIR. For Vector ESIR, the ten target classes are treated as slices and nine directions are retained, matching the nine-dimensional representations used by MESIR and TSIR. We use the same three classifiers, RF, \(k\)NN, and QDA, and evaluate classification performance using accuracy, precision, recall, F1 score, and AUC.
\begin{table}[!htbp]
\centering
\caption{Classification performance on the MSTAR dataset.}
\label{tab:mstar_results}
\resizebox{\linewidth}{!}{%
\begin{tabular}{llccccc}
\toprule
Classifier & Method & Accuracy & Precision & Recall & F1 & AUC \\
\midrule
RF & MESIR
& \textbf{0.767 (0.011)}
& \textbf{0.761 (0.011)}
& \textbf{0.759 (0.011)}
& \textbf{0.759 (0.011)}
& \textbf{0.970 (0.002)} \\
& TSIR
& 0.734 (0.012)
& 0.723 (0.013)
& 0.721 (0.012)
& 0.720 (0.012)
& 0.962 (0.002) \\
& Vector ESIR
& 0.298 (0.019)
& 0.318 (0.018)
& 0.280 (0.019)
& 0.262 (0.022)
& 0.750 (0.010) \\
\midrule
kNN & MESIR
& \textbf{0.715 (0.012)}
& \textbf{0.709 (0.013)}
& \textbf{0.702 (0.013)}
& \textbf{0.702 (0.013)}
& \textbf{0.940 (0.004)} \\
& TSIR
& 0.706 (0.011)
& 0.694 (0.012)
& 0.690 (0.012)
& 0.690 (0.012)
& 0.935 (0.005) \\
& Vector ESIR
& 0.420 (0.017)
& 0.414 (0.017)
& 0.416 (0.017)
& 0.412 (0.017)
& 0.677 (0.009) \\
\midrule
QDA & MESIR
& \textbf{0.681 (0.012)}
& \textbf{0.675 (0.013)}
& \textbf{0.667 (0.013)}
& \textbf{0.663 (0.013)}
& \textbf{0.943 (0.003)} \\
& TSIR
& 0.663 (0.012)
& 0.649 (0.013)
& 0.645 (0.012)
& 0.640 (0.012)
& 0.938 (0.003) \\
& Vector ESIR
& 0.160 (0.010)
& 0.178 (0.014)
& 0.158 (0.010)
& 0.144 (0.011)
& 0.533 (0.006) \\
\bottomrule
\end{tabular}%
}
\end{table}

Table~\ref{tab:mstar_results} reports the means and standard deviations of the classification metrics across \(50\) random splits. MESIR achieves the best performance across all three classifiers. Both matrix-based methods outperform Vector ESIR under the current settings. These results demonstrate the effectiveness of MESIR for dimension reduction and subsequent classification on the MSTAR data, which exhibit pronounced heavy-tailed behavior.

\section{Conclusion and Future Work}
We proposed matrix elliptical sliced inverse regression (MESIR) for sufficient dimension reduction with matrix-valued predictors. Under matrix elliptical distributions, MESIR combines eigenvalue-calibrated marginal Kendall's tau matrices with inverse-regression Kendall's tau matrices to recover row and column reduction subspaces while preserving matrix structure and reducing sensitivity to large-magnitude observations. We established convergence rates, subspace consistency, and consistent structural dimension selection under weak moment conditions, allowing the predictor dimensions and number of slices to diverge. We also extended the framework to tensor-valued predictors. Simulations and applications to COVID-19 CT and MSTAR images demonstrate competitive performance, with advantages becoming more pronounced under heavier tails. Future work includes extending the Kendall-based construction to SAVE \citep{dennis2000save} and directional regression \citep{li2007directional}, relaxing the elliptical assumption,
and developing methods for settings where the mode dimensions are comparable to or exceed the sample size.

%%%%%%%% REFERANCE %%%%%%%%
\bibliographystyle{biometrika}
\bibliography{Ref}

%================================================
% Main text theorem
%================================================

%================================================
% Appendix proof
%================================================
\newpage
\appendix
Section~\ref{supp_Supplementary} reports additional simulation
results for column subspace recovery.
Sections~\ref{supp_lemma} and~\ref{supp_proof_of_matrix} provide
technical lemmas and proofs of the MESIR results, respectively.
Section~\ref{subsec:tensor_mesir} extends the method to higher-order
tensor predictors, with estimation procedures in
Section~\ref{subsec:tesir}, theoretical results in
Section~\ref{sub:TESIR Theoretical properties}, simulations in
Section~\ref{subsec:tensor_simulation} and proofs in
Section~\ref{supp_tensor_proof}.

\section{Supplementary Simulations}
\label{supp_Supplementary}
\label{simulation results}
The column subspace recovery results for MESIR are presented in
Tables~\ref{tab:A1_A3_column_compare} and
\ref{tab:B1_B3_column_compare}. These results exhibit patterns
consistent with those observed for the row and CTSs in
section~\ref{sec:MESIR_simulation}. MESIR maintains competitive recovery performance under
light-tailed distributions and shows more stable recovery as the tails
become heavier, with its advantage over TSIR becoming more apparent under
$t_2$ and Cauchy predictors. MESIR-Auto achieves performance close to
MESIR in most settings.

\begin{table}[!htbp]
\centering
\caption{Column subspace recovery for Models A1--A3.}
\label{tab:A1_A3_column_compare}
\footnotesize
\setlength{\tabcolsep}{5pt}
\renewcommand{\arraystretch}{1}
\begin{tabular}{lc|ccc|ccc}
\toprule
Model & Metric & \multicolumn{3}{c|}{$n=500$} & \multicolumn{3}{c}{$n=1000$} \\
& & TSIR & MESIR & MESIR-Auto & TSIR & MESIR & MESIR-Auto \\
\midrule
\multicolumn{2}{c}{} & \multicolumn{6}{c}{\textbf{Normal Distribution}} \\
\midrule
\multirow{2}{*}{A1} & $R_2^2$ & \textbf{0.958 (0.012)} & 0.949 (0.015) & 0.949 (0.015) & \textbf{0.979 (0.006)} & 0.976 (0.007) & 0.976 (0.007) \\
 & $D_2$ & \textbf{0.202 (0.029)} & 0.222 (0.033) & 0.222 (0.033) & \textbf{0.143 (0.019)} & 0.155 (0.021) & 0.155 (0.021) \\
\multirow{2}{*}{A2} & $R_2^2$ & \textbf{0.958 (0.013)} & 0.951 (0.014) & 0.951 (0.014) & \textbf{0.979 (0.007)} & 0.977 (0.007) & 0.977 (0.007) \\
 & $D_2$ & \textbf{0.203 (0.033)} & 0.220 (0.032) & 0.220 (0.032) & \textbf{0.143 (0.023)} & 0.150 (0.024) & 0.150 (0.024) \\
\multirow{2}{*}{A3} & $R_2^2$ & \textbf{0.959 (0.012)} & 0.951 (0.015) & 0.951 (0.015) & \textbf{0.980 (0.007)} & 0.977 (0.007) & 0.977 (0.007) \\
 & $D_2$ & \textbf{0.200 (0.031)} & 0.218 (0.036) & 0.218 (0.036) & \textbf{0.140 (0.023)} & 0.150 (0.024) & 0.150 (0.024) \\
\midrule
\multicolumn{2}{c}{} & \multicolumn{6}{c}{\textbf{$t_3$ Distribution}} \\
\midrule
\multirow{2}{*}{A1} & $R_2^2$ & 0.872 (0.169) & \textbf{0.905 (0.035)} & \textbf{0.905 (0.035)} & 0.937 (0.098) & \textbf{0.954 (0.016)} & \textbf{0.954 (0.016)} \\
 & $D_2$ & 0.329 (0.142) & \textbf{0.303 (0.054)} & \textbf{0.303 (0.054)} & 0.232 (0.094) & \textbf{0.210 (0.038)} & \textbf{0.210 (0.038)} \\
\multirow{2}{*}{A2} & $R_2^2$ & 0.902 (0.101) & \textbf{0.913 (0.028)} & \textbf{0.913 (0.028)} & 0.949 (0.035) & \textbf{0.959 (0.012)} & \textbf{0.959 (0.012)} \\
 & $D_2$ & 0.298 (0.097) & \textbf{0.290 (0.048)} & \textbf{0.290 (0.048)} & 0.218 (0.062) & \textbf{0.201 (0.030)} & \textbf{0.201 (0.030)} \\
\multirow{2}{*}{A3} & $R_2^2$ & 0.890 (0.095) & \textbf{0.900 (0.033)} & \textbf{0.900 (0.033)} & 0.947 (0.024) & \textbf{0.954 (0.014)} & \textbf{0.954 (0.014)} \\
 & $D_2$ & 0.321 (0.086) & \textbf{0.312 (0.050)} & \textbf{0.312 (0.050)} & 0.226 (0.047) & \textbf{0.212 (0.034)} & \textbf{0.212 (0.034)} \\
\midrule
\multicolumn{2}{c}{} & \multicolumn{6}{c}{\textbf{$t_2$ Distribution}} \\
\midrule
\multirow{2}{*}{A1} & $R_2^2$ & 0.642 (0.328) & \textbf{0.830 (0.060)} & 0.817 (0.100) & 0.780 (0.289) & \textbf{0.920 (0.031)} & \textbf{0.920 (0.031)} \\
 & $D_2$ & 0.545 (0.248) & \textbf{0.407 (0.071)} & 0.420 (0.113) & 0.408 (0.232) & \textbf{0.278 (0.053)} & \textbf{0.278 (0.053)} \\
\multirow{2}{*}{A2} & $R_2^2$ & 0.679 (0.318) & \textbf{0.828 (0.077)} & 0.810 (0.120) & 0.795 (0.278) & \textbf{0.924 (0.031)} & \textbf{0.924 (0.031)} \\
 & $D_2$ & 0.515 (0.238) & \textbf{0.405 (0.087)} & 0.426 (0.137) & 0.393 (0.225) & \textbf{0.270 (0.055)} & \textbf{0.270 (0.055)} \\
\multirow{2}{*}{A3} & $R_2^2$ & 0.641 (0.345) & \textbf{0.828 (0.066)} & 0.810 (0.118) & 0.742 (0.325) & \textbf{0.920 (0.030)} & \textbf{0.920 (0.030)} \\
 & $D_2$ & 0.542 (0.257) & \textbf{0.408 (0.076)} & 0.428 (0.136) & 0.440 (0.256) & \textbf{0.278 (0.052)} & \textbf{0.278 (0.052)} \\
\midrule
\multicolumn{2}{c}{} & \multicolumn{6}{c}{\textbf{Cauchy Distribution}} \\
\midrule
\multirow{2}{*}{A1} & $R_2^2$ & 0.054 (0.069) & \textbf{0.284 (0.231)} & 0.236 (0.184) & 0.047 (0.054) & \textbf{0.567 (0.233)} & 0.501 (0.255) \\
 & $D_2$ & 0.972 (0.037) & \textbf{0.833 (0.148)} & 1.075 (0.305) & 0.976 (0.028) & \textbf{0.636 (0.170)} & 0.776 (0.325) \\
\multirow{2}{*}{A2} & $R_2^2$ & 0.050 (0.073) & \textbf{0.294 (0.231)} & 0.255 (0.191) & 0.053 (0.070) & \textbf{0.577 (0.222)} & 0.493 (0.231) \\
 & $D_2$ & 0.974 (0.039) & \textbf{0.828 (0.146)} & 1.035 (0.289) & 0.973 (0.037) & \textbf{0.630 (0.162)} & 0.790 (0.309) \\
\multirow{2}{*}{A3} & $R_2^2$ & 0.059 (0.079) & \textbf{0.300 (0.232)} & 0.255 (0.179) & 0.043 (0.060) & \textbf{0.562 (0.242)} & 0.509 (0.246) \\
 & $D_2$ & 0.969 (0.043) & \textbf{0.823 (0.149)} & 1.041 (0.276) & 0.978 (0.032) & \textbf{0.639 (0.174)} & 0.762 (0.306) \\
\bottomrule
\end{tabular}
\end{table}

\begin{table}[!htbp]
\centering
\caption{Column subspace recovery for Models B1--B3.}
\label{tab:B1_B3_column_compare}
\footnotesize
\setlength{\tabcolsep}{5pt}
\renewcommand{\arraystretch}{1}
\begin{tabular}{lc|ccc|ccc}
\toprule
Model & Metric & \multicolumn{3}{c|}{$n=500$} & \multicolumn{3}{c}{$n=1000$} \\
& & TSIR & MESIR & MESIR-Auto & TSIR & MESIR & MESIR-Auto \\
\midrule
\multicolumn{2}{c}{} & \multicolumn{6}{c}{\textbf{Normal Distribution}} \\
\midrule
\multirow{2}{*}{B1} & $R_2^2$ & \textbf{0.882 (0.030)} & 0.828 (0.054) & 0.824 (0.078) & \textbf{0.944 (0.013)} & 0.928 (0.017) & 0.928 (0.017) \\
 & $D_2$ & \textbf{0.341 (0.044)} & 0.410 (0.057) & 0.437 (0.092) & \textbf{0.235 (0.028)} & 0.266 (0.032) & 0.266 (0.032) \\
\multirow{2}{*}{B2} & $R_2^2$ & \textbf{0.883 (0.026)} & 0.823 (0.042) & 0.814 (0.076) & \textbf{0.943 (0.013)} & 0.927 (0.018) & 0.927 (0.018) \\
 & $D_2$ & \textbf{0.341 (0.038)} & 0.418 (0.050) & 0.440 (0.085) & \textbf{0.237 (0.027)} & 0.269 (0.033) & 0.269 (0.033) \\
\multirow{2}{*}{B3} & $R_2^2$ & \textbf{0.877 (0.029)} & 0.821 (0.044) & 0.810 (0.083) & \textbf{0.940 (0.013)} & 0.923 (0.017) & 0.923 (0.017) \\
 & $D_2$ & \textbf{0.348 (0.040)} & 0.420 (0.052) & 0.450 (0.093) & \textbf{0.244 (0.026)} & 0.276 (0.029) & 0.276 (0.029) \\
\midrule
\multicolumn{2}{c}{} & \multicolumn{6}{c}{\textbf{$t_3$ Distribution}} \\
\midrule
\multirow{2}{*}{B1} & $R_2^2$ & 0.683 (0.187) & \textbf{0.735 (0.097)} & 0.728 (0.128) & 0.880 (0.096) & \textbf{0.889 (0.031)} & 0.889 (0.031) \\
 & $D_2$ & 0.542 (0.150) & \textbf{0.508 (0.085)} & 0.548 (0.118) & 0.335 (0.092) & \textbf{0.330 (0.046)} & 0.337 (0.059) \\
\multirow{2}{*}{B2} & $R_2^2$ & 0.681 (0.177) & \textbf{0.733 (0.087)} & 0.719 (0.132) & \textbf{0.892 (0.043)} & 0.889 (0.027) & 0.888 (0.027) \\
 & $D_2$ & 0.546 (0.145) & \textbf{0.510 (0.080)} & 0.561 (0.125) & \textbf{0.323 (0.060)} & 0.331 (0.040) & 0.334 (0.046) \\
\multirow{2}{*}{B3} & $R_2^2$ & 0.620 (0.164) & 0.714 (0.088) & \textbf{0.714 (0.132)} & 0.869 (0.060) & \textbf{0.879 (0.033)} & 0.874 (0.054) \\
 & $D_2$ & 0.602 (0.134) & \textbf{0.529 (0.079)} & 0.575 (0.105) & 0.354 (0.072) & \textbf{0.345 (0.046)} & 0.354 (0.065) \\
\midrule
\multicolumn{2}{c}{} & \multicolumn{6}{c}{\textbf{$t_2$ Distribution}} \\
\midrule
\multirow{2}{*}{B1} & $R_2^2$ & 0.349 (0.184) & 0.603 (0.132) & \textbf{0.619 (0.162)} & 0.572 (0.212) & \textbf{0.827 (0.057)} & 0.814 (0.095) \\
 & $D_2$ & 0.798 (0.121) & \textbf{0.621 (0.105)} & 0.661 (0.119) & 0.633 (0.166) & \textbf{0.411 (0.065)} & 0.442 (0.098) \\
\multirow{2}{*}{B2} & $R_2^2$ & 0.342 (0.134) & 0.601 (0.132) & \textbf{0.621 (0.163)} & 0.628 (0.205) & \textbf{0.826 (0.053)} & 0.819 (0.091) \\
 & $D_2$ & 0.806 (0.088) & \textbf{0.623 (0.105)} & 0.664 (0.128) & 0.587 (0.164) & \textbf{0.412 (0.062)} & 0.440 (0.104) \\
\multirow{2}{*}{B3} & $R_2^2$ & 0.336 (0.156) & \textbf{0.590 (0.129)} & 0.582 (0.159) & 0.561 (0.226) & \textbf{0.823 (0.064)} & 0.807 (0.104) \\
 & $D_2$ & 0.809 (0.101) & \textbf{0.633 (0.099)} & 0.682 (0.132) & 0.641 (0.171) & \textbf{0.415 (0.070)} & 0.448 (0.116) \\
\midrule
\multicolumn{2}{c}{} & \multicolumn{6}{c}{\textbf{Cauchy Distribution}} \\
\midrule
\multirow{2}{*}{B1} & $R_2^2$ & 0.148 (0.087) & 0.363 (0.120) & \textbf{0.364 (0.137)} & 0.182 (0.102) & \textbf{0.584 (0.117)} & 0.555 (0.159) \\
 & $D_2$ & 0.922 (0.048) & \textbf{0.794 (0.079)} & 0.830 (0.130) & 0.903 (0.058) & \textbf{0.639 (0.091)} & 0.708 (0.134) \\
\multirow{2}{*}{B2} & $R_2^2$ & 0.142 (0.082) & 0.346 (0.136) & \textbf{0.371 (0.162)} & 0.165 (0.104) & \textbf{0.578 (0.131)} & 0.543 (0.155) \\
 & $D_2$ & 0.925 (0.045) & \textbf{0.804 (0.088)} & 0.825 (0.136) & 0.912 (0.059) & \textbf{0.641 (0.104)} & 0.707 (0.137) \\
\multirow{2}{*}{B3} & $R_2^2$ & 0.143 (0.086) & \textbf{0.371 (0.129)} & 0.369 (0.159) & 0.158 (0.102) & \textbf{0.581 (0.133)} & 0.557 (0.164) \\
 & $D_2$ & 0.924 (0.048) & \textbf{0.789 (0.084)} & 0.825 (0.137) & 0.916 (0.057) & \textbf{0.639 (0.102)} & 0.698 (0.142) \\
\bottomrule
\end{tabular}
\end{table}

\clearpage

\section{Technical Lemmas}
\label{supp_lemma}

\begin{lemma}
\label{lem:Kr0_concentration_main}
Define
\[
\begin{aligned}
\mathbf K_{r,0}
=\mathbb E\left[
\frac{(\mathbf X-\mathbf X')(\mathbf X-\mathbf X')^\top}
{\|\mathbf X-\mathbf X'\|_F^2}
\right],\quad
\mathbf K_{c,0}
=\mathbb E\left[
\frac{(\mathbf X-\mathbf X')^\top(\mathbf X-\mathbf X')}
{\|\mathbf X-\mathbf X'\|_F^2}
\right].
\end{aligned}
\]
Suppose that
\[
\begin{aligned}
n\ge\frac{16}{3}\Big\{
&[r^*(\mathbf K_{r,0})+1]
(\log p+\log(1/\alpha_r))\\
&\vee [r^*(\mathbf K_{c,0})+1]
(\log q+\log(1/\alpha_c))
\Big\}.
\end{aligned}
\]
Then, with probabilities at least $1-\alpha_r$
and $1-\alpha_c$, respectively,
\[
\begin{aligned}
\|\widehat{\mathbf K}_{r,0}-\mathbf K_{r,0}\|_2
&\le \|\mathbf K_{r,0}\|_2
\sqrt{\frac{16[r^*(\mathbf K_{r,0})+1]
(\log p+\log(1/\alpha_r))}{3n}},\\
\|\widehat{\mathbf K}_{c,0}-\mathbf K_{c,0}\|_2
&\le \|\mathbf K_{c,0}\|_2
\sqrt{\frac{16[r^*(\mathbf K_{c,0})+1]
(\log q+\log(1/\alpha_c))}{3n}},
\end{aligned}
\]
where $r^*(\mathbf K_{r,0})
=\operatorname{tr}(\mathbf K_{r,0})/
\|\mathbf K_{r,0}\|_2$ and
$r^*(\mathbf K_{c,0})
=\operatorname{tr}(\mathbf K_{c,0})/
\|\mathbf K_{c,0}\|_2$.

Since $\mathbf K_r=p\mathbf K_{r,0}$,
$\mathbf K_c=q\mathbf K_{c,0}$,
$\widehat{\mathbf K}_r=p\widehat{\mathbf K}_{r,0}$
and $\widehat{\mathbf K}_c=q\widehat{\mathbf K}_{c,0}$,
the corresponding bounds are
\[
\begin{aligned}
\|\widehat{\mathbf K}_r-\mathbf K_r\|_2
&\le \|\mathbf K_r\|_2
\sqrt{\frac{16[r^*(\mathbf K_r)+1]
(\log p+\log(1/\alpha_r))}{3n}},\\
\|\widehat{\mathbf K}_c-\mathbf K_c\|_2
&\le \|\mathbf K_c\|_2
\sqrt{\frac{16[r^*(\mathbf K_c)+1]
(\log q+\log(1/\alpha_c))}{3n}}.
\end{aligned}
\]
\end{lemma}

\begin{proof}

Set
\(h_{r,0}(\mathbf X_i,\mathbf X_j)
=(\mathbf X_i-\mathbf X_j)(\mathbf X_i-\mathbf X_j)^\top/
\|\mathbf X_i-\mathbf X_j\|_F^2\)
and
\(\mathbf H_{ij}^{(r)}
=h_{r,0}(\mathbf X_i,\mathbf X_j)-\mathbf K_{r,0}\).
Then
\(\mathbf K_{r,0}
=\mathbb E\{h_{r,0}(\mathbf X_1,\mathbf X_2)\}\)
and
\[
\widehat{\mathbf K}_{r,0}
=\frac{2}{n(n-1)}
\sum_{1\le i<j\le n}
h_{r,0}(\mathbf X_i,\mathbf X_j).
\]
Since
\(\|h_{r,0}(\mathbf X_i,\mathbf X_j)\|_2
=\|\mathbf X_i-\mathbf X_j\|_2^2/
\|\mathbf X_i-\mathbf X_j\|_F^2\le1\),
we have
\(\|\mathbf H_{ij}^{(r)}\|_2
\le1+\|\mathbf K_{r,0}\|_2=:R_{1,r}\).
Moreover, as
\(h_{r,0}(\mathbf X_i,\mathbf X_j)^2
\preceq h_{r,0}(\mathbf X_i,\mathbf X_j)\),
\[
\begin{aligned}
\left\|\mathbb E\{(\mathbf H_{ij}^{(r)})^2\}\right\|_2
&\le
\left\|\mathbb E h_{r,0}(\mathbf X_i,\mathbf X_j)^2
\right\|_2+\|\mathbf K_{r,0}^2\|_2\\
&\le \|\mathbf K_{r,0}\|_2
+\|\mathbf K_{r,0}\|_2^2=:R_{2,r}.
\end{aligned}
\]
Since $\operatorname{tr}(\mathbf K_{r,0})=1$,
we obtain
\(R_{2,r}
=\|\mathbf K_{r,0}\|_2^2
\{1+r^*(\mathbf K_{r,0})\}\)
and
\(R_{2,r}/R_{1,r}=\|\mathbf K_{r,0}\|_2\).
By Theorem 8.1 of \citet{Han02012018},
for $t\le\|\mathbf K_{r,0}\|_2$,
\[
\mathbb P\left(
\|\widehat{\mathbf K}_{r,0}-\mathbf K_{r,0}\|_2
\ge t\right)
\le p\exp\left(-\frac{3nt^2}{16R_{2,r}}\right).
\]
Taking
\(t=\|\mathbf K_{r,0}\|_2
\sqrt{16\{1+r^*(\mathbf K_{r,0})\}
\{\log p+\log(1/\alpha_r)\}/(3n)}\),
which ensures \(t\le\|\mathbf K_{r,0}\|_2\),
and the preceding probability is at most \(\alpha_r\).
The column bound follows by transposition; multiplying by \(p\) and \(q\)
gives the bounds for the scaled Kendall's tau matrices.

\end{proof}

\begin{lemma}
\label{lem:Eigenvalue}
Suppose that \(\mathbf X\sim\mathcal{MED}_{p\times q}
(\mathbf M,\boldsymbol\Sigma\otimes\boldsymbol\Omega,\varphi)\).
Then \(\mathbf K_{r,0}\) and \(\mathbf K_{c,0}\) share eigenbases with
\(\boldsymbol\Sigma\) and \(\boldsymbol\Omega\), respectively, and their
ordered eigenvalues satisfy
\begin{equation}
\label{eq:Kr0_sigma}
\lambda_i(\mathbf K_{r,0})
=\mathbb E\left[
\frac{\lambda_i(\boldsymbol\Sigma)\sum_{j=1}^q\lambda_j(\boldsymbol\Omega)Z_{ij}^2}
{\sum_{a=1}^p\sum_{b=1}^q\lambda_a(\boldsymbol\Sigma)\lambda_b(\boldsymbol\Omega)Z_{ab}^2}
\right],\quad i=1,\ldots,p,
\end{equation}
\begin{equation}
\label{eq:Kc0_omiga}
\lambda_j(\mathbf K_{c,0})
=\mathbb E\left[
\frac{\lambda_j(\boldsymbol\Omega)\sum_{i=1}^p\lambda_i(\boldsymbol\Sigma)Z_{ij}^2}
{\sum_{a=1}^p\sum_{b=1}^q\lambda_a(\boldsymbol\Sigma)\lambda_b(\boldsymbol\Omega)Z_{ab}^2}
\right],\quad j=1,\ldots,q,
\end{equation}
where $Z_{ij}\stackrel{\mathrm{i.i.d.}}{\sim}N(0,1)$.
Moreover, $\lambda_i(K_r)=p\lambda_i(K_{r,0})$,
$\lambda_a(\Sigma_k)=\lambda_b(\Sigma_k)
\iff\lambda_a(K_k)=\lambda_b(K_k)$, and
$\lambda_a(\Sigma_k)>\lambda_b(\Sigma_k)
\iff\lambda_a(K_k)>\lambda_b(K_k)$.
\end{lemma}
\begin{proof}
Write
\(\boldsymbol\Sigma=\mathbf U_r\boldsymbol\Lambda_r\mathbf U_r^\top\)
and \(\boldsymbol\Omega=\mathbf U_c\boldsymbol\Lambda_c\mathbf U_c^\top\),
where \(\boldsymbol\Lambda_r=\operatorname{diag}(\lambda_{r,1},\ldots,\lambda_{r,p})\)
and \(\boldsymbol\Lambda_c=\operatorname{diag}(\lambda_{c,1},\ldots,\lambda_{c,q})\)
contain eigenvalues in nonincreasing order. Let \(\mathbf Z=(Z_{ij})\). For an independent copy
\(\widetilde{\mathbf X}\), the angular representation gives
\[
\frac{\mathbf X-\widetilde{\mathbf X}}{\|\mathbf X-\widetilde{\mathbf X}\|_F}
\stackrel d=
\frac{\mathbf U_r\boldsymbol\Lambda_r^{1/2}\mathbf Z
\boldsymbol\Lambda_c^{1/2}\mathbf U_c^\top}
{\|\boldsymbol\Lambda_r^{1/2}\mathbf Z\boldsymbol\Lambda_c^{1/2}\|_F}.
\]
Define \(S_i=\sum_{j=1}^q\lambda_{c,j}Z_{ij}^2\),
\(T_j=\sum_{i=1}^p\lambda_{r,i}Z_{ij}^2\), and
\(D=\sum_i\lambda_{r,i}S_i=\sum_j\lambda_{c,j}T_j\).
The variables \(S_1,\ldots,S_p\) are i.i.d.\ and positive,
as are \(T_1,\ldots,T_q\).

Substitution into the Kendall's tau kernel yields
\[
\mathbf K_{r,0}=\mathbf U_r\,
\mathbb E\left[
\frac{\boldsymbol\Lambda_r^{1/2}\mathbf Z\boldsymbol\Lambda_c
\mathbf Z^\top\boldsymbol\Lambda_r^{1/2}}{D}
\right]\mathbf U_r^\top.
\]
Changing the sign of any row of \(\mathbf Z\) leaves \(D\) unchanged,
so the off-diagonal entries of the expectation vanish. Its diagonal
entries are \(\lambda_{r,i}\mathbb E(S_i/D)\).
To verify their ordering, fix \(i\ne j\) and put
\(R=\sum_{a\ne i,j}\lambda_{r,a}S_a\).
Exchangeability of \((S_i,S_j)\) gives
\[
\begin{aligned}
\mathbb E\frac{\lambda_{r,i}S_i-\lambda_{r,j}S_j}{D}
&=\frac{\lambda_{r,i}-\lambda_{r,j}}{2}\,
\mathbb E\left[
\frac{R(S_i+S_j)+2(\lambda_{r,i}+\lambda_{r,j})S_iS_j}
{(R+\lambda_{r,i}S_i+\lambda_{r,j}S_j)
 (R+\lambda_{r,i}S_j+\lambda_{r,j}S_i)}
\right].
\end{aligned}
\]
The factor inside the expectation is positive, establishing the ordering
and equality claims and hence \eqref{eq:Kr0_sigma}.
Transposition gives \eqref{eq:Kc0_omiga} and the column analogue.
\[
\lambda_i(\mathbf K_r)
=
p\,\mathbb E\!\left[
\frac{
\lambda_i(\boldsymbol\Sigma)\sum_{j=1}^{q}
\lambda_j(\boldsymbol\Omega)Z_{ij}^{2}
}{
\sum_{a=1}^{p}\sum_{b=1}^{q}
\lambda_a(\boldsymbol\Sigma)\lambda_b(\boldsymbol\Omega)Z_{ab}^{2}
}
\right],
\quad
\lambda_j(\mathbf K_c)
=
q\,\mathbb E\!\left[
\frac{
\lambda_j(\boldsymbol\Omega)\sum_{i=1}^{p}
\lambda_i(\boldsymbol\Sigma)Z_{ij}^{2}
}{
\sum_{a=1}^{p}\sum_{b=1}^{q}
\lambda_a(\boldsymbol\Sigma)\lambda_b(\boldsymbol\Omega)Z_{ab}^{2}
}
\right].
\]
\end{proof}

\begin{lemma}
\label{thm:tensor_kendall_concentration}
 Let \(\mathcal X\in\mathbb R^{p_1\times\cdots\times p_K}\)
be a random tensor, and let \(\mathcal X_1,\ldots,\mathcal X_n\)
be an i.i.d.\ sample from its distribution.
Let \(\widetilde{\mathcal X}\) be an independent copy of \(\mathcal X\).
For each \(k=1,\ldots,K\), denote the mode-\(k\) unfoldings of
\(\mathcal X\), \(\widetilde{\mathcal X}\), and \(\mathcal X_i\) by
\(\mathcal X_{(k)}\), \(\widetilde{\mathcal X}_{(k)}\), and
\(\mathcal X_{i,(k)}\), respectively. These matrices belong to
\(\mathbb R^{p_k\times p_{-k}}\), where
\(p_{-k}=\prod_{j\neq k}p_j\).
Define
\[
\mathbf K_{k,0}
=
\mathbb E\left[
\frac{
(\mathcal X_{(k)}-\widetilde{\mathcal X}_{(k)})
(\mathcal X_{(k)}-\widetilde{\mathcal X}_{(k)})^\top
}{
\|\mathcal X_{(k)}-\widetilde{\mathcal X}_{(k)}\|_F^2
}
\right],
\widehat{\mathbf K}_{k,0}
=
\frac{2}{n(n-1)}
\sum_{1\le i<i'\le n}
\frac{
(\mathcal X_{i,(k)}-\mathcal X_{i',(k)})
(\mathcal X_{i,(k)}-\mathcal X_{i',(k)})^\top
}{
\|\mathcal X_{i,(k)}-\mathcal X_{i',(k)}\|_F^2}.
\]
If
\(
n
\ge
\frac{16}{3}
\bigl[
r^*(\mathbf K_{k,0})+1
\bigr]
\bigl\{\log p_k+\log(1/\alpha_k)\bigr\},
\)
thus with probability at least \(1-\alpha_k\),
\[
\left\|
\widehat{\mathbf K}_{k,0}
-
\mathbf K_{k,0}
\right\|_2
\le
\|\mathbf K_{k,0}\|_2
\sqrt{
\frac{16}{3}
\frac{
\bigl[
r^*(\mathbf K_{k,0})+1
\bigr]
\bigl\{\log p_k+\log(1/\alpha_k)\bigr\}
}{n}
}.
\]
\end{lemma}
\begin{proof}

For each fixed \(k\), the unfoldings \(\mathcal X_{i,(k)}\) form an i.i.d.\ sample of \(p_k\times p_{-k}\) matrices. The row Kendall's tau
matrix and its sample estimator are exactly \(\mathbf K_{k,0}\) and
\(\widehat{\mathbf K}_{k,0}\). Applying
Lemma~\ref{lem:Kr0_concentration_main} with
\((p,q,\alpha_r)=(p_k,p_{-k},\alpha_k)\) proves the result.

\end{proof}

\begin{lemma}
\label{lem:tensor_eigenvalue}
Suppose that \(\mathcal X\sim\mathcal{TED}_{p_1\times\cdots\times p_K}
(\mathcal M,\boldsymbol\Sigma_1\otimes\cdots\otimes\boldsymbol\Sigma_K,\varphi)\).
Then, for each \(k\), \(\mathbf K_{k,0}\) and \(\boldsymbol\Sigma_k\)
share an eigenbasis, and
\[
\lambda_a(\mathbf K_{k,0})
=\mathbb E\left[
\frac{\lambda_a(\boldsymbol\Sigma_k)
\sum_{j=1}^{p_{-k}}\lambda_j(\boldsymbol\Sigma_{-k})Z_{aj}^2}
{\sum_{b=1}^{p_k}\sum_{j=1}^{p_{-k}}
\lambda_b(\boldsymbol\Sigma_k)\lambda_j(\boldsymbol\Sigma_{-k})Z_{bj}^2}
\right],\quad a=1,\ldots,p_k,
\]
where \(Z_{aj}\stackrel{\mathrm{i.i.d.}}{\sim}N(0,1)\).
Moreover, $\lambda_a(K_k)=p_k\lambda_a(K_{k,0})$,
$\lambda_a(\Sigma_k)=\lambda_b(\Sigma_k)
\iff\lambda_a(K_k)=\lambda_b(K_k)$, and
$\lambda_a(\Sigma_k)>\lambda_b(\Sigma_k)
\iff\lambda_a(K_k)>\lambda_b(K_k)$.
\end{lemma}
\begin{proof}
For each mode \(k\), put \(p_{-k}=\prod_{\ell\ne k}p_\ell\) and
\(\boldsymbol\Sigma_{-k}=\bigotimes_{\ell\ne k}\boldsymbol\Sigma_\ell\),
with the factor order corresponding to the mode-\(k\) unfolding.
Write \(\boldsymbol\Sigma_k=\mathbf U_k\boldsymbol\Lambda_k\mathbf U_k^\top\)
and \(\boldsymbol\Sigma_{-k}=\mathbf U_{-k}\boldsymbol\Lambda_{-k}\mathbf U_{-k}^\top\),
where \(\boldsymbol\Lambda_k=\operatorname{diag}(\lambda_{k,1},\ldots,\lambda_{k,p_k})\)
contains eigenvalues in nonincreasing order and
\(\boldsymbol\Lambda_{-k}=\operatorname{diag}(\omega_1,\ldots,\omega_{p_{-k}})\).
For a \(p_k\times p_{-k}\) matrix \(\mathbf Z=(Z_{aj})\) with i.i.d.\ standard
normal entries,
\[
\frac{\mathcal X_{(k)}-\widetilde{\mathcal X}_{(k)}}
{\|\mathcal X_{(k)}-\widetilde{\mathcal X}_{(k)}\|_F}
\stackrel d=
\frac{\mathbf U_k\boldsymbol\Lambda_k^{1/2}\mathbf Z
\boldsymbol\Lambda_{-k}^{1/2}\mathbf U_{-k}^\top}
{\|\boldsymbol\Lambda_k^{1/2}\mathbf Z\boldsymbol\Lambda_{-k}^{1/2}\|_F}.
\]
Let \(\tau_a=\sum_{j=1}^{p_{-k}}\omega_jZ_{aj}^2\).
The variables \(\tau_1,\ldots,\tau_{p_k}\) are i.i.d.\ and positive.
The mode-\(k\) unfolding is matrix elliptical with row and column scatter
matrices \(\boldsymbol\Sigma_k\) and \(\boldsymbol\Sigma_{-k}\).
Apply Lemma~\ref{lem:Eigenvalue} with
\((p,q,\boldsymbol\Sigma,\boldsymbol\Omega)
=(p_k,p_{-k},\boldsymbol\Sigma_k,\boldsymbol\Sigma_{-k})\) and $S_a=\tau_a$ gives the stated spectral identities and
eigenvalue ordering.
\end{proof}

%================================================
% Slice-based Kendall error
%================================================

\section{Proofs of MESIR Results}
\label{supp_proof_of_matrix}

\subsection{Proof of Proposition~\ref{prop:Kxs_error_main}}

\begin{proof}

Notice that
{
\begin{align*}
\left\|
\widehat{\mathbf K}_{r}^{\mathrm{IR}}
-
\mathbf K^{\mathrm{IR}}_{r}
\right\|_2
&=
\Biggl\|
\widetilde{\mathbf K}_{r}^{\mathrm{IR}}
-
\mathbf K^{\mathrm{IR}}_{r}
+
\mathbf K^{\mathrm{or}}_{r}
-
\widetilde{\mathbf K}_{r}^{\mathrm{IR}}
+
\mathbf K^{\mathrm{S}}_{r}
-
\mathbf K^{\mathrm{or}}_{r}
+
\widehat{\mathbf K}_{r}^{\mathrm{IR}}
-
\mathbf K^{\mathrm{S}}_{r}
\Biggr\|_2
\\
&=
\Biggl\|
\frac{2}{n(n-1)}
\sum_{i'<i}
\frac{
(\mathbf m_i-\mathbf m_{i'})
(\mathbf m_i-\mathbf m_{i'})^\top
}{
\|\mathbf m_i-\mathbf m_{i'}\|_F^2/p
}
-
\mathbb E\left[
\frac{
\{\mathbf m(Y)-\mathbf m(\widetilde Y)\}
\{\mathbf m(Y)-\mathbf m(\widetilde Y)\}^\top
}{
\|\mathbf m(Y)-\mathbf m(\widetilde Y)\|_F^2/p
}
\right]
\\
&+
\frac{2}{H(H-1)}
\sum_{s'<s}
\frac{
(\mathbf m_s-\mathbf m_{s'})
(\mathbf m_s-\mathbf m_{s'})^\top
}{
\|\mathbf m_s-\mathbf m_{s'}\|_F^2/p
}
-
\frac{2}{n(n-1)}
\sum_{i'<i}
\frac{
(\mathbf m_i-\mathbf m_{i'})
(\mathbf m_i-\mathbf m_{i'})^\top
}{
\|\mathbf m_i-\mathbf m_{i'}\|_F^2/p
}
\\
&+
\frac{2}{H(H-1)}
\sum_{s'<s}
\frac{
(\mathbf M_s-\mathbf M_{s'})
(\mathbf M_s-\mathbf M_{s'})^\top
}{
\|\mathbf M_s-\mathbf M_{s'}\|_F^2/p
}
-
\frac{2}{H(H-1)}
\sum_{s'<s}
\frac{
(\mathbf m_s-\mathbf m_{s'})
(\mathbf m_s-\mathbf m_{s'})^\top
}{
\|\mathbf m_s-\mathbf m_{s'}\|_F^2/p
}
\\
&+
\frac{2}{H(H-1)}
\sum_{s'<s}
\frac{
(\widehat{\mathbf M}_s-\widehat{\mathbf M}_{s'})
(\widehat{\mathbf M}_s-\widehat{\mathbf M}_{s'})^\top
}{
\|\widehat{\mathbf M}_s-\widehat{\mathbf M}_{s'}\|_F^2/p
}
-
\frac{2}{H(H-1)}
\sum_{s'<s}
\frac{
(\mathbf M_s-\mathbf M_{s'})
(\mathbf M_s-\mathbf M_{s'})^\top
}{
\|\mathbf M_s-\mathbf M_{s'}\|_F^2/p
}
\Biggr\|_2
\\
&:=
\|{\mathbf A}_1+{\mathbf A}_2+{\mathbf A}_3+{\mathbf A}_4\|_2
\\
&\le
\|{\mathbf A}_1\|_2+\|{\mathbf A}_2\|_2
+\|{\mathbf A}_3\|_2+\|{\mathbf A}_4\|_2,
\end{align*}
}
where
\(
\mathbf m_i=\mathbf m(Y_i),
\mathbf m_{sj}=\mathbf m(Y_{sj}),
\mathbf m_s=\frac1l\sum_{j=1}^l\mathbf m_{sj},
\)
and
\(
\mathbf M_s
=
\mathbb E(\mathbf X\mid Y\in J_s),
\widehat{\mathbf M}_s
=
\frac1l\sum_{j=1}^l\mathbf X_{sj}.
\)
For $\mathbf A_1$, Lemma~\ref{lem:Kr0_concentration_main} and
Assumption~\ref{assump:bounded_K} give
\[
\|\mathbf A_1\|_2
=O_p\!\left\{
\|\mathbf K_r^{\mathrm{IR}}\|_2
\sqrt{\frac{r^*(\mathbf K_r^{\mathrm{IR}})\log p}{n}}
\right\}
=O_p\!\left(\sqrt{\frac{p\log p}{n}}\right),
\qquad
r^*(\mathbf K_r^{\mathrm{IR}})
=\frac{\operatorname{tr}(\mathbf K_r^{\mathrm{IR}})}
{\|\mathbf K_r^{\mathrm{IR}}\|_2}\le p.
\]

For ${{\mathbf A}_2}$,
\begin{align*}
\widetilde{\mathbf K}_{r}^{\mathrm{IR}}/p
&=
\frac{2}{n(n-1)}
\left\{
\sum_{s'<s}
\sum_{j=1}^l\sum_{k=1}^l
\frac{
(\mathbf m_{sj}-\mathbf m_{s'k})
(\mathbf m_{sj}-\mathbf m_{s'k})^\top
}{
\|\mathbf m_{sj}-\mathbf m_{s'k}\|_F^2
}
\right.
\\
&\qquad\qquad\left.
+
\sum_{s=1}^H\sum_{j<k}
\frac{
(\mathbf m_{sj}-\mathbf m_{sk})
(\mathbf m_{sj}-\mathbf m_{sk})^\top
}{
\|\mathbf m_{sj}-\mathbf m_{sk}\|_F^2
}
\right\}
\\
&=
\frac{
l^2\binom{H}{2}
}{
\binom{n}{2}
}
\frac{1}{
l^2\binom{H}{2}
}
\sum_{s'<s}
\sum_{j=1}^l\sum_{k=1}^l
\frac{
(\mathbf m_{sj}-\mathbf m_{s'k})
(\mathbf m_{sj}-\mathbf m_{s'k})^\top
}{
\|\mathbf m_{sj}-\mathbf m_{s'k}\|_F^2
}
\\
&\quad+
\frac{
H\binom{l}{2}
}{
\binom{n}{2}
}
\frac{1}{
H\binom{l}{2}
}
\sum_{s=1}^H\sum_{j<k}
\frac{
(\mathbf m_{sj}-\mathbf m_{sk})
(\mathbf m_{sj}-\mathbf m_{sk})^\top
}{
\|\mathbf m_{sj}-\mathbf m_{sk}\|_F^2
}
\\
&=: {\mathbf B}_1+{\mathbf B}_2 .
\end{align*}

Consequently,
\begin{align*}
{\mathbf A}_2/p
&=
\frac{2}{H(H-1)}
\sum_{s'<s}
\frac{
(\mathbf m_s-\mathbf m_{s'})
(\mathbf m_s-\mathbf m_{s'})^\top
}{
\|\mathbf m_s-\mathbf m_{s'}\|_F^2
}
-
{\mathbf B}_1-{\mathbf B}_2                                                      \\
&=: {\mathbf B}_3-{\mathbf B}_1-{\mathbf B}_2 ,
\end{align*}
\[
\|{\mathbf A}_2\|_2/p
\le
\|{\mathbf B}_3-{\mathbf B}_1\|_2+\|{\mathbf B}_2\|_2 .
\]

Since
\(
\left\|
\frac{\mathbf U\mathbf U^\top}{\|\mathbf U\|_F^2}
\right\|_2
=
\frac{\|\mathbf U\|_2^2}{\|\mathbf U\|_F^2}
\le1,
\)
we have
\({
\|{\mathbf B}_2\|_2
\le
\frac{
H\binom{l}{2}
}{
\binom{n}{2}
}
=
O(H^{-1}).
}\)
Furthermore,
{
\begin{align*}
\|{\mathbf B}_3-{\mathbf B}_1\|_2
&=
\Biggl\|
\frac{1}{
l^2\binom{H}{2}
}
\sum_{s'<s}\sum_{j=1}^l\sum_{k=1}^l
\left\{
\frac{
(\mathbf m_{sj}-\mathbf m_{s'j})
(\mathbf m_{sk}-\mathbf m_{s'k})^\top
}{
\|\mathbf m_s-\mathbf m_{s'}\|_F^2
}
\right.
\\
&\qquad\qquad\left.
-
\frac{
(\mathbf m_{sj}-\mathbf m_{s'k})
(\mathbf m_{sj}-\mathbf m_{s'k})^\top
}{
\|\mathbf m_{sj}-\mathbf m_{s'k}\|_F^2
}
\right\}
\\
&\qquad+
\left(
1-
\frac{
l^2\binom{H}{2}
}{
\binom{n}{2}
}
\right)
\frac{1}{
l^2\binom{H}{2}
}
\sum_{s'<s}\sum_{j=1}^l\sum_{k=1}^l
\frac{
(\mathbf m_{sj}-\mathbf m_{s'k})
(\mathbf m_{sj}-\mathbf m_{s'k})^\top
}{
\|\mathbf m_{sj}-\mathbf m_{s'k}\|_F^2
}
\Biggr\|_2
\\
&=: \|{\mathbf D}_1+{\mathbf D}_2\|_2
\\
&\le
\|{\mathbf D}_1\|_2+\|{\mathbf D}_2\|_2 .
\end{align*}
}

Write
\(
\mathbf u_{sj,s'k}
=
\mathbf m_{sj}-\mathbf m_{s'k},
\mathbf v_{ss'}
=
\mathbf m_s-\mathbf m_{s'},
\mathbf V_{ss'}
=
\mathbf M_s-\mathbf M_{s'}.
\)
Since
\(
\mathbf v_{ss'}
=
\frac{1}{l^2}
\sum_{j=1}^l\sum_{k=1}^l
\mathbf u_{sj,s'k},
\)
we can write
{
\begin{align*}
{\mathbf D}_1
&=
\frac{1}{l^2\binom{H}{2}}
\sum_{s'<s}\sum_{j=1}^l\sum_{k=1}^l
\left\{
\frac{
\mathbf v_{ss'}\mathbf v_{ss'}^\top
}{
\|\mathbf v_{ss'}\|_F^2
}
-
\frac{
\mathbf u_{sj,s'k}\mathbf u_{sj,s'k}^\top
}{
\|\mathbf u_{sj,s'k}\|_F^2
}
\right\}.
\end{align*}
}

By adding 
$\mathbf V_{ss'}\mathbf V_{ss'}^\top/\|\mathbf V_{ss'}\|_F^2$ and Assumption~\ref{assump1},
we obtain
\begin{align*}
\|\mathbf D_1\|_F
&\le
\frac{1}{l^2\binom{H}{2}}
\sum_{s'<s}\sum_{j=1}^l\sum_{k=1}^l
\Biggl\{
\left\|
\frac{\mathbf v_{ss'}\mathbf v_{ss'}^\top}
     {\|\mathbf v_{ss'}\|_F^2}
-
\frac{\mathbf V_{ss'}\mathbf V_{ss'}^\top}
     {\|\mathbf V_{ss'}\|_F^2}
\right\|_F
+
\left\|
\frac{\mathbf u_{sj,s'k}\mathbf u_{sj,s'k}^\top}
     {\|\mathbf u_{sj,s'k}\|_F^2}
-
\frac{\mathbf V_{ss'}\mathbf V_{ss'}^\top}
     {\|\mathbf V_{ss'}\|_F^2}
\right\|_F
\Biggr\}
\\
&\le
\frac{2}{l^2\binom{H}{2}}
\sum_{s'<s}\sum_{j=1}^l\sum_{k=1}^l
\Biggl\{
\left\|
\frac{\mathbf v_{ss'}}{\|\mathbf v_{ss'}\|_F}
-
\frac{\mathbf V_{ss'}}{\|\mathbf V_{ss'}\|_F}
\right\|_F
+
\left\|
\frac{\mathbf u_{sj,s'k}}{\|\mathbf u_{sj,s'k}\|_F}
-
\frac{\mathbf V_{ss'}}{\|\mathbf V_{ss'}\|_F}
\right\|_F
\Biggr\}
\\
&\le
\frac{2}{l^2\binom{H}{2}}
\sum_{s'<s}\sum_{j=1}^l\sum_{k=1}^l
\Biggl\{
\frac{
\|\mathbf v_{ss'}-\mathbf V_{ss'}\|_F
+
\bigl|\|\mathbf V_{ss'}\|_F-\|\mathbf v_{ss'}\|_F\bigr|
}{
\|\mathbf V_{ss'}\|_F
}
\\
&\qquad\qquad+
\frac{
\|\mathbf u_{sj,s'k}-\mathbf V_{ss'}\|_F
+
\bigl|\|\mathbf V_{ss'}\|_F-\|\mathbf u_{sj,s'k}\|_F\bigr|
}{
\|\mathbf V_{ss'}\|_F
}
\Biggr\}
\\
&\le
\frac{4}{l^2\binom{H}{2}}
\sum_{s'<s}\sum_{j=1}^l\sum_{k=1}^l
\frac{
\|\mathbf v_{ss'}-\mathbf V_{ss'}\|_F
+
\|\mathbf u_{sj,s'k}-\mathbf V_{ss'}\|_F
}{
\|\mathbf V_{ss'}\|_F
}
\\
&\le
\frac{4C}{l^2\binom{H}{2}}
\sum_{s'<s}\sum_{j=1}^l\sum_{k=1}^l
\Bigl\{
\|\mathbf v_{ss'}-\mathbf V_{ss'}\|_F
+
\|\mathbf u_{sj,s'k}-\mathbf V_{ss'}\|_F
\Bigr\}.
\end{align*}

Let
\(
R_H
=
\max_s\max_{1\le j\le l}
\|\mathbf m_{sj}-\mathbf M_s\|_F.
\)
Since
\(
\mathbf m_s
=
\frac1l\sum_{j=1}^l\mathbf m_{sj},
\)
we have
\(
\max_s\|\mathbf m_s-\mathbf M_s\|_F
\le R_H.
\)
Consequently,
\(
\|\mathbf v_{ss'}-\mathbf V_{ss'}\|_F
\le 2R_H,
\|\mathbf u_{sj,s'k}-\mathbf V_{ss'}\|_F
\le 2R_H.
\)
It follows that
\({
\|{\mathbf D}_1\|_2
\le 16CR_H.
}\)

Therefore, using Assumption~\ref{assump2}
and the Assumption~\ref{assump1}, we obtain
\begin{align*}
\|\mathbf m_{sj}-\mathbf M_s\|_F
&=
\left\|
\mathbb E\left[
\mathbf m(Y_{sj})-\mathbf m(Y)
\mid Y\in J_s
\right]
\right\|_F
\\
&\le
\mathbb E\left[
\|\mathbf m(Y_{sj})-\mathbf m(Y)\|_F
\mid Y\in J_s
\right]
\\
&\le
C\sqrt{pq}\,
\max_{1\le j,k\le l}|Y_{sj}-Y_{sk}|.
\end{align*}
Thus,
\(
R_H
=
O_p(\sqrt{pq}H^{-1}),
\)
which gives
\({
\|{\mathbf D}_1\|_2
=
O_p(\sqrt{pq}H^{-1}).
}\)

Moreover,
{
\[
\|{\mathbf D}_2\|_2
\le
\left|
1-
\frac{
l^2\binom{H}{2}
}{
\binom{n}{2}
}
\right|
=
\frac{
H\binom{l}{2}
}{
\binom{n}{2}
}
=
O(H^{-1}).
\]
}
Combining the above bounds, we obtain
{
\[
\|{\mathbf A}_2\|_2
\le
p\{\|{\mathbf D}_1\|_2+\|{\mathbf D}_2\|_2+\|{\mathbf B}_2\|_2\}
=
O_p(p^{3/2}q^{1/2}H^{-1}).
\]
}

For $\mathbf A_3$, let
$\boldsymbol\Delta_s=\mathbf m_s-\mathbf M_s$.
The preceding bound for $R_H$ gives
$\max_s\|\boldsymbol\Delta_s\|_F\le R_H
=O_p(\sqrt{pq}H^{-1})$.
Since $\mathbf v_{ss'}-\mathbf V_{ss'}
=\boldsymbol\Delta_s-\boldsymbol\Delta_{s'}$, Assumption~\ref{assump1} yield
\begin{align*}
\frac{\|\mathbf A_3\|_2}{p}
&\le \frac{2}{H(H-1)}\sum_{s<s'}
\left\|
\frac{\mathbf V_{ss'}\mathbf V_{ss'}^\top}{\|\mathbf V_{ss'}\|_F^2}
-\frac{\mathbf v_{ss'}\mathbf v_{ss'}^\top}{\|\mathbf v_{ss'}\|_F^2}
\right\|_2\\
&\le \frac{8C}{H(H-1)}
\sum_{s<s'}\|\boldsymbol\Delta_s-\boldsymbol\Delta_{s'}\|_F\\
&\le \frac{8C}{H}\sum_{s=1}^H\|\boldsymbol\Delta_s\|_F
\le 8CR_H
=O_p(\sqrt{pq}H^{-1}).
\end{align*}
Hence $\|\mathbf A_3\|_2=O_p(p^{3/2}q^{1/2}H^{-1})$.

For ${{\mathbf A}_4}$, denote 
\(
\mathbf e_s=\widehat{\mathbf M}_s-\mathbf m_s
=\frac1l\sum_{j=1}^l(\mathbf X_{sj}-\mathbf m_{sj}),
\)
\(
\widehat{\mathbf V}_{ss'}
=\widehat{\mathbf M}_s-\widehat{\mathbf M}_{s'}.
\)
Then
\(
\widehat{\mathbf V}_{ss'}-\mathbf V_{ss'}
=(\mathbf e_s-\mathbf e_{s'})
+(\boldsymbol\Delta_s-\boldsymbol\Delta_{s'}).
\)
Consequently,
\[
\begin{aligned}
\frac{2}{H(H-1)}
\sum_{s<s'}
\|\widehat{\mathbf V}_{ss'}-\mathbf V_{ss'}\|_F
&\le
\frac2H\sum_{s=1}^H\|\mathbf e_s\|_F
+\frac2H\sum_{s=1}^H\|\boldsymbol\Delta_s\|_F\\
&\le
2\left\{\frac1H\sum_{s=1}^H\|\mathbf e_s\|_F^2\right\}^{1/2}
+2\max_s\|\boldsymbol\Delta_s\|_F\\
&=
O_p\!\left(\sqrt{\frac{pqH}{n}}\right)
+O_p(\sqrt{pq}H^{-1}).
\end{aligned}
\]

By Assumption~\ref{assump3}, \(
\|\widehat{\mathbf M}_s\|_F = O_p(\sqrt{pq}).
\) Therefore,
\(
\|\widehat{\mathbf V}_{ss'}\|_F
=
\|\widehat{\mathbf M}_s-\widehat{\mathbf M}_{s'}\|_F
\le
\|\widehat{\mathbf M}_s\|_F+\|\widehat{\mathbf M}_{s'}\|_F
=
O_p(\sqrt{pq}).
\) Moreover,
\[
\begin{aligned}
\|\mathbf V_{ss'}\|_F
&\le\|\mathbf V_{ss'}-\widehat{\mathbf V}_{ss'}\|_F
+\|\widehat{\mathbf V}_{ss'}\|_F\\
&\le\|\mathbf e_s\|_F+\|\mathbf e_{s'}\|_F
+\|\boldsymbol\Delta_s\|_F+\|\boldsymbol\Delta_{s'}\|_F
+\|\widehat{\mathbf V}_{ss'}\|_F\\
&=O_p\!\left(\sqrt{\frac{pqH}{n}}\right)
+O_p(\sqrt{pq}H^{-1})+O_p(\sqrt{pq})\\
&=O_p(\sqrt{pq}),
\end{aligned}
\]

Similar to the bound of ${{\mathbf A}_3}$, we have
\[
\begin{aligned}
\|{\mathbf A}_4\|_2/p
&=
\left\|
\frac{2}{H(H-1)}
\sum_{s<s'}
\left[
\frac{\widehat{\mathbf V}_{ss'}
      \widehat{\mathbf V}_{ss'}^\top}
     {\|\widehat{\mathbf V}_{ss'}\|_F^2}
-
\frac{\mathbf V_{ss'}\mathbf V_{ss'}^\top}
     {\|\mathbf V_{ss'}\|_F^2}
\right]
\right\|_2\\
&\le
\frac{8}{H(H-1)}
\sum_{s<s'}
\frac{\|\widehat{\mathbf V}_{ss'}-\mathbf V_{ss'}\|_F}
     {\|\mathbf V_{ss'}\|_F}\\
&\le
4\max_{s<s'}
\frac{\|\mathbf V_{ss'}\|_F+
      \|\widehat{\mathbf V}_{ss'}\|_F}
     {\|\mathbf V_{ss'}\|_F^2}
\frac{2}{H(H-1)}
\sum_{s<s'}
\|\widehat{\mathbf V}_{ss'}-\mathbf V_{ss'}\|_F.
\end{aligned}
\]

Then we have
{
\[
\begin{aligned}
\|{\mathbf A}_4\|_2
&=
p\,O_p((pq)^{-1/2})
\left\{
O_p\!\left(\sqrt{\frac{pqH}{n}}\right)
+O_p(\sqrt{pq}H^{-1})
\right\}\\
&=
O_p(pH^{1/2}n^{-1/2})
+O_p(pH^{-1}).
\end{aligned}
\]
}

Combining ${{\mathbf A}_1,{\mathbf A}_2,{\mathbf A}_3,{\mathbf A}_4}$,
\[
\left\|
\widehat{\mathbf K}_{r}^{\mathrm{IR}}
-
\mathbf K^{\mathrm{IR}}_{r}
\right\|_2
=
O_p(p^{3/2}q^{1/2}H^{-1})
+
O_p(pH^{1/2}n^{-1/2}).
\]
\end{proof}

\subsection{Proof of Proposition~\ref{cr:thm:Cr}}

\begin{proof}

Assume $p\ge2$.
Write ${{\mathbf D}_{ij}=\mathbf X_i-\mathbf X_j}$ and $N=\binom n2$, and let
{
\[
 \mathbf K_r=p\E\frac{{\mathbf D}{\mathbf D}^\top}{\ff {\mathbf D}^2},\qquad
 \widehat{\mathbf K}_r=\frac pN\sum_{i<j}\frac{{\mathbf D}_{ij}{\mathbf D}_{ij}^\top}{\ff{{\mathbf D}_{ij}}^2}
 =\Uh\operatorname{diag}(\widehat\kappa_{r,1},\ldots,\widehat\kappa_{r,p})\Uh^\top,
\]
}
where ${{\mathbf D}=\mathbf X-\widetilde{\mathbf X}}$,
and $\Uh=(\widehat{\mathbf u}_{r,1},\ldots,\widehat{\mathbf u}_{r,p})$.
Denote
\(
 \lambda_{r,0,k}=\widehat{\mathbf u}_{r,k}^\top\boldsymbol{\Sigma}\widehat{\mathbf u}_{r,k},
 {\boldsymbol{\lambda}}_{r,0}=(\lambda_{r,0,1},\ldots,\lambda_{r,0,p})^\top,
 \)
 \(\mathbf C_0=\Uh\operatorname{diag}({\boldsymbol{\lambda}}_{r,0})\Uh^\top.
\)
Notice that
\begin{equation}\label{cr:eq:decomp}
 \op{\widehat{\mathbf K}_r^{\mathrm{cal}}-\boldsymbol{\Sigma}}\le\ff{\widehat{\mathbf K}_r^{\mathrm{cal}}-\boldsymbol{\Sigma}}
 \le\underbrace{\ff{\widehat{\mathbf K}_r^{\mathrm{cal}}-\mathbf C_0}}_{\text{spectral-repair error}}
   +\underbrace{\ff{\mathbf C_0-\boldsymbol{\Sigma}}}_{\text{Kendall's tau eigenbasis error}}.
\end{equation}
We bound the two terms separately.

\medskip\noindent\textbf{Part I: Kendall's tau eigenbasis error.}
By Lemma~\ref{lem:Eigenvalue} \eqref{eq:Kr0_sigma} , $\mathbf K_r$ and $\boldsymbol{\Sigma}$ share an eigenbasis, and their corresponding eigenvalues $\kappa_{r,i}=\lambda_{i}(\mathbf K_r)$ and $\lambda_{r,i}=\lambda_{i}(\boldsymbol{\Sigma})$ satisfy
\[
 \kappa_{r,i}=p\E\frac{\lambda_{r,i}\tau_i}{\sum_a\lambda_{r,a}\tau_a},
 \qquad
 \tau_i=\sum_{\ell=1}^q\omega_\ell Z_{i\ell}^2,
\]
where $\omega_\ell$ are the eigenvalues of $\boldsymbol{\Omega}$, and $\{\tau_i\}_{i=1}^{p}$ are i.i.d. and positive.
Write $P_i=\tau_i/\sum_a\tau_a$; then $(P_1,\ldots,P_p)$ is exchangeable.
Fix $i\ne j$ and write
\(
 D_1=\sum_a\lambda_{r,a}P_a,
 D_2=\sum_{a\ne i,j}\lambda_{r,a}P_a
     +\lambda_{r,i}P_j+\lambda_{r,j}P_i,
 T_{ij}=P_i+P_j-(P_i-P_j)^2.
\)
Then we have
\[
\begin{aligned}
 \frac{\kappa_{r,i}-\kappa_{r,j}}{p}
 &=\frac12\E\left\{
 \frac{\lambda_{r,i}P_i-\lambda_{r,j}P_j}{D_1}
 +\frac{\lambda_{r,i}P_j-\lambda_{r,j}P_i}{D_2}\right\}\\
 &=\frac{\lambda_{r,i}-\lambda_{r,j}}{2}\E
 \frac{(P_i+P_j)\sum_{a\ne i,j}\lambda_{r,a}P_a
       +2(\lambda_{r,i}+\lambda_{r,j})P_iP_j}{D_1D_2}.
\end{aligned}
\]

Under Assumption~\ref{assump4} ${c_\Sigma}\le D_1,D_2\le {C_\Sigma}$, and 
\(
 (P_i+P_j)\sum_{a\ne i,j}\lambda_{r,a}P_a
       +2(\lambda_{r,i}+\lambda_{r,j})P_iP_j
\ge {c_\Sigma}\{(P_i+P_j)(1-P_i-P_j)+4P_iP_j\}
 =c_\Sigma T_{ij}\ge0.
\)
The Gaussian representation of $\{\tau_i\}_{i=1}^p$,
together with Jensen's inequality and the spherical
fourth-moment identities, gives
$\mathbb E(P_i-P_j)^2\le 4/\{p(p+2)\}$.
Since $\mathbb E P_i=1/p$, we obtain
$\mathbb E T_{ij}\ge 2/(p+2)$.
Thus, for $p\ge2$ and $\lambda_{r,i}\ne\lambda_{r,j}$,
\[
 \frac{|\kappa_{r,i}-\kappa_{r,j}|}{|\lambda_{r,i}-\lambda_{r,j}|}
 \ge\frac{p c_\Sigma}{2{C_\Sigma}^2}\E\{P_i+P_j-(P_i-P_j)^2\}
 \ge\frac{p {c_\Sigma}}{{C_\Sigma}^2(p+2)}
 \ge\frac{{c_\Sigma}}{2{C_\Sigma}^2}.
\]
When $\lambda_{r,i}=\lambda_{r,j}$, exchangeability gives $\kappa_{r,i}=\kappa_{r,j}$.
Hence there exists a scalar Lipschitz function $f$ satisfying
\[
 f(\kappa_{r,i})=\lambda_{r,i},\qquad f(\mathbf K_r)=\boldsymbol{\Sigma},\qquad
 |f(x)-f(y)|\le L|x-y|,\quad L=\frac{2{C_\Sigma}^2}{{c_\Sigma}}.
\]
For symmetric matrices $\mathbf Q_1,\mathbf Q_2$ with
eigenpairs $(a_i,\mathbf u_i)$ and $(b_j,\mathbf v_j)$,
respectively, the Lipschitz property of $f$ gives
\[
\|f(\mathbf Q_1)-f(\mathbf Q_2)\|_F^2
=
\sum_{i,j}\{f(a_i)-f(b_j)\}^2
(\mathbf u_i^\top\mathbf v_j)^2
\le L^2\|\mathbf Q_1-\mathbf Q_2\|_F^2.
\]
On the other hand, denote
\({
 {\mathbf A}=\Uh^\top f(\mathbf K_r)\Uh=\Uh^\top\boldsymbol{\Sigma}\Uh=(A_{ij})_{p\times p},
}\) \(
 \mathbf C_0
 =\Uh\operatorname{diag}(A_{11},\ldots,A_{pp})\Uh^\top
 =\Uh\operatorname{diag}\bigl(
   \widehat{\mathbf u}_{r,1}^\top\boldsymbol{\Sigma}\widehat{\mathbf u}_{r,1},\ldots,
   \widehat{\mathbf u}_{r,p}^\top\boldsymbol{\Sigma}\widehat{\mathbf u}_{r,p}
   \bigr)\Uh^\top,
\) \({
 \ff{\mathbf C_0-f(\mathbf K_r)}^2
 =\ff{\operatorname{diag}(A_{11},\ldots,A_{pp})-{\mathbf A}}^2
 =\sum_{i\ne j}A_{ij}^2.
}\)
For any ${{\mathbf d}=(d_1,\ldots,d_p)^\top\in\RR^p}$, let
${{\mathbf C}({\mathbf d})=\Uh\operatorname{diag}({\mathbf d})\Uh^\top}$. Then
{
\begin{align*}
 \ff{{\mathbf C}({\mathbf d})-f(\mathbf K_r)}^2
=\ff{\operatorname{diag}({\mathbf d})-{\mathbf A}}^2
 &=\sum_{i=1}^p(d_i-A_{ii})^2+\sum_{i\ne j}A_{ij}^2\\
 &\ge\sum_{i\ne j}A_{ij}^2
 =\ff{\mathbf C_0-f(\mathbf K_r)}^2.
\end{align*}
}

Take ${{\mathbf C}({\mathbf d})=f(\widehat{\mathbf K}_r)=\Uh f(\operatorname{diag}(\widehat\kappa_{r,1},\ldots,\widehat\kappa_{r,p}))\Uh^\top}$, then we have
\[
 \ff{\mathbf C_0-f(\mathbf K_r)}
 \le\ff{f(\widehat{\mathbf K}_r)-f(\mathbf K_r)}
 \le L\ff{\widehat{\mathbf K}_r-\mathbf K_r}
\]
We next control $\widehat{\mathbf K}_r-\mathbf K_r$. Denote
$h_K(\mathbf x,\mathbf y)=p(\mathbf x-\mathbf y)(\mathbf x-\mathbf y)^\top/\|\mathbf x-\mathbf y\|_F^2$,
$b_n=\lfloor n/2\rfloor$, $\E h_K(\mathbf X_1,\mathbf X_2)=\mathbf K_r$. For any permutation $\pi$, let
\({
 {\mathbf V}_\pi=\frac1{b_n}\sum_{a=1}^{b_n}
 h_K(\mathbf X_{\pi(2a-1)},\mathbf X_{\pi(2a)}), 
 \widehat{\mathbf K}_r=\frac1{n!}\sum_\pi {\mathbf V}_\pi.
}\)
Then we have
\begin{equation}\label{cr:eq:Krate}
\begin{aligned}
 \E\ff{\widehat{\mathbf K}_r-\mathbf K_r}^2
 &=\E\left\|\frac1{n!}\sum_\pi({\mathbf V}_\pi-\mathbf K_r)\right\|_F^2\\
 &\le\frac1{n!}\sum_\pi\E\|{\mathbf V}_\pi-\mathbf K_r\|_F^2\\
 &=\frac1{b_n^2}\sum_{a=1}^{b_n}
   \E\|h_K(\mathbf X_1,\mathbf X_2)-\mathbf K_r\|_F^2\\
 &=\frac1{b_n}\left\{\E\|h_K(\mathbf X_1,\mathbf X_2)\|_F^2-\|\mathbf K_r\|_F^2\right\}\\
 &\le\frac1{b_n}\E\|h_K(\mathbf X_1,\mathbf X_2)\|_F^2\\
 &=\frac{p^2}{b_n}\E\frac{\|{\mathbf D}{\mathbf D}^\top\|_F^2}{\|{\mathbf D}\|_F^4}
 \le\frac{p^2}{b_n}\E\frac{\{\tr({\mathbf D}{\mathbf D}^\top)\}^2}{\|{\mathbf D}\|_F^4}
 =\frac{p^2}{b_n}\le\frac{Cp^2}{n}.
\end{aligned}
\end{equation}

For any $t>0$, Markov's inequality further gives
\[
 \Pr\!\left\{\|\widehat{\mathbf K}_r-\mathbf K_r\|_F>t\frac p{\sqrt n}\right\}
 \le\frac{n}{t^2p^2}\E\|\widehat{\mathbf K}_r-\mathbf K_r\|_F^2
 \le\frac C{t^2},
 \qquad
 \|\widehat{\mathbf K}_r-\mathbf K_r\|_F=O_p(p/\sqrt n).
\]
Consequently,
\begin{equation}\label{cr:eq:basis}
 \ff{\mathbf C_0-\boldsymbol{\Sigma}}=\ff{\mathbf C_0-f(\mathbf K_r)}
 \le\ff{f(\widehat{\mathbf K}_r)-f(\mathbf K_r)}
 \le L\ff{\widehat{\mathbf K}_r-\mathbf K_r}
 =O_p(p/\sqrt n).
\end{equation}

\medskip\noindent\textbf{Part II: spectral-repair eigenvalue error.}
First express the calibration as a convex optimization problem. Define
${s_{ij,k}=\|{\mathbf e}_k^\top\Uh^\top {\mathbf D}_{ij}\|_2^2}$ and
\[
 L_n({\boldsymbol{\theta}})=\one^\top{\boldsymbol{\theta}}+
 \frac pN\sum_{i<j}\log
 \frac{\sum_k s_{ij,k}e^{-\theta_k}}{\sum_k s_{ij,k}},
\]
where $\mathbf e_k$is the $k$th standard basis vector and
$\mathbf 1_p=(1,\ldots,1)^\top\in\mathbb R^p$.
Since ${L_n({\boldsymbol{\theta}}+c\one)=L_n({\boldsymbol{\theta}})}$,
we remove the common-shift ambiguity by restricting optimization to
${\mathcal H_p=\{{\boldsymbol{\theta}}:\one^\top{\boldsymbol{\theta}}=0\}}$.
Define the weights
${w_{ij,k}({\boldsymbol{\theta}})=s_{ij,k}e^{-\theta_k}/
(\sum_a s_{ij,a}e^{-\theta_a})}$ and write
${{\mathbf w}_{ij}({\boldsymbol{\theta}})
=(w_{ij,1}({\boldsymbol{\theta}}),\ldots,
w_{ij,p}({\boldsymbol{\theta}}))^\top}$.
Direct differentiation gives
\begin{align}
 \nabla L_n({\boldsymbol{\theta}})
 &=\one-\frac pN\sum_{i<j}{\mathbf w}_{ij}({\boldsymbol{\theta}}),
 \label{cr:eq:grad}\\
 \nabla^2L_n({\boldsymbol{\theta}})
 &=\frac pN\sum_{i<j}
 \left\{\operatorname{diag}({\mathbf w}_{ij}({\boldsymbol{\theta}}))
 -{\mathbf w}_{ij}({\boldsymbol{\theta}})
 {\mathbf w}_{ij}({\boldsymbol{\theta}})^\top\right\}.
 \label{cr:eq:Hessian}
\end{align}
Since ${\sum_k w_{ij,k}({\boldsymbol{\theta}})=1}$,
${\one^\top\nabla L_n({\boldsymbol{\theta}})
=p-(p/N)\sum_{i<j}\sum_k w_{ij,k}({\boldsymbol{\theta}})
=p-(p/N)N=0}$.
Therefore, the gradient itself belongs to $\mathcal H_p$.

Let ${\widehat{\boldsymbol{\theta}}}$ be the exact minimizer
of $L_n$ over $\mathcal H_p$, and denote
\begin{equation}\label{cr:eq:estimator}
g_k({\boldsymbol{\theta}})=\frac{pe^{\theta_k}}{\sum_a e^{\theta_a}},
 \qquad
 \widehat{\boldsymbol{\lambda}}_r=g(\widehat{\boldsymbol{\theta}}),
 \qquad
 \widehat{\mathbf K}_r^{\mathrm{cal}}
 =\Uh\operatorname{diag}(\widehat{\boldsymbol{\lambda}}_r)\Uh^\top.
\end{equation}
At a constrained minimizer, ${\nabla L_n({\boldsymbol{\theta}})=0}$.

Writing ${{\boldsymbol{\lambda}}_r=g({\boldsymbol{\theta}})}$ gives
${w_{ij,k}({\boldsymbol{\theta}})=(s_{ij,k}/\lambda_{r,k})/
(\sum_a s_{ij,a}/\lambda_{r,a})}$.
Thus ${\nabla L_n({\boldsymbol{\theta}})=0}$ is equivalent to
\[
 \lambda_{r,k}=\frac pN\sum_{i<j}
 \frac{s_{ij,k}}{\sum_a s_{ij,a}/\lambda_{r,a}},
 \qquad \sum_k\lambda_{r,k}=p,
\]
which is the row calibration equation.
Hence \eqref{cr:eq:estimator} defines the corresponding calibrated estimator.

\smallskip\noindent\emph{(i) Reference standardization and gradient control.}

Let ${{\boldsymbol{\theta}}_0=\log{\boldsymbol{\lambda}}_{r,0}}$,
${{\boldsymbol{\theta}}_0^c={\boldsymbol{\theta}}_0-p^{-1}(\one^\top{\boldsymbol{\theta}}_0)\one}$,
and write
{
\[
 {\boldsymbol{\Phi}}_0=\operatorname{diag}({\boldsymbol{\lambda}}_{r,0}),\qquad
 {\mathbf B}={\boldsymbol{\Phi}}_0^{-1/2}\Uh^\top\boldsymbol{\Sigma}^{1/2},\qquad
 {\mathbf R}={\mathbf B}{\mathbf B}^\top
   ={\boldsymbol{\Phi}}_0^{-1/2}\Uh^\top\boldsymbol{\Sigma}\Uh{\boldsymbol{\Phi}}_0^{-1/2}.
\]
}
Since $\sum_k\lambda_{r,0,k}=p$, we have ${g({\boldsymbol{\theta}}_0^c)={\boldsymbol{\lambda}}_{r,0}}$.
Moreover, $\lambda_{r,0,k}=\widehat{\mathbf u}_k^\top\boldsymbol{\Sigma}\widehat{\mathbf u}_k$
implies $R_{kk}=1$. Using ${\mathbf C_0=\Uh{\boldsymbol{\Phi}}_0\Uh^\top}$, we obtain
\({
 {\mathbf R}-{\mathbf I}_p
 ={\boldsymbol{\Phi}}_0^{-1/2}
   \bigl(\Uh^\top\boldsymbol{\Sigma}\Uh-{\boldsymbol{\Phi}}_0\bigr){\boldsymbol{\Phi}}_0^{-1/2}
 ={\boldsymbol{\Phi}}_0^{-1/2}\Uh^\top(\boldsymbol{\Sigma}-\mathbf C_0)\Uh{\boldsymbol{\Phi}}_0^{-1/2}.
}\)
Assumption~\ref{assump4} and \(
 \lambda_{\min}(\boldsymbol{\Sigma})
 \le \widehat{\mathbf u}_k^\top\boldsymbol{\Sigma}\widehat{\mathbf u}_k
 =\lambda_{r,0,k}
 \le\lambda_{\max}(\boldsymbol{\Sigma})
\) give
${c_\Sigma}\le\lambda_{r,0,k}\le {C_\Sigma}$, so
${\op{{\boldsymbol{\Phi}}_0^{-1/2}}^2\le {c_\Sigma}^{-1}}$.
Together with \eqref{cr:eq:basis}, this yields

\begin{equation}\label{cr:eq:R}
\begin{aligned}
 \ff{{\mathbf R}-{\mathbf I}_p}
 &\le\op{{\boldsymbol{\Phi}}_0^{-1/2}}^2
      \ff{\Uh^\top(\boldsymbol{\Sigma}-\mathbf C_0)\Uh}\\
 &\le {c_\Sigma}^{-1}\ff{\boldsymbol{\Sigma}-\mathbf C_0}
 =O_p(p/\sqrt n).
\end{aligned}
\end{equation}

Furthermore, by Assumption~\ref{assump4} and \({{\boldsymbol{\Phi}}_0^{-1/2}\Uh^\top\boldsymbol{\Sigma}\Uh{\boldsymbol{\Phi}}_0^{-1/2}
 ={\mathbf R}}\), we have
{
\[
\frac{c_\Sigma}{C_\Sigma}
\leq \lambda_{\min}({\mathbf R})
\leq \lambda_{\max}({\mathbf R})
\leq \frac{C_\Sigma}{c_\Sigma}.
\]
}

Denote
\[
 {\mathbf G}_{ij}
 =\frac{\boldsymbol{\Sigma}^{-1/2}{\mathbf D}_{ij}{\mathbf D}_{ij}^\top\boldsymbol{\Sigma}^{-1/2}}
        {\tr(\boldsymbol{\Sigma}^{-1}{\mathbf D}_{ij}{\mathbf D}_{ij}^\top)},
 \qquad
 {\mathbf M}_n=\frac pN\sum_{i<j}{\mathbf G}_{ij}.
\]
Then $\mathbf G_{ij}\succeq0$,
$\operatorname{tr}(\mathbf G_{ij})=1$, and
$\|\mathbf G_{ij}\|_F\le1$.
The representation
$\mathbf G_{ij}\overset d=
\mathbf Z\boldsymbol\Omega\mathbf Z^\top/
\operatorname{tr}(\mathbf Z\boldsymbol\Omega\mathbf Z^\top)$
implies orthogonal invariance, so
$\mathbb E\mathbf G_{ij}=\mathbf I_p/p$ and
$\mathbb E\mathbf M_n=\mathbf I_p$.

Since $\mathbf M_n$ is an order-two U-statistic,
the argument for \eqref{cr:eq:Krate} yields
\[
\begin{aligned}
 \E\ff{{\mathbf M}_n-{\mathbf I}_p}^2
 &\le\frac1{b_n}\E\ff{p{\mathbf G}_{12}-{\mathbf I}_p}^2\\
 &=\frac1{b_n}
   \left\{p^2\E\ff{{\mathbf G}_{12}}^2-\ff{{\mathbf I}_p}^2\right\}\\
 &\le\frac{p^2}{b_n}\E\ff{{\mathbf G}_{12}}^2
 \le\frac{p^2}{b_n}
 \le\frac{Cp^2}{n},
 \qquad b_n=\lfloor n/2\rfloor.
\end{aligned}
\]

Markov's inequality and $p/\sqrt n\to0$ imply
\begin{equation}\label{cr:eq:M}
\eta_n:=\|\mathbf M_n-\mathbf I_p\|_F
=O_p(p/\sqrt n)=o_p(1),
\qquad
\|\mathbf M_n\|_2
\le1+\|\mathbf M_n-\mathbf I_p\|_F
=1+\eta_n.
\end{equation}

We next express the gradient at $\theta_0^c$.
Let $\mathbf O=\mathbf R^{-1/2}\mathbf B$, which is orthogonal
since $\mathbf B\mathbf B^\top=\mathbf R$, and write
$\overline{\mathbf G}_{ij}=\mathbf O\mathbf G_{ij}\mathbf O^\top$.
Define
\[
\mathbf S_n(\mathbf R)
=\frac pN\sum_{i<j}
\frac{\overline{\mathbf G}_{ij}}
     {\operatorname{tr}(\mathbf R\overline{\mathbf G}_{ij})},
\qquad
\mathbf F_n(\mathbf R)
=\mathbf R^{1/2}\mathbf S_n(\mathbf R)\mathbf R^{1/2}.
\]
Since $\mathbf B=\mathbf R^{1/2}\mathbf O$, the definitions of
$\mathbf B$ and $\mathbf G_{ij}$ give
\[
\mathbf F_n(\mathbf R)
=\frac pN\sum_{i<j}
\frac{\mathbf B\mathbf G_{ij}\mathbf B^\top}
     {\operatorname{tr}(\mathbf B\mathbf G_{ij}\mathbf B^\top)}
=\frac pN\sum_{i<j}
\frac{
\boldsymbol\Phi_0^{-1/2}\widehat{\mathbf U}_r^\top
\mathbf D_{ij}\mathbf D_{ij}^\top\widehat{\mathbf U}_r
\boldsymbol\Phi_0^{-1/2}
}{
\operatorname{tr}\!\left(
\boldsymbol\Phi_0^{-1/2}\widehat{\mathbf U}_r^\top
\mathbf D_{ij}\mathbf D_{ij}^\top\widehat{\mathbf U}_r
\boldsymbol\Phi_0^{-1/2}
\right)
}.
\]
The $k$th diagonal entry of each normalized summand is
$(s_{ij,k}/\lambda_{r,0,k})/
(\sum_a s_{ij,a}/\lambda_{r,0,a})=w_{ij,k}(\theta_0^c)$.
Hence $\operatorname{diagvec}\{\mathbf F_n(\mathbf R)\}
=(p/N)\sum_{i<j}\mathbf w_{ij}(\theta_0^c)$, and
\eqref{cr:eq:grad} yields
\[
\nabla L_n(\theta_0^c)
=\mathbf 1_p-\operatorname{diagvec}\{\mathbf F_n(\mathbf R)\},
\]
where $\operatorname{diagvec}$ extracts the diagonal entries
as a column vector. Thus
$\|\nabla L_n(\theta_0^c)\|_2
\le\|\mathbf F_n(\mathbf R)-\mathbf I_p\|_F$.

To bound the right-hand side, write
$\overline{\mathbf M}_n=\mathbf O\mathbf M_n\mathbf O^\top$.
Orthogonality gives
$\operatorname{tr}(\overline{\mathbf G}_{ij})=1$,
$\|\overline{\mathbf M}_n-\mathbf I_p\|_F=\eta_n$, and
$\|\overline{\mathbf M}_n\|_2=\|\mathbf M_n\|_2$.

Next bound ${{\mathbf S}_n({\mathbf R})-\overline {\mathbf M}_n}$,
\[
\begin{aligned}
 {\mathbf S}_n({\mathbf R})-\overline {\mathbf M}_n
 &=\frac pN\sum_{i<j}\overline {\mathbf G}_{ij}
   \left\{\frac1{\tr({\mathbf R}\overline {\mathbf G}_{ij})}-1\right\}\\
 &=-\frac pN\sum_{i<j}\overline {\mathbf G}_{ij}
   \frac{\tr\{({\mathbf R}-{\mathbf I}_p)\overline {\mathbf G}_{ij}\}}
        {\tr({\mathbf R}\overline {\mathbf G}_{ij})}.
\end{aligned}
\]

Denote $c_*={c_\Sigma}/{C_\Sigma}$ and $C_*={C_\Sigma}/{c_\Sigma}$. Since ${\tr\overline {\mathbf G}_{ij}=1}$,  ${\overline {\mathbf G}_{ij}\succeq0}$,
\({
 \tr({\mathbf R}\overline {\mathbf G}_{ij})
 \ge c_*\tr\overline {\mathbf G}_{ij}=c_*.
}\) For any symmetric matrix $\mathbf H$, the Cauchy--Schwarz
inequality gives
\begin{align*}
\frac pN\sum_{i<j}
\{\operatorname{tr}(\mathbf H\overline{\mathbf G}_{ij})\}^2
&=
\frac pN\sum_{i<j}
\left\langle
\overline{\mathbf G}_{ij}^{1/2},
\mathbf H\overline{\mathbf G}_{ij}^{1/2}
\right\rangle_F^2
\\
&\le
\frac pN\sum_{i<j}
\|\overline{\mathbf G}_{ij}^{1/2}\|_F^2
\|\mathbf H\overline{\mathbf G}_{ij}^{1/2}\|_F^2
\\
&=
\frac pN\sum_{i<j}
\operatorname{tr}(\mathbf H^2\overline{\mathbf G}_{ij})
=
\operatorname{tr}(\mathbf H^2\overline{\mathbf M}_n)
\\
&\le
\|\overline{\mathbf M}_n\|_2\operatorname{tr}(\mathbf H^2)
=
\|\mathbf M_n\|_2\|\mathbf H\|_F^2.
\end{align*}

Applying Cauchy--Schwarz again gives

\[
\begin{aligned}
 &\left|\tr\{{\mathbf H}({\mathbf S}_n({\mathbf R})-\overline {\mathbf M}_n)\}\right|\\
 &\quad\le c_*^{-1}\frac pN\sum_{i<j}
 |\tr({\mathbf H}\overline {\mathbf G}_{ij})|\,
 |\tr\{({\mathbf R}-{\mathbf I}_p)\overline {\mathbf G}_{ij}\}|\\
 &\quad\le c_*^{-1}
 \left[\frac pN\sum_{i<j}\{\tr({\mathbf H}\overline {\mathbf G}_{ij})\}^2\right]^{1/2}
 \left[\frac pN\sum_{i<j}
 \{\tr(({\mathbf R}-{\mathbf I}_p)\overline {\mathbf G}_{ij})\}^2\right]^{1/2}\\
 &\quad\le c_*^{-1}\op{{\mathbf M}_n}\ff {\mathbf H}\ff{{\mathbf R}-{\mathbf I}_p}.
 \end{aligned}
\]

If ${{\mathbf S}_n({\mathbf R})=\overline {\mathbf M}_n}$, the bound is immediate.
Otherwise, choose 
\({
 {\mathbf H}=({\mathbf S}_n({\mathbf R})-\overline {\mathbf M}_n)/\ff{{\mathbf S}_n({\mathbf R})-\overline {\mathbf M}_n},
}\)
The preceding inequality gives
\begin{equation}
\begin{aligned}
 \ff{{\mathbf S}_n({\mathbf R})-\overline {\mathbf M}_n}
 \le c_*^{-1}\op{{\mathbf M}_n}\ff{{\mathbf R}-{\mathbf I}_p}.
 \label{cr:eq:scoreidentity}
 \end{aligned}
\end{equation}

Overall, since \({
 {\mathbf F}_n({\mathbf R})-{\mathbf I}_p
 ={}{\mathbf R}^{1/2}({\mathbf S}_n({\mathbf R})-\overline {\mathbf M}_n){\mathbf R}^{1/2}
+{\mathbf R}^{1/2}(\overline {\mathbf M}_n-{\mathbf I}_p){\mathbf R}^{1/2}
 +({\mathbf R}-{\mathbf I}_p),
}\)
together with $\|\mathbf R\|_2\le C_*$,
\eqref{cr:eq:R}, \eqref{cr:eq:M}, and \eqref{cr:eq:scoreidentity}, gives
\begin{equation}\label{cr:eq:scorerate}
\begin{aligned}
\|\nabla L_n(\boldsymbol\theta_0^c)\|_2
&=\|\operatorname{diagvec}\{\mathbf I_p-\mathbf F_n(\mathbf R)\}\|_2
\le\|\mathbf F_n(\mathbf R)-\mathbf I_p\|_F\\
&\le C_*\|\mathbf S_n(\mathbf R)-\overline{\mathbf M}_n\|_F
+C_*\|\overline{\mathbf M}_n-\mathbf I_p\|_F
+\|\mathbf R-\mathbf I_p\|_F\\
&\le\{1+C_*c_*^{-1}\|\mathbf M_n\|_2\}\|\mathbf R-\mathbf I_p\|_F
+C_*\eta_n
=O_p(p/\sqrt n).
\end{aligned}
\end{equation}

\smallskip\noindent\emph{(ii) Positive Hessian curvature and local strong convexity.}

Let ${{\mathbf G}}$ have the same distribution as ${{\mathbf G}_{ij}}$. Diagonalizing $\boldsymbol{\Omega}$, we can write
{
\[
 {\mathbf G}\overset d=
 \frac{\sum_{\ell=1}^q\omega_\ell {\mathbf z}_\ell {\mathbf z}_\ell^\top}
      {\sum_{\ell=1}^q\omega_\ell\|{\mathbf z}_\ell\|_2^2},
 \qquad {\mathbf z}_\ell\overset{\mathrm{iid}}\sim N_p(\mathbf0,{\mathbf I}_p),
\]
}
where $\omega_\ell$ are the eigenvalues of $\boldsymbol\Omega$.
Let $r_\ell=\|\mathbf z_\ell\|_2$ and
$\mathbf v_\ell=\mathbf z_\ell/r_\ell$. Then
$\mathbf G\overset d=\sum_{\ell=1}^q
\alpha_\ell\mathbf v_\ell\mathbf v_\ell^\top$,
where $\alpha_\ell=\omega_\ell r_\ell^2/
\sum_t\omega_t r_t^2$.
The directions $\mathbf v_1,\ldots,\mathbf v_q$ are i.i.d.\
uniform on $\mathbb S^{p-1}$ and independent of
$\boldsymbol\alpha=(\alpha_1,\ldots,\alpha_q)^\top$.

For any fixed symmetric matrix ${{\mathbf H}}$,
${\tr({\mathbf H}{\mathbf G})=\sum_\ell\alpha_\ell {\mathbf v}_\ell^\top {\mathbf H}{\mathbf v}_\ell}$. Then we have
{
\[
\begin{aligned}
 \E\!\left[\{\tr({\mathbf H}{\mathbf G})\}^2\mid{\boldsymbol{\alpha}}\right]
 &=\E\!\left[
   \left(\sum_\ell\alpha_\ell {\mathbf v}_\ell^\top {\mathbf H}{\mathbf v}_\ell\right)^2
   \,\middle|\,{\boldsymbol{\alpha}}\right]\\
 &\le\sum_\ell\alpha_\ell
       \E\!\left[({\mathbf v}_\ell^\top {\mathbf H}{\mathbf v}_\ell)^2\mid{\boldsymbol{\alpha}}\right]\\
 &=\sum_\ell\alpha_\ell
   \frac{(\tr {\mathbf H})^2+2\ff {\mathbf H}^2}{p(p+2)}
 =\frac{(\tr {\mathbf H})^2+2\ff {\mathbf H}^2}{p(p+2)}.
\end{aligned}
\]
}
Taking expectations about ${{\boldsymbol{\alpha}}}$ again yields
{
\[
 \E\{\tr({\mathbf H}{\mathbf G})\}^2
 \le\frac{(\tr {\mathbf H})^2+2\ff {\mathbf H}^2}{p(p+2)}.
\]
}
In particular, when ${\operatorname{tr}({\mathbf H})=0}$,
${p\mathbb E\{\operatorname{tr}({\mathbf H}{\mathbf G})\}^2
\le 2\|{\mathbf H}\|_F^2/(p+2)}$.

For a linear operator $\mathcal A$ on the real symmetric
$p\times p$ matrices, define
$\|\mathcal A\|_{\mathrm{HS}}^2
=\sum_{a=1}^{d_p}\|\mathcal A[\mathbf E_a]\|_F^2$,
where $\{\mathbf E_a\}_{a=1}^{d_p}$ is a Frobenius-orthonormal
basis and $d_p=p(p+1)/2$.
Define $\mathcal A_G[\mathbf H]=\tr(\mathbf H\mathbf G)\mathbf G$,
$\mathcal T_n=(p/N)\sum_{i<j}\mathcal A_{G_{ij}}$, and
$\mathcal T=p\mathbb E\mathcal A_G$.
Since $\|\mathcal A_G\|_{\mathrm{HS}}^2=\|\mathbf G\|_F^4\le1$,
the U-statistic second-moment bound gives
\[
\mathbb E\|\mathcal T_n-\mathcal T\|_{\mathrm{HS}}^2
\le\frac1{b_n}
\mathbb E\|p\mathcal A_{G_{12}}-\mathcal T\|_{\mathrm{HS}}^2
\le\frac{p^2}{b_n}
\le\frac{Cp^2}{n}.
\]
Thus $\zeta_n:=\|\mathcal T_n-\mathcal T\|_{\mathrm{HS}}
=O_p(p/\sqrt n)=o_p(1)$ when $p=o(n^{1/2})$.
For every symmetric matrix $\mathbf H$,
$|\langle\mathbf H,(\mathcal T_n-\mathcal T)[\mathbf H]\rangle_F|
\le\zeta_n\|\mathbf H\|_F^2$.
Combining the bound above, for all symmetric $\mathbf H$ with
$\tr(\mathbf H)=0$,
\begin{equation}\label{cr:eq:T}
\begin{aligned}
\frac pN\sum_{i<j}\{\tr(\mathbf H\mathbf G_{ij})\}^2
&=\langle\mathbf H,\mathcal T[\mathbf H]\rangle_F
+\langle\mathbf H,(\mathcal T_n-\mathcal T)[\mathbf H]\rangle_F\\
&\le\left(\frac2{p+2}+\zeta_n\right)\|\mathbf H\|_F^2.
\end{aligned}
\end{equation}

By \eqref{cr:eq:Hessian}, for ${{\mathbf v}\in\mathcal H_p}$,
\({
{\mathbf v}^\top\nabla^2L_n({\boldsymbol{\theta}}_0^c){\mathbf v}
=\sum_k v_k^2[{\mathbf F}_n({\mathbf R})]_{kk}
-\frac pN\sum_{i<j}
\left\{\sum_kv_kw_{ij,k}({\boldsymbol{\theta}}_0^c)\right\}^2,
}\)
where the first term is bounded below by
${(1-d_n)\|{\mathbf v}\|_2^2}$, with
${d_n=\|{\mathbf F}_n({\mathbf R})-{\mathbf I}_p\|_2=O_p(p/\sqrt n)}$.
Let ${{\mathbf D}_v=\operatorname{diag}({\mathbf v})}$ and
${{\mathbf H}_v={\mathbf R}^{1/2}{\mathbf D}_v{\mathbf R}^{1/2}}$. Since $R_{kk}=1$ and
${\one^\top {\mathbf v}=0}$, we have

\[
\operatorname{tr}({\mathbf H}_v)=0,\qquad
\|{\mathbf H}_v\|_F\le\lambda_{\max}({\mathbf R})\|{\mathbf v}\|_2,\qquad
\sum_kv_kw_{ij,k}({\boldsymbol{\theta}}_0^c)
=\frac{\operatorname{tr}({\mathbf H}_v\overline {\mathbf G}_{ij})}
{\operatorname{tr}({\mathbf R}\overline {\mathbf G}_{ij})}.
\]

Together with \eqref{cr:eq:T}, these bounds yield

\[
\frac pN\sum_{i<j}
\left\{\sum_kv_kw_{ij,k}({\boldsymbol{\theta}}_0^c)\right\}^2
\le
\left(\frac{\lambda_{\max}({\mathbf R})}{\lambda_{\min}({\mathbf R})}\right)^2
\left(\frac{2}{p+2}+\zeta_n\right)\|{\mathbf v}\|_2^2.
\]

Thus ${\ff{{\mathbf R}-{\mathbf I}_p}=o_p(1)}$, $d_n,\zeta_n=o_p(1)$ and $2/(p+2)\le1/2$ imply

\begin{equation}\label{cr:eq:H0}
 {\mathbf v}^\top\nabla^2L_n({\boldsymbol{\theta}}_0^c){\mathbf v}\ge\tfrac14\|{\mathbf v}\|_2^2,
 \qquad\forall {\mathbf v}\in\mathcal H_p,
\end{equation}
Moreover, for any ${{\mathbf h}\in\mathcal H_p}$ with ${\|{\mathbf h}\|_\infty\le\delta}$, then

\[
 w_{ij,k}({\boldsymbol{\theta}}_0^c+{\mathbf h})=
 \frac{w_{ij,k}({\boldsymbol{\theta}}_0^c)e^{-h_k}}{\sum_a w_{ij,a}({\boldsymbol{\theta}}_0^c)e^{-h_a}}
 \ge e^{-2\delta}w_{ij,k}({\boldsymbol{\theta}}_0^c).
\]

With \({
 {\mathbf v}^\top\nabla^2L_n({\boldsymbol{\theta}}){\mathbf v}
 =\frac pN\sum_{i<j}\sum_{k<a}w_{ij,k}({\boldsymbol{\theta}})w_{ij,a}({\boldsymbol{\theta}})(v_k-v_a)^2,
}\)
we extend \eqref{cr:eq:H0} to the whole neighborhood:
\begin{equation}\label{cr:eq:local}
 {\mathbf v}^\top\nabla^2L_n({\boldsymbol{\theta}}_0^c+{\mathbf h}){\mathbf v}\ge c_\delta\|{\mathbf v}\|_2^2,
 \quad c_\delta=\tfrac14e^{-4\delta},\quad
 {\mathbf v},{\mathbf h}\in\mathcal H_p,\quad\|{\mathbf h}\|_\infty\le\delta.
\end{equation}

\smallskip\noindent\emph{(iii) From gradient error to log-parameter error.}
For ${{\mathbf h}\in\mathcal H_p}$ with ${\|{\mathbf h}\|_\infty\le\delta}$,
Taylor's integral formula and \eqref{cr:eq:local} give

\begin{align*}
 L_n({\boldsymbol{\theta}}_0^c+{\mathbf h})-L_n({\boldsymbol{\theta}}_0^c)
 &=\nabla L_n({\boldsymbol{\theta}}_0^c)^\top {\mathbf h}
 +\int_0^1(1-t){\mathbf h}^\top\nabla^2L_n({\boldsymbol{\theta}}_0^c+t{\mathbf h}){\mathbf h}\,dt\\
 &\ge-\|\nabla L_n({\boldsymbol{\theta}}_0^c)\|_2\|{\mathbf h}\|_2
       +\tfrac12c_\delta\|{\mathbf h}\|_2^2.
\end{align*}

Fix $\delta>0$. Since $\|\nabla L_n(\theta_0^c)\|_2=o_p(1)$,
the preceding bound is strictly positive uniformly over
$\{\mathbf h\in\mathcal H_p:\|\mathbf h\|_2=\delta\}$
with probability tending to one.
Hence the minimum over the corresponding closed ball is attained
in its interior. By convexity and local strong convexity,
this minimizer is global and unique on $\mathcal H_p$.
Using $L_n(\widehat\theta)\le L_n(\theta_0^c)$ then yields
\begin{equation}\label{cr:eq:thetarate}
\|\widehat{\boldsymbol\theta}-\boldsymbol\theta_0^c\|_2
\le\frac{2}{c_\delta}\|\nabla L_n(\boldsymbol\theta_0^c)\|_2
=O_p(p/\sqrt n).
\end{equation}

\smallskip\noindent
\emph{(iv) Calibrated eigenvalue error.}
For \eqref{cr:eq:estimator},
\[
Dg({\boldsymbol{\theta}})=\operatorname{diag}(g({\boldsymbol{\theta}}))-p^{-1}g({\boldsymbol{\theta}})g({\boldsymbol{\theta}})^\top,
\qquad
0\preceq Dg({\boldsymbol{\theta}})\preceq\operatorname{diag}(g({\boldsymbol{\theta}})).
\]

Since ${g_k({\boldsymbol{\theta}})\le e^{2\delta}\lambda_{r,0,k}
\le e^{2\delta}C_\Sigma}$ whenever
${\|{\boldsymbol{\theta}}-{\boldsymbol{\theta}}_0^c\|_\infty\le\delta}$,
$g$ is Lipschitz on this neighborhood.
Using $g(\boldsymbol\theta_0^c)=\boldsymbol\lambda_{r,0}$ and \eqref{cr:eq:thetarate}, we obtain
\begin{equation}\label{cr:eq:repair}
\begin{aligned}
\|\widehat{\mathbf K}_r^{\mathrm{cal}}-\mathbf C_0\|_F
&=\|\Uh\operatorname{diag}(\widehat{\boldsymbol\lambda}_r
-\boldsymbol\lambda_{r,0})\Uh^\top\|_F
=\|\widehat{\boldsymbol\lambda}_r-\boldsymbol\lambda_{r,0}\|_2\\
&\le e^{2\delta}C_\Sigma
\|\widehat{\boldsymbol\theta}-\boldsymbol\theta_0^c\|_2
=O_p(p/\sqrt n).
\end{aligned}
\end{equation}

Substituting \eqref{cr:eq:basis} and \eqref{cr:eq:repair} into the initial decomposition \eqref{cr:eq:decomp} gives
\[
 \ff{\widehat{\mathbf K}_r^{\mathrm{cal}}-\boldsymbol{\Sigma}}
 \le\ff{\widehat{\mathbf K}_r^{\mathrm{cal}}-\mathbf C_0}+\ff{\mathbf C_0-\boldsymbol{\Sigma}}
 =O_p(p/\sqrt n).
\]

\end{proof}

\subsection{Proof of Theorem~\ref{thm:mesir_consistency_main}}
\begin{proof}

By Assumption~\ref{assump4} and Proposition~\ref{cr:thm:Cr}, if $p=o(n^{1/2})$, 
$\|\widehat{\mathbf K}_r^{\mathrm{cal}}-\boldsymbol\Sigma\|_2
=O_p(p/\sqrt n)=o_p(1)$.
Weyl's inequality gives
\[
\lambda_{\min}(\widehat{\mathbf K}_r^{\mathrm{cal}})
\ge c_\Sigma-\|\widehat{\mathbf K}_r^{\mathrm{cal}}-\boldsymbol\Sigma\|_2
=c_\Sigma-o_p(1),
\]
so $\|(\widehat{\mathbf K}_r^{\mathrm{cal}})^{-1}\|_2=O_p(1)$.
The inverse identity then yields
\[
\begin{aligned}
\|(\widehat{\mathbf K}_r^{\mathrm{cal}})^{-1}-\boldsymbol\Sigma^{-1}\|_2
&=\|\boldsymbol\Sigma^{-1}
(\boldsymbol\Sigma-\widehat{\mathbf K}_r^{\mathrm{cal}})
(\widehat{\mathbf K}_r^{\mathrm{cal}})^{-1}\|_2\\
&\le\|\boldsymbol\Sigma^{-1}\|_2
\|\widehat{\mathbf K}_r^{\mathrm{cal}}-\boldsymbol\Sigma\|_2
\|(\widehat{\mathbf K}_r^{\mathrm{cal}})^{-1}\|_2
=O_p(p/\sqrt n).
\end{aligned}
\]
Combining these bounds with Assumption~\ref{assump:bounded_K}
and Proposition~\ref{prop:Kxs_error_main}, we obtain
\begin{equation}\label{eq:mesir_theorem_bound}
\begin{aligned}
&\|(\widehat{\mathbf K}_r^{\mathrm{cal}})^{-1}
\widehat{\mathbf K}_r^{\mathrm{IR}}
-\boldsymbol\Sigma^{-1}\mathbf K_r^{\mathrm{IR}}\|_2\\
&\quad\le
\|(\widehat{\mathbf K}_r^{\mathrm{cal}})^{-1}-\boldsymbol\Sigma^{-1}\|_2
\|\mathbf K_r^{\mathrm{IR}}\|_2
+\|(\widehat{\mathbf K}_r^{\mathrm{cal}})^{-1}\|_2
\|\widehat{\mathbf K}_r^{\mathrm{IR}}-\mathbf K_r^{\mathrm{IR}}\|_2\\
&\quad=O_p(pn^{-1/2})
+O_p(p^{3/2}q^{1/2}H^{-1})
+O_p(pH^{1/2}n^{-1/2})\\
&\quad=O_p(p^{3/2}q^{1/2}H^{-1})
+O_p(pH^{1/2}n^{-1/2}),
\end{aligned}
\end{equation}
where the calibration term is absorbed since $H\ge1$.

\end{proof}

\subsection{Proof of Corollary~\ref{cor:mesir_consistency_row}}

\begin{proof}

Let $a_{n,r}=p^{3/2}q^{1/2}H^{-1}+pH^{1/2}n^{-1/2}$, and let $\boldsymbol\Gamma_1$ and $\widehat{\boldsymbol\Gamma}_1$
contain the leading $d_1$
orthonormal eigenvectors of $\mathbf K_r^{\mathrm{IR}}$ and
$\widehat{\mathbf K}_r^{\mathrm{IR}}$,
respectively. By Assumption~\ref{assump:bounded_K},
\(
 \delta_1:=\lambda_{d_1}(\mathbf K_r^{\mathrm{IR}})
 -\lambda_{d_1+1}(\mathbf K_r^{\mathrm{IR}})\ge c_r>0.
\)
Applying the Davis--Kahan theorem in \citep{Han02012018}
to these symmetric matrices gives
\[
 \|\mathbf P_{\widehat{\boldsymbol\Gamma}_1}
 -\mathbf P_{\boldsymbol\Gamma_1}\|_2
 \le \frac{C}{\delta_1}
 \|\widehat{\mathbf K}_r^{\mathrm{IR}}-\mathbf K_r^{\mathrm{IR}}\|_2
 =O_p(a_{n,r}).
\]

Since
$\widehat{\mathbf B}_1=
(\widehat{\mathbf K}_r^{\mathrm{cal}})^{-1}
\widehat{\boldsymbol\Gamma}_1$
and
$\operatorname{Span}(\mathbf B_1)
=\operatorname{Span}(\boldsymbol{\Sigma}^{-1}\boldsymbol\Gamma_1)$,
then we have
\[
\begin{aligned}
 \|\mathbf P_{\widehat{\mathbf B}_1}
 -\mathbf P_{\mathbf B_1}\|_2
 &\le C\left\{
 \|\mathbf P_{\widehat{\boldsymbol\Gamma}_1}
 -\mathbf P_{\boldsymbol\Gamma_1}\|_2
 +\|(\widehat{\mathbf K}_r^{\mathrm{cal}})^{-1}
 -\boldsymbol{\Sigma}^{-1}\|_2
 \right\}\\
 &\le C\left\{
 \delta_1^{-1}
 \|\widehat{\mathbf K}_r^{\mathrm{IR}}-\mathbf K_r^{\mathrm{IR}}\|_2
 +\|\widehat{\mathbf K}_r^{\mathrm{cal}}
 -\boldsymbol{\Sigma}\|_2
 \right\}\\
 &=O_p(a_{n,r})+O_p(p/\sqrt n)\\
 &=O_p(p^{3/2}q^{1/2}H^{-1}
 +pH^{1/2}n^{-1/2}).
\end{aligned}
\]
\end{proof}

\subsection{Proof of Theorem~\ref{thm:mesir_dimension_selection_consistency}}

\begin{proof}
We prove the result for $\widehat d_1$; the proof for $\widehat d_2$ is analogous. Let
\[
\Delta_{n,r}
=
\left\|
\widehat{\mathbf K}_{r}^{\mathrm{IR}}
-
\mathbf K^{\mathrm{IR}}_{r}
\right\|_2 .
\]
By Proposition~\ref{prop:Kxs_error_main} and Weyl's inequality,
\begin{align*}
\Delta_{n,r}
&=
O_p(a_{n,r}),\qquad
\max_{1\le j\le p}
\left|
\widehat\lambda_j^{(r)}
-
\lambda_j^{(r)}
\right|
\le
\Delta_{n,r},\qquad
\tau_{n,r}
=
a_{n,r}^{\gamma}\to0,
\qquad
\frac{\Delta_{n,r}}{\tau_{n,r}}
=
O_p(a_{n,r}^{1-\gamma})
=
o_p(1).
\end{align*}
Hence, for
\(\mathcal E_n=\{\Delta_{n,r}\le\tau_{n,r}\}\),
we have \(P(\mathcal E_n)\to1\). Recall that
\[
R_{n,j}^{(r)}
=
\frac{
\widehat\lambda_j^{(r)}+\tau_{n,r}
}{
\widehat\lambda_{j+1}^{(r)}+\tau_{n,r}
},
\qquad
1\le j\le d_{1,\max}.
\]

On $\mathcal E_n$, Assumption~\ref{assump:bounded_K} and Weyl's inequality imply
\(c_r-\Delta_{n,r}\le\widehat\lambda_j^{(r)}
\le C_r+\Delta_{n,r}\) for \(1\le j\le d_1\),
whereas
\(0\le\widehat\lambda_j^{(r)}\le\Delta_{n,r}\) for \(j>d_1\).
Therefore,
\begin{align*}
\max_{1\le j<d_1}R_{n,j}^{(r)}
&\le
\frac{
C_r+\Delta_{n,r}+\tau_{n,r}
}{
c_r-\Delta_{n,r}+\tau_{n,r}
}
\le
\frac{C_r+2\tau_{n,r}}{c_r}
=
O(1),\\
R_{n,d_1}^{(r)}
&\ge
\frac{
c_r-\Delta_{n,r}+\tau_{n,r}
}{
\Delta_{n,r}+\tau_{n,r}
}
\ge
\frac{c_r}{2\tau_{n,r}}
\longrightarrow\infty,\\
\max_{d_1<j\le d_{1,\max}}R_{n,j}^{(r)}
&\le
\frac{
\Delta_{n,r}+\tau_{n,r}
}{
\tau_{n,r}
}
=
1+\frac{\Delta_{n,r}}{\tau_{n,r}}
=
O_p(1).
\end{align*}

Thus,
\[
R_{n,d_1}^{(r)}
\xrightarrow{p}\infty,
\qquad
\max_{\substack{1\le j\le d_{1,\max}\\ j\ne d_1}}
R_{n,j}^{(r)}
=
O_p(1),
\]
which gives
\[
P\left\{
R_{n,d_1}^{(r)}
>
\max_{\substack{1\le j\le d_{1,\max}\\ j\ne d_1}}
R_{n,j}^{(r)}
\right\}
\to1.
\]
By the definition of $\widehat d_1$,
\(P(\widehat d_1=d_1)\to1\). The same argument gives
\(P(\widehat d_2=d_2)\to1\).
\end{proof}
\section{Extension}
\label{subsec:tensor_mesir}

\subsection{Tensor elliptical sliced inverse regression}
\label{subsec:tesir}

We extend the matrix elliptical sliced inverse regression (MESIR) framework in Section \ref{subsec:mesir} to a general $K$-mode tensor predictor
$\mathcal X\in\mathbb R^{p_1\times\cdots\times p_K}$ and a univariate response $Y \in \mathbb{R}^{1}.$ Tensor sufficient dimension reduction seeks to reduce the dimensions of all modes simultaneously while retaining all regression information about \(Y\) and preserving the multiway structure of \(\mathcal X\). Let $\mathbf B_k\in\mathbb R^{p_k\times d_k}$,
$k=1,\ldots,K$, be semi-orthogonal matrices.
We consider
\begin{equation}
\label{TESIR-core}
Y\perp\!\!\!\perp \mathcal X
\mid
\mathcal X\times_1\mathbf B_1^{\top}
\times_2\mathbf B_2^{\top}
\cdots
\times_K\mathbf B_K^{\top}.
\end{equation}
When \(\operatorname{Span}(\mathbf B_k)\), \(k=1,\ldots,K\), are the smallest mode-\(k\) reduction subspaces satisfying this conditional independence relation, we define
\[
\mathcal S_{Y\mid\mathcal X}
=
\operatorname{Span}
(\mathbf B_K\otimes\cdots\otimes\mathbf B_1)
\]
as the central tensor subspace, where \(d_k\) is the structural dimension along mode \(k\). The proposed tensor elliptical sliced inverse regression (TESIR) estimates the mode-\(k\) reduction subspaces through the mode-\(k\) unfoldings
$\mathcal X_{(k)}\in\mathbb R^{p_k\times p_{-k}}$,
where $p_{-k}=\prod_{j\ne k}p_j$. The key theoretical ingredient is a mode-wise linearity property that links the inverse regression of each unfolding to the corresponding mode-\(k\) reduction subspace. The following proposition establishes this property under the tensor elliptical model.

\begin{proposition} 
\label{pro:tensor_linear_condition}
Suppose that
\(
\mathcal X
\sim
\mathcal{TED}_{p_1\times\cdots\times p_K}
\left(
\mathcal M,
\boldsymbol\Sigma_1\otimes\cdots\otimes\boldsymbol\Sigma_K,
\varphi
\right),
\)
then for any column-full-rank
\({\mathbf A_k}\in\mathbb R^{p_k\times d_k}\) $(d_k\le p_k)$, {we have}
\[
\mathbb E
\left(
\mathcal X_{(k)}-\mathcal M_{(k)}
\mid
{\mathbf A_k}^\top\mathcal X_{(k)}
\right)
=
\boldsymbol\Sigma_k{\mathbf A_k}
({\mathbf A_k}^\top\boldsymbol\Sigma_k{\mathbf A_k})^{-1}
{\mathbf A_k}^\top
(\mathcal X_{(k)}-\mathcal M_{(k)}) .
\]

\end{proposition}
For each \(k\), the reduced tensor in \eqref{TESIR-core} is a function of \(\mathbf B_k^\top\mathcal X_{(k)}\). Hence,
\[
Y\perp\!\!\!\perp\mathcal X_{(k)}
\mid
\mathbf B_k^\top\mathcal X_{(k)}.
\]
Applying Proposition~\ref{pro:tensor_linear_condition} to the mode-\(k\) reduction basis \(\mathbf B_k\), we obtain
\[
\begin{aligned}
\mathbb E(\mathcal X_{(k)}-\mathcal M_{(k)}\mid Y)
&=\mathbb E\left\{
\mathbb E(\mathcal X_{(k)}-\mathcal M_{(k)}
\mid Y,\mathbf B_k^\top\mathcal X_{(k)})\mid Y
\right\}\\
&=\mathbb E\left\{
\mathbb E(\mathcal X_{(k)}-\mathcal M_{(k)}
\mid\mathbf B_k^\top\mathcal X_{(k)})\mid Y
\right\}\\
&=\boldsymbol\Sigma_k\mathbf B_k
(\mathbf B_k^\top\boldsymbol\Sigma_k\mathbf B_k)^{-1}
\mathbf B_k^\top
\mathbb E(\mathcal X_{(k)}-\mathcal M_{(k)}\mid Y)\\
&=\mathbf P^{\top}_{\mathbf B_k(\boldsymbol\Sigma_k)}
\mathbb E(\mathcal X_{(k)}-\mathcal M_{(k)}\mid Y).
\end{aligned}
\]

Therefore, the mode-\(k\) inverse regression curve satisfies
\begin{equation}\label{eq:tensor_inverse_regression_spaces}
\operatorname{Span}
\left\{
\mathbb E(\mathcal X_{(k)}\mid Y)
-
\mathbb E(\mathcal X_{(k)})
\right\}
\subseteq
\operatorname{Span}
(\boldsymbol\Sigma_k\mathbf B_k).
\end{equation}

Following the Kendall's tau matrices construction of \citet{he2025new},
we define the marginal and inverse-regression
Kendall's tau matrices for each mode $k$.
Let $(\widetilde{\mathcal X},\widetilde Y)$
be an independent copy of $(\mathcal X,Y)$. The mode-\(k\) marginal Kendall's tau matrix is defined as
\[
\mathbf K_k
=
\mathbb E
\left[
\frac{
(\mathcal X_{(k)}-\widetilde{\mathcal X}_{(k)})
(\mathcal X_{(k)}-\widetilde{\mathcal X}_{(k)})^\top
}
{
\|\mathcal X_{(k)}-\widetilde{\mathcal X}_{(k)}\|_F^2/p_k
}
\right].
\]
The inverse-regression Kendall's tau matrix is defined as
\[
\mathbf K_{k}^{\mathrm{IR}}
=
\mathbb E
\left[
\frac{
\{
\mathbb E(\mathcal X_{(k)}\mid Y)
-
\mathbb E(\widetilde{\mathcal X}_{(k)}
\mid \widetilde Y)
\}
\{
\mathbb E(\mathcal X_{(k)}\mid Y)
-
\mathbb E(\widetilde{\mathcal X}_{(k)}
\mid \widetilde Y)
\}^{\top}
}
{
\|
\mathbb E(\mathcal X_{(k)}\mid Y)
-
\mathbb E(\widetilde{\mathcal X}_{(k)}
\mid \widetilde Y)
\|_F^2/p_k
}
\right].
\]

The inverse-regression Kendall's tau matrix
$\mathbf K_k^{\mathrm{IR}}$ shares the column space
of $\operatorname{Cov}\{\mathbb E(\mathcal X_{(k)}\mid Y)\}$. Together with
\eqref{eq:tensor_inverse_regression_spaces},
this gives
\[
\operatorname{Span}(\mathbf K_k^{\mathrm{IR}})
\subseteq\operatorname{Span}(\boldsymbol\Sigma_k\mathbf B_k).
\]
To resolve the scale indeterminacy, we impose
the normalization
$\operatorname{tr}(\boldsymbol\Sigma_k)=p_k$. Under the tensor elliptical model,
$\mathbf K_k$ and $\boldsymbol\Sigma_k$
share eigenvectors,
but the corresponding eigenvalues are
generally not proportional. We therefore retain the marginal
Kendall's tau eigenvectors and calibrate
the scatter eigenvalues using
\begin{equation}
\label{eq:tensor_population_scatter_calibration}
\boldsymbol\Sigma_k
=
p_k\mathbb E\left\{
\frac{
(\mathcal X_{(k)}-\widetilde{\mathcal X}_{(k)})
(\mathcal X_{(k)}-\widetilde{\mathcal X}_{(k)})^\top
}{
\operatorname{tr}\left[
(\mathcal X_{(k)}-\widetilde{\mathcal X}_{(k)})^\top
\boldsymbol\Sigma_k^{-1}
(\mathcal X_{(k)}-\widetilde{\mathcal X}_{(k)})
\right]
}
\right\}.
\end{equation}

Given i.i.d.\ observations \(\{(\mathcal X_i,Y_i)\}_{i=1}^n\), the sample mode-\(k\) marginal Kendall's tau matrix is
\begin{equation}\label{eq:tesir_sample_marginal}
\widehat{\mathbf K}_k
=\frac{2}{n(n-1)}
\sum_{1\le i<i'\le n}
\frac{(\mathcal X_{i,(k)}-\mathcal X_{i',(k)})
(\mathcal X_{i,(k)}-\mathcal X_{i',(k)})^\top}
{\|\mathcal X_{i,(k)}-\mathcal X_{i',(k)}\|_F^2/p_k}.
\end{equation}

Let $\widehat{\mathbf U}_k=(\widehat{\mathbf u}_{k,1},\ldots,
\widehat{\mathbf u}_{k,p_k})$ be an orthogonal eigenvector matrix
of $\widehat{\mathbf K}_k$, and set
$s_{ij,a}^{(k)}=\|\widehat{\mathbf u}_{k,a}^{\top}
(\mathcal X_{i,(k)}-\mathcal X_{j,(k)})\|_2^2$.
Using $\widehat{\mathbf U}_k$ as the fixed eigenbasis, the
empirical diagonal equations corresponding to
\eqref{eq:tensor_population_scatter_calibration} are
\[
\widehat\lambda_{k,a}
=\frac{2p_k}{n(n-1)}\sum_{i<j}
\frac{s_{ij,a}^{(k)}}
{\sum_{b=1}^{p_k}s_{ij,b}^{(k)}/\widehat\lambda_{k,b}}
\quad \text{s.t.} \quad \sum_{a=1}^{p_k}\widehat\lambda_{k,a}=p_k,
\quad a=1,\ldots,p_k,
\]
A positive solution to the empirical calibration equations
defines the eigenvalue-calibrated marginal Kendall's tau matrix
\[
\widehat{\mathbf K}_k^{\mathrm{cal}}
=\widehat{\mathbf U}_k
\operatorname{diag}(\widehat\lambda_{k,1},\ldots,
\widehat\lambda_{k,p_k})
\widehat{\mathbf U}_k^\top,
\]
which estimates $\boldsymbol\Sigma_k$.
The solution is obtained using the iterative procedure
in Algorithm~\ref{alg:tesir_calibration}.

To estimate \(\mathbf K_k^{\mathrm{IR}}\), we partition the range of \(Y\) into \(H\) slices \(J_1,\ldots,J_H\). For each slice, we estimate the conditional mean $\mathbb E(\mathcal X\mid Y\in J_s)$ by $\widehat{\mathcal M}_s
=n_s^{-1}\sum_{i:Y_i\in J_s}\mathcal X_i$,
where \(n_s\) is the number of observations in \(J_s\). We then unfold these tensor slice means along mode \(k\) and construct
\begin{equation}\label{eq:tesir_sample_ir}
\widehat{\mathbf K}_{k}^{\mathrm{IR}}
=\frac{2}{H(H-1)}
\sum_{1\le s<s'\le H}
\frac{
\{(\widehat{\mathcal M}_s)_{(k)}-(\widehat{\mathcal M}_{s'})_{(k)}\}
\{(\widehat{\mathcal M}_s)_{(k)}-(\widehat{\mathcal M}_{s'})_{(k)}\}^\top}
{\|(\widehat{\mathcal M}_s)_{(k)}-(\widehat{\mathcal M}_{s'})_{(k)}\|_F^2/p_k}.
\end{equation}

 Algorithm~\ref{alg:tensor_esir} summarizes the TESIR procedure for estimating the mode-\(k\) reduction subspaces and the resulting central tensor subspace.

\begin{algorithm}
\caption{TESIR for mode-wise reduction subspace estimation}
\label{alg:tensor_esir}
\begin{algorithmic}[1]

\Require Observations \(\{(\mathcal X_i,Y_i)\}_{i=1}^n\), number of slices \(H\), structural dimensions \((d_1,\ldots,d_K)\).

\State Partition the responses into \(H\) slices \(\{J_s\}_{s=1}^{H}\) and compute the tensor slice means \(\{\widehat{\mathcal M}_s\}_{s=1}^{H}\).

\For{$k=1,\ldots,K$}

\State Compute the sample marginal Kendall's tau matrix
$\widehat{\mathbf K}_k$ in \eqref{eq:tesir_sample_marginal}
and the sample inverse-regression Kendall's tau matrix
$\widehat{\mathbf K}_k^{\mathrm{IR}}$ in \eqref{eq:tesir_sample_ir}. Take $\widehat{\mathbf K}_k$ 
as input to Algorithm~\ref{alg:tesir_calibration}
to obtain the eigenvalue-calibrated marginal Kendall's tau matrix
$\widehat{\mathbf K}_k^{\mathrm{cal}}$.

\State Take the leading $d_k$ orthonormal eigenvectors
of $\widehat{\mathbf K}_k^{\mathrm{IR}}$,
denoted by $\widehat{\boldsymbol\Gamma}_k$.

\State Compute
$\widehat{\mathbf B}_k
=
(\widehat{\mathbf K}_k^{\mathrm{cal}})^{-1}
\widehat{\boldsymbol\Gamma}_k$.

\EndFor

\Ensure $\widehat{\mathbf B}_1,\ldots,\widehat{\mathbf B}_K$
and $\widehat{\mathcal S}_{Y\mid\mathcal X}
=
\operatorname{Span}
(\widehat{\mathbf B}_K\otimes\cdots\otimes
\widehat{\mathbf B}_1)$.

\end{algorithmic}
\end{algorithm}
The resulting reduction predictor
$\mathcal X\times_1\widehat{\mathbf B}_1^\top
\cdots\times_K\widehat{\mathbf B}_K^\top$
is a $d_1\times\cdots\times d_K$ tensor.
When $K=2$, the TESIR formulation reduces to the  MESIR formulation in Section~\ref{subsec:mesir}.

\begin{algorithm}[!htbp]
\caption{Iterative eigenvalue calibration of mode-wise marginal Kendall's tau matrices}
\label{alg:tesir_calibration}
\begin{algorithmic}[1]

\Require Observations $\{\mathcal X_i\}_{i=1}^n$,
$\{\widehat{\mathbf K}_k\}_{k=1}^K$, iterative
tolerance $\varepsilon>0$, and maximum iterations $m_{\max}$.

\For{$k=1,\ldots,K$}

\State Compute the orthogonal eigenvector matrix
$\widehat{\mathbf U}_k
=(\widehat{\mathbf u}_{k,1},\ldots,
\widehat{\mathbf u}_{k,p_k})$
of $\widehat{\mathbf K}_k$, and set
$s_{ij,a}^{(k)}
=\|\widehat{\mathbf u}_{k,a}^{\top}
(\mathcal X_{i,(k)}-\mathcal X_{j,(k)})\|_2^2$.

\State Initialize $\lambda_{k,a}^{(0)}>0$,
$k=1,\ldots,p_k$, using the eigenvalues of
$\widehat{\mathbf K}_k$, with small eigenvalues
replaced by positive constants.

\State For $m=0,\ldots,m_{\max}-1$, compute
\[
\widetilde\lambda_{k,a}^{(m+1)}
=\frac{2p_k}{n(n-1)}
\sum_{i<j}
\frac{s_{ij,a}^{(k)}}
{\sum_{b=1}^{p_k}s_{ij,b}^{(k)}/\lambda_{k,b}^{(m)}},
\quad a=1,\ldots,p_k.
\]

\State If
$\sqrt{\sum_{a=1}^{p_k}
(\widetilde\lambda_{k,a}^{(m+1)}/
\lambda_{k,a}^{(m)}-1)^2}
\leq\varepsilon$,
stop; otherwise, set
\[
\lambda_{k,a}^{(m+1)}
=\frac{p_k\widetilde\lambda_{k,a}^{(m+1)}}
{\sum_{b=1}^{p_k}\widetilde\lambda_{k,b}^{(m+1)}}, \quad a=1,\ldots,p_k,
\]
and repeat.

\State Upon convergence, set
$\widehat\lambda_{k,a}=\lambda_{k,a}^{(m)}$
and form
$\widehat{\mathbf K}_k^{\mathrm{cal}}
=\widehat{\mathbf U}_k
\operatorname{diag}(\widehat\lambda_{k,1},
\ldots,\widehat\lambda_{k,p_k})
\widehat{\mathbf U}_k^\top$.

\EndFor

\Ensure $\{\widehat{\mathbf K}_k^{\mathrm{cal}}\}_{k=1}^K$.

\end{algorithmic}
\end{algorithm}

\noindent\textit{Structural dimension selection.}
In practice, the structural dimensions \(d_1,\ldots,d_K\) need to be estimated from the data. We extend the eigenvalue-ratio procedure in Section~\ref{subsec:mesir_model_selection} to select the structural dimension for each mode. 

Specifically, define
\[
a_{n,k}
=
p_k^{3/2}p_{-k}^{1/2}H^{-1}
+
p_kH^{1/2}n^{-1/2},
\qquad
\tau_{n,k}=a_{n,k}^{\gamma},
\quad 0<\gamma<1.
\]
For a prespecified maximum dimension $d_{k,\max}<p_k$,
we estimate $d_k$ by
\begin{equation}
\label{eq:TESIR-Auto}
\widehat d_k
=
\underset{1\le j\le d_{k,\max}}{\arg\max}
\frac{\widehat{\lambda}_j^{(k)}+\tau_{n,k}}
{\widehat{\lambda}_{j+1}^{(k)}+\tau_{n,k}},
\qquad k=1,\ldots,K.
\end{equation}
Here \(a_{n,k}\) is the mode-\(k\) estimation error rate established in Appendix~\ref{sub:TESIR Theoretical properties}. After obtaining \(\widehat d_1,\ldots,\widehat d_K\), we apply Algorithm ~\ref{alg:tensor_esir} using the selected dimensions. Specifically, for each mode \(k\), we take the leading \(\widehat d_k\) eigenvectors of \(\widehat{\mathbf K}_k^{\mathrm{IR}}\) to form \(\widehat{\boldsymbol\Gamma}_k\), and estimate the reduction basis by \(\widehat{\mathbf B}_k=(\widehat{\mathbf K}_k^{\mathrm{cal}})^{-1}\widehat{\boldsymbol\Gamma}_k,\ k=1,\ldots,K\).

\subsection{Theoretical analysis}
\label{sub:TESIR Theoretical properties}
In this section, we establish the asymptotic properties of TESIR. The analysis focuses on convergence of the calibrated marginal and sample inverse-regression Kendall's tau matrices for each mode, recovery of the corresponding reduction subspaces, and consistency of structural dimension selection. For theoretical convenience, we assume \(n=Hl\), with each slice containing exactly \(l\) observations, where both \(H\) and \(l\) may diverge with \(n\). We present the results for a general mode \(k\in\{1,\ldots,K\}\).

Denote the inverse regression curve by
$\mathcal M(Y)=\mathbb E(\mathcal X\mid Y)
\in\mathbb R^{p_1\times\cdots\times p_K}$.
We write $\mathcal X=\mathcal M(Y)+\mathcal E$,
where $\mathcal E$ is the residual tensor.
For the sample $\{(\mathcal X_i,Y_i)\}_{i=1}^n$, let
\[
\mathcal X_i
=\mathcal M(Y_i)+\mathcal E_i
=:\mathcal M_i+\mathcal E_i,
\qquad i=1,\ldots,n.
\]

We first establish the convergence rate of the sample mode-\(k\) inverse-regression Kendall's tau matrix in Proposition~\ref{prop:tensor_slice_kendall_error}. Together with the convergence of the eigenvalue-calibrated marginal Kendall's tau matrix in Proposition~\ref{ck:thm:Ck}, this gives the convergence rate in Theorem~\ref{thm:tensor_mesir_operator_consistency} and consistency of the estimated mode-$k$ reduction subspace in Corollary~\ref{cor:tensor_mode_subspace_consistency}. Finally, Theorem~\ref{theo:TESIR-Auto} establishes consistency of the structural dimension estimators. The following assumptions extend those for MESIR in Section~\ref{section-MESIR Teory} to the tensor setting and serve analogous roles in the theoretical analysis.

\begin{tensorassumption}
\label{ass:tensor_slice_regular}
Let $\mathcal M_s=\mathbb E(\mathcal X\mid Y\in J_s)$,
$s=1,\ldots,H$, and for the $a$th observation
$Y_{sa}$ in slice $J_s$, write
\(
\boldsymbol\mu_{sa}^{(k)}
=[\mathcal M(Y_{sa})]_{(k)},
\boldsymbol\mu_s^{(k)}
=(\mathcal M_s)_{(k)},
\widehat{\boldsymbol\mu}_s^{(k)}
=(\widehat{\mathcal M}_s)_{(k)}.
\)
For each mode \(k=1,\ldots,K\),
assume that the mode-wise inverse regression curves
satisfy the following slice-separation and local
regularity conditions:
\[
\max_{s\neq s'}
\frac{1}{
\|\boldsymbol\mu_s^{(k)}-\boldsymbol\mu_{s'}^{(k)}\|_F
}
<\infty,
\qquad
\max_{s\neq s'}
\max_{1\le a,b\le l}
\frac{
\|\boldsymbol\mu_{sa}^{(k)}-\boldsymbol\mu_{s'b}^{(k)}\|_F
}{
\|\boldsymbol\mu_s^{(k)}-\boldsymbol\mu_{s'}^{(k)}\|_F
}
<\infty,
\]
and
\[
\max_{1\le s\le H}
\max_{1\le a,b\le l}
|Y_{sa}-Y_{sb}|
=
O_p(H^{-1}) .
\]
\end{tensorassumption}

\begin{tensorassumption}
\label{ass:tensor_smoothness}
Let \(\mathcal M(Y)=\mathbb E(\mathcal X\mid Y)= 
\{M_{a_1,\ldots,a_K}(Y)\}_{1\le a_k\le p_k,\;k=1,\ldots,K}\). Assume that each \(M_{a_1,\ldots,a_K}(y)\) is continuously differentiable in $y$, with
\[
\sup_{1\le a_k\le p_k,\;k=1,\ldots,K}\sup_y
\left|\frac{d}{dy}M_{a_1,\ldots,a_K}(y)\right|=O(1).
\]
\end{tensorassumption}

\begin{tensorassumption}
\label{ass:tensor_bounded_inverse_kendall}
Let $\lambda_j^{(k)}
=\lambda_j(\mathbf K_k^{\mathrm{IR}})$,
$j=1,\ldots,p_k$.
Assume that there exist constants
$0<c_k<C_k<\infty$ such that, for each
$k=1,\ldots,K$,
\(
c_k\le\lambda_{d_k}^{(k)}
\le\cdots\le\lambda_1^{(k)}
\le C_k
\).
\end{tensorassumption}

\begin{tensorassumption}
\label{ass:tensor_moment}
Assume that \(
\mathcal X
\sim
\mathcal{TED}_{p_1\times\cdots\times p_K}
\left(
\mathcal M,
\boldsymbol\Sigma_1\otimes\cdots\otimes\boldsymbol\Sigma_K,
\varphi
\right)
\) and that, for some constant \(m\ge2\),
\[
\sup_{a_1,\ldots,a_K}
\mathbb E
\left|
X_{a_1,\ldots,a_K}
\right|^m
<\infty .
\]
\end{tensorassumption}

\begin{tensorassumption}
\label{ass:tensor_kendall_scaling}
For each \(k=1,\ldots,K\), assume that there exist positive constants
\(
0<c_{\Sigma_k}\le C_{\Sigma_k}<\infty,
\) such that $0<c_{\Sigma_k}
\le
\lambda_{p_k}(\boldsymbol\Sigma_k)
\le
\lambda_1(\boldsymbol\Sigma_k)
\le C_{\Sigma_k}
<\infty$.
\end{tensorassumption}

We first quantify the estimation error of the sample inverse-regression Kendall's tau matrix
$\widehat{\mathbf K}_k^{\mathrm{IR}}$,
constructed from the mode-$k$ unfoldings of the slice means
$\{\widehat{\mathcal M}_s\}_{s=1}^H$,
relative to its population counterpart
$\mathbf K_k^{\mathrm{IR}}$.

\begin{proposition}
\label{prop:tensor_slice_kendall_error}
Under Assumptions~\ref{ass:tensor_slice_regular}--\ref{ass:tensor_moment},
for each \(k=1,\ldots,K\), if \(p_k=o(n^{1/2})\), we have
\[
\left\|
\widehat{\mathbf K}_{k}^{\mathrm{IR}}
-
\mathbf K_{k}^{\mathrm{IR}}
\right\|_2
=
O_p\!\left(
p_k^{3/2}p_{-k}^{1/2}H^{-1}
\right)
+
O_p\!\left(
p_kH^{1/2}n^{-1/2}
\right).
\]
\end{proposition}

When $K=2$, Proposition~\ref{prop:tensor_slice_kendall_error} reduces to the row- and
column-wise convergence rates of MESIR in Proposition~\ref{prop:Kxs_error_main}.
When $K=1$, it further reduces to the vector-valued
setting considered by \citet{chen2022high},
with convergence rate $O_p\left(
p_1^{3/2}H^{-1}
+p_1H^{1/2}n^{-1/2}
\right)$.

We next establish the convergence rate of the calibrated marginal
Kendall's tau matrix $\widehat{\mathbf K}_k^{\mathrm{cal}}$ to the
mode-$k$ scatter shape $\boldsymbol\Sigma_k$.
\begin{proposition}\label{ck:thm:Ck}
Suppose that $\mathcal X$ follows the tensor elliptical model,
Assumption~\ref{ass:tensor_kendall_scaling} holds, and
$p_k=o(n^{1/2})$.
Then the eigenvalue-calibrated marginal Kendall's tau matrix
$\widehat{\mathbf K}_k^{\mathrm{cal}}$ satisfies
\[
\left\|\widehat{\mathbf K}_k^{\mathrm{cal}}
-\boldsymbol\Sigma_k\right\|_2
=O_p\!\left(\frac{p_k}{\sqrt n}\right).
\]
\end{proposition}

The column space of
$\boldsymbol\Sigma_k^{-1}\mathbf K_k^{\mathrm{IR}}$
equals the mode-$k$ reduction subspace.
The following theorem establishes the rate at which
$(\widehat{\mathbf K}_k^{\mathrm{cal}})^{-1}
\widehat{\mathbf K}_k^{\mathrm{IR}}$
converges to this population matrix.

\begin{theorem}
\label{thm:tensor_mesir_operator_consistency}
Under the conditions of Proposition~\ref{prop:tensor_slice_kendall_error} and ~\ref{ck:thm:Ck}, for each \(k=1,\ldots,K\), we have  
\[
\left\|
(\widehat{\mathbf K}_k^{\mathrm{cal}})^{-1}\widehat{\mathbf K}_{k}^{\mathrm{IR}}
-
\boldsymbol{\Sigma}_k^{-1} \mathbf K_{k}^{\mathrm{IR}}
\right\|_2
=
O_p(p_k^{3/2}p_{-k}^{1/2}H^{-1})
+
O_p\!\left(
p_kH^{1/2}n^{-1/2}
\right).
\]
\end{theorem}
The calibration error contributes an additional $O_p(p_kn^{-1/2})$
term, which is absorbed into $O_p(p_kH^{1/2}n^{-1/2})$
in the displayed rate.

Let $\mathbf P_{\mathbf B_k}$ and
$\mathbf P_{\widehat{\mathbf B}_k}$ denote the orthogonal projections onto the true and estimated
reduction subspaces, respectively. Applying a variant
of the Davis--Kahan theorem \citep{vu2013minimax} to the
inverse-regression Kendall's tau matrices, and controlling the
calibrated inverse transformations using
Proposition~\ref{ck:thm:Ck}, yields the following corollary.

\begin{corollary}
\label{cor:tensor_mode_subspace_consistency}
Under the conditions of Proposition~\ref{prop:tensor_slice_kendall_error} and ~\ref{ck:thm:Ck},
for each \(k=1,\ldots,K\), we have
\[
\left\|
\mathbf P_{\widehat{\mathbf B}_k}
-
\mathbf P_{\mathbf B_k}
\right\|_2
=
O_p\!\left(
p_k^{3/2}p_{-k}^{1/2}H^{-1}
\right)
+
O_p\!\left(
p_kH^{1/2}n^{-1/2}
\right).
\]
\end{corollary}

For fixed $K$, since the projection matrix onto the central tensor
subspace can be written as
$\mathbf P_{\mathcal S_{Y\mid\mathcal X}}
=\mathbf P_{\mathbf B_K}\otimes\cdots\otimes
\mathbf P_{\mathbf B_1}$,
Corollary~\ref{cor:tensor_mode_subspace_consistency}, applied to each mode, together with the
continuity of the Kronecker product, implies that
\[
\left\|
\mathbf P_{\widehat{\mathcal S}_{Y\mid\mathcal X}}
-
\mathbf P_{\mathcal S_{Y\mid\mathcal X}}
\right\|_2
\xrightarrow{p}0.
\]

The preceding results establish consistency of the estimated CTS when \(d_k,k=1,\ldots,K\)
are known. We next establish consistency of the eigenvalue-ratio estimators in Section~ \ref{subsec:tesir} when
the structural dimensions are unknown.
\begin{theorem}
\label{theo:TESIR-Auto}
Suppose that Assumptions~\ref{ass:tensor_slice_regular}--\ref{ass:tensor_moment} hold and that
$a_{n,k}=o(1)$ for each $k=1,\ldots,K$.
Let $\widehat d_k$ be defined as in
\eqref{eq:TESIR-Auto}.
Then, for fixed $K$,
\[
\mathbb P(\widehat d_k=d_k)\to 1,
\qquad k=1,\ldots,K.
\]
\end{theorem}

\subsection{Simulation studies}
\label{subsec:tensor_simulation}

The simulation study evaluates the performance of TESIR
for third-order tensor predictors
$\mathcal X\in\mathbb R^{10\times10\times10}$
under different tail behaviours, focusing on the recovery
of the mode-wise reduction subspaces and the central tensor subspace, as well as the impact of estimating the structural dimensions. We also consider the following methods:
(1) \emph{TESIR}, implemented with the true structural dimensions;
(2) \emph{TESIR-Auto}, which estimates the mode-wise dimensions
using the eigenvalue-ratio procedure
introduced in Appendix~\ref{subsec:tesir};
(3) \emph{TSIR} of
\citet{ding2015tensor};
and (4) \emph{Vector ESIR} of \citet{chen2022high},
applied to the vectorized predictor.
TESIR, TESIR-Auto and TSIR directly estimate the
mode-wise reduction subspaces by exploiting the
tensor structure, whereas Vector ESIR is evaluated only for recovery of the central tensor subspace. We assess subspace recovery using the squared multiple correlation coefficient $R_k^2$ and the projection distance $D_k$, following the definitions in Section~\ref{sec:MESIR_simulation}. For third-order tensor predictors, $k=1,2,3$ indexes the three modes, with $\mathcal B_k\subseteq\mathbb R^{p_k}$ denoting the true mode-$k$ reduction subspace and $\widehat{\mathcal B}_k=\operatorname{Span}(\widehat{\mathbf B}_k)$ its estimate. The mode-wise measures $R_k^2$ and $D_k$ are calculated for each of the three modes. We also evaluate recovery of the central tensor subspace by applying the same measures to $\mathcal B=\operatorname{Span}(\mathbf B_3\otimes\mathbf B_2\otimes\mathbf B_1)$ and $\widehat{\mathcal B}=\operatorname{Span}(\widehat{\mathbf B}_3\otimes\widehat{\mathbf B}_2\otimes\widehat{\mathbf B}_1)$. Larger $R^2$ and smaller $D$ indicate more accurate subspace recovery.

\paragraph{Simulation settings.}
We generate the predictor tensors independently according to
\[
\mathcal X=\xi^{1/2}\mathcal U,
\qquad
\operatorname{vec}(\mathcal U)
\sim N_{p_1p_2p_3}
(\mathbf 0,\mathbf I_{p_1p_2p_3}),
\]
where $\xi$ is independent of $\mathcal U$. The mode-wise scatter matrices are set to $(\boldsymbol\Sigma_1,\boldsymbol\Sigma_2,\boldsymbol\Sigma_3)=(\mathbf I_{p_1},\mathbf I_{p_2},\mathbf I_{p_3})$.
As in the matrix experiments, we consider the normal,
$t_3$, $t_2$, and Cauchy distributions by varying
the scale factor $\xi$.

Throughout the tensor experiments, we set
$p_1=p_2=p_3=10$ and consider sample sizes
$n\in\{500,1000\}$.
In all models, the response error is $\sigma\varepsilon$,
where $\varepsilon\sim N(0,1)$ is independent of
$\mathcal X$ and $\sigma=0.05$.
We use $H=10$ response slices of approximately equal size,
and each experiment is replicated 100 times.
We construct nonlinear response models using selected
entries of the reduced tensor predictor.
For Models B4--B6, let
\(
d_1=d_2=d_3=2,
\mathbf B_1=\mathbf B_2=\mathbf B_3
=(\mathbf e_1,\mathbf e_2)
\) and
\(
\mathcal T
=\mathcal X\times_1\mathbf B_1^\top
\times_2\mathbf B_2^\top
\times_3\mathbf B_3^\top
\in\mathbb R^{2\times2\times2},
\)
where $T_{ijk}$ denote its entries. Define
\[
\begin{aligned}
r &= \|\mathcal T\|_F,\quad
a_1=T_{111}+T_{222},\quad
a_2=T_{211}+T_{121}+T_{112}-T_{222},\quad
u_1=\frac{|a_2|}{|a_1|+|a_2|},\\
a_3 &= T_{111}+T_{221}-T_{212}-T_{122},\quad
a_4=T_{211}-T_{121}+T_{112}-T_{222},\quad
u_2=\frac{|a_4|}{|a_3|+|a_4|},\\
a_5 &= T_{111}-T_{221}-T_{212}+T_{122},\quad
a_6=T_{211}+T_{121}-T_{112}-T_{222},\quad
u_3=\frac{|a_6|}{|a_5|+|a_6|},\\
s &= \left(\sum_{i,j,k=1}^{2}|T_{ijk}|^{3/4}\right)^{4/3}.
\end{aligned}
\]
We consider the following three regression models:
\[
\begin{aligned}
\text{B4:}\quad Y
&=r\,\operatorname{sgn}(a_1)(1+40u_1^{10})
+\sigma\varepsilon,\\
\text{B5:}\quad Y
&=r\,\operatorname{sgn}(a_3)\exp(4u_2^4)
+\sigma\varepsilon,\\
\text{B6:}\quad Y
&=s\,\operatorname{sgn}(a_5)(1+10u_3^5)
+\sigma\varepsilon.
\end{aligned}
\] 

The three models share the same reduction structure
but differ in the nonlinear relationship between
$Y$ and $\mathcal T$.
The true mode-wise reduction subspaces are all
$\operatorname{Span}(\mathbf e_1,\mathbf e_2)$,
so that $d_1=d_2=d_3=2$.
TESIR and TSIR are implemented using the true structural
dimensions, whereas TESIR-Auto uses the dimensions
selected from the data.
For Vector ESIR, we retain $d_1d_2d_3=8$ directions
for the central tensor subspace.

\paragraph{Simulation results.}
Tables~\ref{tab:B4_B6_mode1_compare}--\ref{tab:B4_B6_mode3_compare} report the mode-wise subspace recovery results, while Table~\ref{tab:B4_B6_kron_compare} reports recovery of the central tensor subspace. Each entry is averaged over \(100\) replications, with standard deviations reported in parentheses.

When the structural dimensions are known, TSIR generally provides better subspace recovery under normally distributed predictors. Under \(t_3\) predictors, TESIR becomes competitive and shows some advantage for Models B4--B6 when the sample size is relatively small. As the tails become heavier, the advantage of TESIR becomes more pronounced. Under \(t_2\) and Cauchy predictors, TESIR generally achieves larger \(R^2\) values and smaller $D$ than TSIR for both the mode-wise reduction subspaces and the central tensor subspace, while Vector ESIR is generally less accurate in recovering the latter. Although recovery of the central tensor subspace remains challenging under the Cauchy distribution, TESIR retains a clear relative advantage over the competing methods.

TESIR-Auto estimates the mode-wise structural dimensions
from the data, whereas TESIR uses the true dimensions.
Across most settings, TESIR-Auto achieves subspace recovery
comparable to TESIR in terms of both $R^2$ and projection
distance $D$, indicating that estimating the structural
dimensions incurs relatively little loss in recovery accuracy.
Some deterioration is observed under small sample sizes
and extremely heavy-tailed distributions, where dimension
selection becomes more challenging. Overall, TESIR remains competitive under light-tailed
distributions, while its relative advantage over the
competing methods becomes more pronounced as the tails grow
heavier, consistent with the findings for MESIR in Section~\ref{sec:MESIR_simulation}.
Although the $t_2$ and Cauchy distributions do not satisfy
the moment condition in Assumption~\ref{ass:tensor_moment}, TESIR still exhibits
favourable subspace recovery in these settings.

\begin{table}[!htbp]
\centering
\caption{Mode-1 subspace recovery performance for Models B4--B6}
\label{tab:B4_B6_mode1_compare}
\footnotesize
\setlength{\tabcolsep}{5pt}
\renewcommand{\arraystretch}{1}
\begin{tabular}{lc|ccc|ccc}
\toprule
Model & Metric & \multicolumn{3}{c|}{$n=500$} & \multicolumn{3}{c}{$n=1000$} \\
& & TSIR & TESIR & TESIR-Auto & TSIR & TESIR & TESIR-Auto \\
\midrule
\multicolumn{2}{c}{} & \multicolumn{6}{c}{\textbf{Normal Distribution}} \\
\midrule
\multirow{2}{*}{B4} & $R_{1}^2$ & \textbf{0.934 (0.068)} & 0.726 (0.139) & 0.737 (0.178) & \textbf{0.976 (0.010)} & 0.913 (0.036) & 0.917 (0.048) \\
 & $D_{1}$ & \textbf{0.243 (0.083)} & 0.508 (0.127) & 0.568 (0.151) & \textbf{0.150 (0.031)} & 0.290 (0.059) & 0.320 (0.100) \\
\multirow{2}{*}{B5} & $R_{1}^2$ & \textbf{0.931 (0.060)} & 0.685 (0.126) & 0.683 (0.177) & \textbf{0.973 (0.008)} & 0.910 (0.037) & 0.898 (0.075) \\
 & $D_{1}$ & \textbf{0.252 (0.076)} & 0.550 (0.114) & 0.613 (0.124) & \textbf{0.161 (0.026)} & 0.294 (0.058) & 0.324 (0.102) \\
\multirow{2}{*}{B6} & $R_{1}^2$ & \textbf{0.931 (0.052)} & 0.700 (0.125) & 0.711 (0.153) & \textbf{0.977 (0.008)} & 0.915 (0.033) & 0.913 (0.052) \\
 & $D_{1}$ & \textbf{0.255 (0.067)} & 0.536 (0.115) & 0.592 (0.118) & \textbf{0.151 (0.025)} & 0.286 (0.054) & 0.304 (0.085) \\
\midrule
\multicolumn{2}{c}{} & \multicolumn{6}{c}{\textbf{$t_3$ Distribution}} \\
\midrule
\multirow{2}{*}{B4} & $R_{1}^2$ & 0.563 (0.263) & 0.589 (0.154) & \textbf{0.595 (0.169)} & \textbf{0.918 (0.115)} & 0.873 (0.049) & 0.840 (0.125) \\
 & $D_{1}$ & \textbf{0.624 (0.220)} & 0.630 (0.123) & 0.666 (0.122) & \textbf{0.256 (0.129)} & 0.350 (0.067) & 0.425 (0.141) \\
\multirow{2}{*}{B5} & $R_{1}^2$ & 0.493 (0.212) & \textbf{0.573 (0.142)} & 0.566 (0.182) & \textbf{0.859 (0.196)} & 0.855 (0.058) & 0.837 (0.111) \\
 & $D_{1}$ & 0.693 (0.162) & \textbf{0.644 (0.111)} & 0.698 (0.137) & \textbf{0.320 (0.198)} & 0.374 (0.074) & 0.427 (0.132) \\
\multirow{2}{*}{B6} & $R_{1}^2$ & 0.460 (0.189) & 0.549 (0.130) & \textbf{0.568 (0.165)} & \textbf{0.844 (0.205)} & 0.843 (0.081) & 0.833 (0.113) \\
 & $D_{1}$ & 0.721 (0.146) & \textbf{0.664 (0.100)} & 0.695 (0.114) & \textbf{0.339 (0.203)} & 0.385 (0.095) & 0.430 (0.127) \\
\midrule
\multicolumn{2}{c}{} & \multicolumn{6}{c}{\textbf{$t_2$ Distribution}} \\
\midrule
\multirow{2}{*}{B4} & $R_{1}^2$ & 0.297 (0.143) & \textbf{0.532 (0.143)} & 0.529 (0.185) & 0.451 (0.244) & \textbf{0.792 (0.099)} & 0.783 (0.138) \\
 & $D_{1}$ & 0.834 (0.089) & \textbf{0.676 (0.109)} & 0.718 (0.121) & 0.712 (0.203) & \textbf{0.445 (0.103)} & 0.513 (0.119) \\
\multirow{2}{*}{B5} & $R_{1}^2$ & 0.248 (0.140) & 0.525 (0.138) & \textbf{0.538 (0.193)} & 0.397 (0.223) & \textbf{0.800 (0.087)} & 0.766 (0.160) \\
 & $D_{1}$ & 0.863 (0.086) & \textbf{0.682 (0.103)} & 0.725 (0.139) & 0.759 (0.163) & \textbf{0.437 (0.094)} & 0.525 (0.149) \\
\multirow{2}{*}{B6} & $R_{1}^2$ & 0.256 (0.143) & 0.532 (0.163) & \textbf{0.544 (0.208)} & 0.356 (0.163) & \textbf{0.791 (0.103)} & 0.771 (0.147) \\
 & $D_{1}$ & 0.858 (0.087) & \textbf{0.673 (0.123)} & 0.712 (0.135) & 0.795 (0.110) & \textbf{0.445 (0.105)} & 0.512 (0.139) \\
\midrule
\multicolumn{2}{c}{} & \multicolumn{6}{c}{\textbf{Cauchy Distribution}} \\
\midrule
\multirow{2}{*}{B4} & $R_{1}^2$ & 0.205 (0.089) & 0.435 (0.141) & \textbf{0.455 (0.171)} & 0.215 (0.111) & \textbf{0.711 (0.140)} & 0.701 (0.174) \\
 & $D_{1}$ & 0.890 (0.050) & \textbf{0.745 (0.097)} & 0.765 (0.129) & 0.884 (0.064) & \textbf{0.522 (0.127)} & 0.593 (0.143) \\
\multirow{2}{*}{B5} & $R_{1}^2$ & 0.197 (0.101) & 0.452 (0.147) & \textbf{0.466 (0.195)} & 0.210 (0.108) & \textbf{0.721 (0.122)} & 0.704 (0.178) \\
 & $D_{1}$ & 0.894 (0.058) & \textbf{0.733 (0.106)} & 0.778 (0.145) & 0.887 (0.062) & \textbf{0.516 (0.114)} & 0.601 (0.137) \\
\multirow{2}{*}{B6} & $R_{1}^2$ & 0.218 (0.121) & \textbf{0.483 (0.157)} & 0.462 (0.175) & 0.228 (0.108) & 0.710 (0.128) & \textbf{0.725 (0.164)} \\
 & $D_{1}$ & 0.881 (0.072) & \textbf{0.710 (0.114)} & 0.768 (0.139) & 0.877 (0.063) & \textbf{0.527 (0.113)} & 0.577 (0.123) \\
\bottomrule
\end{tabular}
\end{table}

\begin{table}[!htbp]
\centering
\caption{Mode-2 subspace recovery performance for Models B4--B6}
\label{tab:B4_B6_mode2_compare}
\footnotesize
\setlength{\tabcolsep}{5pt}
\renewcommand{\arraystretch}{1}
\begin{tabular}{lc|ccc|ccc}
\toprule
Model & Metric & \multicolumn{3}{c|}{$n=500$} & \multicolumn{3}{c}{$n=1000$} \\
& & TSIR & TESIR & TESIR-Auto & TSIR & TESIR & TESIR-Auto \\
\midrule
\multicolumn{2}{c}{} & \multicolumn{6}{c}{\textbf{Normal Distribution}} \\
\midrule
\multirow{2}{*}{B4} & $R_{2}^2$ & \textbf{0.934 (0.060)} & 0.734 (0.122) & 0.705 (0.165) & \textbf{0.975 (0.009)} & 0.915 (0.030) & 0.918 (0.028) \\
 & $D_{2}$ & \textbf{0.246 (0.075)} & 0.503 (0.112) & 0.590 (0.139) & \textbf{0.157 (0.029)} & 0.288 (0.049) & 0.306 (0.085) \\
\multirow{2}{*}{B5} & $R_{2}^2$ & \textbf{0.927 (0.061)} & 0.687 (0.131) & 0.690 (0.149) & \textbf{0.973 (0.009)} & 0.912 (0.029) & 0.914 (0.030) \\
 & $D_{2}$ & \textbf{0.259 (0.076)} & 0.547 (0.115) & 0.593 (0.121) & \textbf{0.161 (0.026)} & 0.293 (0.049) & 0.322 (0.091) \\
\multirow{2}{*}{B6} & $R_{2}^2$ & \textbf{0.928 (0.065)} & 0.700 (0.117) & 0.707 (0.172) & \textbf{0.975 (0.008)} & 0.907 (0.043) & 0.909 (0.054) \\
 & $D_{2}$ & \textbf{0.256 (0.079)} & 0.538 (0.104) & 0.599 (0.134) & \textbf{0.154 (0.027)} & 0.298 (0.065) & 0.314 (0.090) \\
\midrule
\multicolumn{2}{c}{} & \multicolumn{6}{c}{\textbf{$t_3$ Distribution}} \\
\midrule
\multirow{2}{*}{B4} & $R_{2}^2$ & 0.588 (0.245) & \textbf{0.606 (0.137)} & 0.592 (0.188) & \textbf{0.921 (0.113)} & 0.855 (0.064) & 0.826 (0.142) \\
 & $D_{2}$ & \textbf{0.609 (0.204)} & 0.617 (0.115) & 0.681 (0.145) & \textbf{0.251 (0.128)} & 0.373 (0.080) & 0.446 (0.155) \\
\multirow{2}{*}{B5} & $R_{2}^2$ & 0.534 (0.205) & 0.569 (0.146) & \textbf{0.576 (0.176)} & \textbf{0.859 (0.192)} & 0.855 (0.072) & 0.845 (0.112) \\
 & $D_{2}$ & 0.664 (0.162) & \textbf{0.646 (0.119)} & 0.683 (0.134) & \textbf{0.322 (0.195)} & 0.371 (0.090) & 0.426 (0.133) \\
\multirow{2}{*}{B6} & $R_{2}^2$ & 0.493 (0.200) & 0.559 (0.140) & \textbf{0.565 (0.182)} & \textbf{0.854 (0.175)} & 0.840 (0.068) & 0.799 (0.139) \\
 & $D_{2}$ & 0.694 (0.158) & \textbf{0.655 (0.113)} & 0.701 (0.142) & \textbf{0.334 (0.186)} & 0.392 (0.082) & 0.468 (0.144) \\
\midrule
\multicolumn{2}{c}{} & \multicolumn{6}{c}{\textbf{$t_2$ Distribution}} \\
\midrule
\multirow{2}{*}{B4} & $R_{2}^2$ & 0.295 (0.121) & \textbf{0.529 (0.165)} & 0.515 (0.181) & 0.473 (0.247) & 0.791 (0.111) & \textbf{0.807 (0.126)} \\
 & $D_{2}$ & 0.836 (0.075) & \textbf{0.674 (0.126)} & 0.737 (0.144) & 0.696 (0.209) & \textbf{0.443 (0.113)} & 0.486 (0.122) \\
\multirow{2}{*}{B5} & $R_{2}^2$ & 0.336 (0.153) & 0.557 (0.143) & \textbf{0.573 (0.189)} & 0.435 (0.214) & \textbf{0.806 (0.095)} & 0.787 (0.141) \\
 & $D_{2}$ & 0.809 (0.098) & \textbf{0.656 (0.114)} & 0.704 (0.144) & 0.734 (0.164) & \textbf{0.429 (0.098)} & 0.489 (0.142) \\
\multirow{2}{*}{B6} & $R_{2}^2$ & 0.277 (0.140) & 0.547 (0.147) & \textbf{0.548 (0.166)} & 0.377 (0.175) & \textbf{0.821 (0.089)} & 0.802 (0.135) \\
 & $D_{2}$ & 0.846 (0.085) & \textbf{0.664 (0.114)} & 0.717 (0.132) & 0.781 (0.118) & \textbf{0.412 (0.097)} & 0.478 (0.131) \\
\midrule
\multicolumn{2}{c}{} & \multicolumn{6}{c}{\textbf{Cauchy Distribution}} \\
\midrule
\multirow{2}{*}{B4} & $R_{2}^2$ & 0.251 (0.108) & 0.449 (0.138) & \textbf{0.465 (0.196)} & 0.262 (0.115) & 0.700 (0.135) & \textbf{0.705 (0.171)} \\
 & $D_{2}$ & 0.863 (0.064) & \textbf{0.736 (0.098)} & 0.765 (0.134) & 0.857 (0.068) & \textbf{0.535 (0.120)} & 0.581 (0.128) \\
\multirow{2}{*}{B5} & $R_{2}^2$ & 0.248 (0.115) & 0.436 (0.150) & \textbf{0.445 (0.185)} & 0.243 (0.132) & 0.716 (0.135) & \textbf{0.717 (0.164)} \\
 & $D_{2}$ & 0.864 (0.067) & \textbf{0.744 (0.102)} & 0.770 (0.124) & 0.867 (0.081) & \textbf{0.518 (0.126)} & 0.572 (0.142) \\
\multirow{2}{*}{B6} & $R_{2}^2$ & 0.249 (0.114) & 0.463 (0.131) & \textbf{0.493 (0.179)} & 0.265 (0.110) & \textbf{0.714 (0.132)} & 0.672 (0.182) \\
 & $D_{2}$ & 0.864 (0.067) & \textbf{0.727 (0.091)} & 0.756 (0.136) & 0.855 (0.067) & \textbf{0.521 (0.120)} & 0.612 (0.158) \\
\bottomrule
\end{tabular}
\end{table}

\begin{table}[!htbp]
\centering
\caption{Mode-3 subspace recovery performance for Models B4--B6}
\label{tab:B4_B6_mode3_compare}
\footnotesize
\setlength{\tabcolsep}{5pt}
\renewcommand{\arraystretch}{1}
\begin{tabular}{lc|ccc|ccc}
\toprule
Model & Metric & \multicolumn{3}{c|}{$n=500$} & \multicolumn{3}{c}{$n=1000$} \\
& & TSIR & TESIR & TESIR-Auto & TSIR & TESIR & TESIR-Auto \\
\midrule
\multicolumn{2}{c}{} & \multicolumn{6}{c}{\textbf{Normal Distribution}} \\
\midrule
\multirow{2}{*}{B4} & $R_{3}^2$ & \textbf{0.938 (0.057)} & 0.735 (0.116) & 0.731 (0.142) & \textbf{0.975 (0.008)} & 0.917 (0.031) & 0.917 (0.044) \\
 & $D_{3}$ & \textbf{0.238 (0.075)} & 0.504 (0.108) & 0.571 (0.112) & \textbf{0.155 (0.025)} & 0.283 (0.053) & 0.310 (0.092) \\
\multirow{2}{*}{B5} & $R_{3}^2$ & \textbf{0.932 (0.064)} & 0.679 (0.136) & 0.679 (0.178) & \textbf{0.975 (0.009)} & 0.914 (0.036) & 0.917 (0.037) \\
 & $D_{3}$ & \textbf{0.248 (0.077)} & 0.554 (0.118) & 0.612 (0.142) & \textbf{0.157 (0.028)} & 0.287 (0.060) & 0.300 (0.081) \\
\multirow{2}{*}{B6} & $R_{3}^2$ & \textbf{0.928 (0.066)} & 0.670 (0.133) & 0.673 (0.178) & \textbf{0.974 (0.009)} & 0.907 (0.038) & 0.902 (0.085) \\
 & $D_{3}$ & \textbf{0.255 (0.080)} & 0.562 (0.117) & 0.617 (0.150) & \textbf{0.158 (0.027)} & 0.299 (0.063) & 0.339 (0.126) \\
\midrule
\multicolumn{2}{c}{} & \multicolumn{6}{c}{\textbf{$t_3$ Distribution}} \\
\midrule
\multirow{2}{*}{B4} & $R_{3}^2$ & 0.558 (0.265) & \textbf{0.614 (0.132)} & 0.556 (0.177) & \textbf{0.919 (0.124)} & 0.851 (0.071) & 0.840 (0.133) \\
 & $D_{3}$ & 0.631 (0.213) & \textbf{0.612 (0.109)} & 0.711 (0.140) & \textbf{0.250 (0.136)} & 0.376 (0.083) & 0.441 (0.148) \\
\multirow{2}{*}{B5} & $R_{3}^2$ & 0.500 (0.211) & 0.569 (0.156) & \textbf{0.573 (0.196)} & \textbf{0.853 (0.201)} & 0.846 (0.084) & 0.809 (0.140) \\
 & $D_{3}$ & 0.687 (0.169) & \textbf{0.645 (0.121)} & 0.703 (0.139) & \textbf{0.327 (0.201)} & 0.381 (0.095) & 0.453 (0.145) \\
\multirow{2}{*}{B6} & $R_{3}^2$ & 0.474 (0.196) & \textbf{0.579 (0.145)} & 0.577 (0.183) & \textbf{0.847 (0.200)} & 0.845 (0.081) & 0.825 (0.133) \\
 & $D_{3}$ & 0.709 (0.152) & \textbf{0.638 (0.116)} & 0.696 (0.138) & \textbf{0.337 (0.199)} & 0.382 (0.094) & 0.454 (0.147) \\
\midrule
\multicolumn{2}{c}{} & \multicolumn{6}{c}{\textbf{$t_2$ Distribution}} \\
\midrule
\multirow{2}{*}{B4} & $R_{3}^2$ & 0.278 (0.147) & 0.515 (0.147) & \textbf{0.542 (0.181)} & 0.447 (0.264) & \textbf{0.803 (0.107)} & 0.795 (0.150) \\
 & $D_{3}$ & 0.845 (0.091) & \textbf{0.688 (0.111)} & 0.714 (0.127) & 0.710 (0.221) & \textbf{0.430 (0.113)} & 0.487 (0.150) \\
\multirow{2}{*}{B5} & $R_{3}^2$ & 0.252 (0.123) & 0.539 (0.144) & \textbf{0.548 (0.192)} & 0.404 (0.214) & \textbf{0.798 (0.107)} & 0.779 (0.145) \\
 & $D_{3}$ & 0.862 (0.073) & \textbf{0.670 (0.112)} & 0.712 (0.130) & 0.755 (0.162) & \textbf{0.436 (0.110)} & 0.496 (0.141) \\
\multirow{2}{*}{B6} & $R_{3}^2$ & 0.274 (0.134) & \textbf{0.527 (0.142)} & 0.523 (0.172) & 0.353 (0.178) & \textbf{0.801 (0.100)} & 0.748 (0.161) \\
 & $D_{3}$ & 0.848 (0.081) & \textbf{0.679 (0.106)} & 0.737 (0.140) & 0.796 (0.121) & \textbf{0.434 (0.105)} & 0.528 (0.149) \\
\midrule
\multicolumn{2}{c}{} & \multicolumn{6}{c}{\textbf{Cauchy Distribution}} \\
\midrule
\multirow{2}{*}{B4} & $R_{3}^2$ & 0.209 (0.100) & \textbf{0.469 (0.153)} & 0.455 (0.172) & 0.229 (0.121) & \textbf{0.718 (0.119)} & 0.701 (0.162) \\
 & $D_{3}$ & 0.887 (0.058) & \textbf{0.721 (0.107)} & 0.766 (0.128) & 0.875 (0.074) & \textbf{0.519 (0.109)} & 0.591 (0.132) \\
\multirow{2}{*}{B5} & $R_{3}^2$ & 0.187 (0.087) & \textbf{0.457 (0.161)} & 0.447 (0.164) & 0.229 (0.111) & 0.713 (0.140) & \textbf{0.720 (0.158)} \\
 & $D_{3}$ & 0.900 (0.049) & \textbf{0.729 (0.113)} & 0.781 (0.137) & 0.875 (0.064) & \textbf{0.520 (0.127)} & 0.565 (0.135) \\
\multirow{2}{*}{B6} & $R_{3}^2$ & 0.204 (0.117) & \textbf{0.448 (0.145)} & 0.435 (0.169) & 0.193 (0.093) & \textbf{0.705 (0.132)} & 0.681 (0.174) \\
 & $D_{3}$ & 0.890 (0.068) & \textbf{0.736 (0.102)} & 0.795 (0.144) & 0.897 (0.053) & \textbf{0.530 (0.120)} & 0.607 (0.145) \\
\bottomrule
\end{tabular}
\end{table}

\begin{table}[!htbp]
\centering
\caption{Central tensor subspace recovery performance for Models B4--B6}
\label{tab:B4_B6_kron_compare}
\footnotesize
\setlength{\tabcolsep}{5pt}
\renewcommand{\arraystretch}{1}
\begin{tabular}{lc|ccc|ccc}
\toprule
Model & Metric & \multicolumn{3}{c|}{$n=500$} & \multicolumn{3}{c}{$n=1000$} \\
& & TSIR & TESIR & TESIR-Auto & TSIR & TESIR & TESIR-Auto \\
\midrule
\multicolumn{2}{c}{} & \multicolumn{6}{c}{\textbf{Normal Distribution}} \\
\midrule
\multirow{2}{*}{B4} & $R^2$ & \textbf{0.826 (0.103)} & 0.394 (0.128) & 0.381 (0.155) & \textbf{0.928 (0.016)} & 0.766 (0.052) & 0.772 (0.061) \\
 & $D$ & \textbf{0.408 (0.086)} & 0.774 (0.083) & 0.816 (0.161) & \textbf{0.267 (0.030)} & 0.481 (0.054) & 0.515 (0.075) \\
\multirow{2}{*}{B5} & $R^2$ & \textbf{0.812 (0.102)} & 0.317 (0.104) & 0.325 (0.151) & \textbf{0.924 (0.015)} & 0.759 (0.051) & 0.753 (0.074) \\
 & $D$ & \textbf{0.425 (0.084)} & 0.824 (0.064) & 0.866 (0.209) & \textbf{0.275 (0.027)} & 0.489 (0.051) & 0.524 (0.078) \\
\multirow{2}{*}{B6} & $R^2$ & \textbf{0.810 (0.101)} & 0.330 (0.108) & 0.346 (0.161) & \textbf{0.928 (0.014)} & 0.753 (0.060) & 0.748 (0.096) \\
 & $D$ & \textbf{0.429 (0.083)} & 0.816 (0.066) & 0.859 (0.262) & \textbf{0.267 (0.026)} & 0.493 (0.059) & 0.531 (0.100) \\
\midrule
\multicolumn{2}{c}{} & \multicolumn{6}{c}{\textbf{$t_3$ Distribution}} \\
\midrule
\multirow{2}{*}{B4} & $R^2$ & \textbf{0.285 (0.273)} & 0.224 (0.103) & 0.199 (0.118) & \textbf{0.808 (0.191)} & 0.636 (0.082) & 0.585 (0.170) \\
 & $D$ & \textbf{0.827 (0.178)} & 0.879 (0.060) & 0.994 (0.303) & \textbf{0.412 (0.151)} & 0.600 (0.067) & 0.689 (0.140) \\
\multirow{2}{*}{B5} & $R^2$ & \textbf{0.191 (0.201)} & 0.184 (0.080) & 0.180 (0.093) & \textbf{0.710 (0.280)} & 0.618 (0.093) & 0.574 (0.149) \\
 & $D$ & \textbf{0.891 (0.126)} & 0.902 (0.045) & 0.983 (0.272) & \textbf{0.495 (0.214)} & 0.613 (0.075) & 0.697 (0.159) \\
\multirow{2}{*}{B6} & $R^2$ & 0.155 (0.181) & 0.181 (0.087) & \textbf{0.185 (0.100)} & \textbf{0.689 (0.277)} & 0.600 (0.100) & 0.552 (0.157) \\
 & $D$ & 0.912 (0.117) & \textbf{0.903 (0.050)} & 0.955 (0.261) & \textbf{0.517 (0.211)} & 0.628 (0.077) & 0.712 (0.146) \\
\midrule
\multicolumn{2}{c}{} & \multicolumn{6}{c}{\textbf{$t_2$ Distribution}} \\
\midrule
\multirow{2}{*}{B4} & $R^2$ & 0.032 (0.044) & \textbf{0.151 (0.083)} & 0.144 (0.075) & 0.183 (0.267) & \textbf{0.502 (0.115)} & 0.499 (0.149) \\
 & $D$ & 0.983 (0.023) & \textbf{0.920 (0.046)} & 0.969 (0.248) & 0.885 (0.182) & \textbf{0.701 (0.082)} & 0.742 (0.116) \\
\multirow{2}{*}{B5} & $R^2$ & 0.027 (0.035) & 0.158 (0.078) & \textbf{0.170 (0.110)} & 0.121 (0.181) & \textbf{0.514 (0.104)} & 0.470 (0.159) \\
 & $D$ & 0.986 (0.018) & \textbf{0.916 (0.043)} & 0.950 (0.232) & 0.930 (0.117) & \textbf{0.693 (0.074)} & 0.773 (0.191) \\
\multirow{2}{*}{B6} & $R^2$ & 0.026 (0.035) & 0.155 (0.083) & \textbf{0.160 (0.101)} & 0.071 (0.105) & \textbf{0.520 (0.112)} & 0.460 (0.150) \\
 & $D$ & 0.987 (0.018) & \textbf{0.918 (0.047)} & 1.020 (0.345) & 0.962 (0.062) & \textbf{0.688 (0.081)} & 0.787 (0.197) \\
\midrule
\multicolumn{2}{c}{} & \multicolumn{6}{c}{\textbf{Cauchy Distribution}} \\
\midrule
\multirow{2}{*}{B4} & $R^2$ & 0.011 (0.009) & 0.092 (0.054) & \textbf{0.098 (0.077)} & 0.014 (0.013) & \textbf{0.359 (0.120)} & 0.348 (0.147) \\
 & $D$ & 0.994 (0.005) & \textbf{0.952 (0.029)} & 1.049 (0.327) & 0.993 (0.006) & \textbf{0.797 (0.076)} & 0.847 (0.220) \\
\multirow{2}{*}{B5} & $R^2$ & 0.009 (0.008) & \textbf{0.092 (0.057)} & 0.091 (0.058) & 0.013 (0.016) & \textbf{0.370 (0.123)} & 0.369 (0.160) \\
 & $D$ & 0.995 (0.004) & \textbf{0.952 (0.030)} & 1.049 (0.312) & 0.993 (0.008) & \textbf{0.790 (0.079)} & 0.846 (0.226) \\
\multirow{2}{*}{B6} & $R^2$ & 0.013 (0.018) & \textbf{0.102 (0.057)} & 0.100 (0.071) & 0.013 (0.012) & \textbf{0.359 (0.120)} & 0.331 (0.146) \\
 & $D$ & 0.994 (0.009) & \textbf{0.947 (0.030)} & 1.095 (0.317) & 0.994 (0.006) & \textbf{0.797 (0.076)} & 0.878 (0.238) \\
\bottomrule
\end{tabular}
\end{table}

\clearpage

\subsection{Proofs of TESIR results}
\label{supp_tensor_proof}

\paragraph{Proof of Proposition~\ref{prop:tensor_slice_kendall_error}}

\begin{proof}
Fix $k\in\{1,\ldots,K\}$. Notice that
\begingroup\scriptsize
{
\begin{align*}
\left\|
\widehat{\mathbf K}_{k}^{\mathrm{IR}}
-
\mathbf K^{\mathrm{IR}}_{k}
\right\|_2
&=
\Biggl\|
\widetilde{\mathbf K}_{k}^{\mathrm{IR}}
-
\mathbf K^{\mathrm{IR}}_{k}
+
\mathbf K^{\mathrm{or}}_{k}
-
\widetilde{\mathbf K}_{k}^{\mathrm{IR}}
+
\mathbf K^{\mathrm{S}}_{k}
-
\mathbf K^{\mathrm{or}}_{k}
+
\widehat{\mathbf K}_{k}^{\mathrm{IR}}
-
\mathbf K^{\mathrm{S}}_{k}
\Biggr\|_2
\\
&=
\Biggl\|
\frac{2}{n(n-1)}
\sum_{i'<i}
\frac{
(\boldsymbol\mu_{i}^{(k)}-\boldsymbol\mu_{i'}^{(k)})
(\boldsymbol\mu_{i}^{(k)}-\boldsymbol\mu_{i'}^{(k)})^\top
}{
\|\boldsymbol\mu_{i}^{(k)}-\boldsymbol\mu_{i'}^{(k)}\|_F^2/p_k
}
-
\mathbb E\left[
\frac{
\{[\mathcal M(Y)]_{(k)}-[\mathcal M(\widetilde Y)]_{(k)}\}
\{[\mathcal M(Y)]_{(k)}-[\mathcal M(\widetilde Y)]_{(k)}\}^\top
}{
\|[\mathcal M(Y)]_{(k)}-[\mathcal M(\widetilde Y)]_{(k)}\|_F^2/p_k
}
\right]
\\
&+
\frac{2}{H(H-1)}
\sum_{s'<s}
\frac{
(\overline{\boldsymbol\mu}_{s}^{(k)}-\overline{\boldsymbol\mu}_{s'}^{(k)})
(\overline{\boldsymbol\mu}_{s}^{(k)}-\overline{\boldsymbol\mu}_{s'}^{(k)})^\top
}{
\|\overline{\boldsymbol\mu}_{s}^{(k)}-\overline{\boldsymbol\mu}_{s'}^{(k)}\|_F^2/p_k
}
-
\frac{2}{n(n-1)}
\sum_{i'<i}
\frac{
(\boldsymbol\mu_{i}^{(k)}-\boldsymbol\mu_{i'}^{(k)})
(\boldsymbol\mu_{i}^{(k)}-\boldsymbol\mu_{i'}^{(k)})^\top
}{
\|\boldsymbol\mu_{i}^{(k)}-\boldsymbol\mu_{i'}^{(k)}\|_F^2/p_k
}
\\
&+
\frac{2}{H(H-1)}
\sum_{s'<s}
\frac{
(\boldsymbol\mu_{s}^{(k)}-\boldsymbol\mu_{s'}^{(k)})
(\boldsymbol\mu_{s}^{(k)}-\boldsymbol\mu_{s'}^{(k)})^\top
}{
\|\boldsymbol\mu_{s}^{(k)}-\boldsymbol\mu_{s'}^{(k)}\|_F^2/p_k
}
-
\frac{2}{H(H-1)}
\sum_{s'<s}
\frac{
(\overline{\boldsymbol\mu}_{s}^{(k)}-\overline{\boldsymbol\mu}_{s'}^{(k)})
(\overline{\boldsymbol\mu}_{s}^{(k)}-\overline{\boldsymbol\mu}_{s'}^{(k)})^\top
}{
\|\overline{\boldsymbol\mu}_{s}^{(k)}-\overline{\boldsymbol\mu}_{s'}^{(k)}\|_F^2/p_k
}
\\
&+
\frac{2}{H(H-1)}
\sum_{s'<s}
\frac{
(\widehat{\boldsymbol\mu}_{s}^{(k)}-\widehat{\boldsymbol\mu}_{s'}^{(k)})
(\widehat{\boldsymbol\mu}_{s}^{(k)}-\widehat{\boldsymbol\mu}_{s'}^{(k)})^\top
}{
\|\widehat{\boldsymbol\mu}_{s}^{(k)}-\widehat{\boldsymbol\mu}_{s'}^{(k)}\|_F^2/p_k
}
-
\frac{2}{H(H-1)}
\sum_{s'<s}
\frac{
(\boldsymbol\mu_{s}^{(k)}-\boldsymbol\mu_{s'}^{(k)})
(\boldsymbol\mu_{s}^{(k)}-\boldsymbol\mu_{s'}^{(k)})^\top
}{
\|\boldsymbol\mu_{s}^{(k)}-\boldsymbol\mu_{s'}^{(k)}\|_F^2/p_k
}
\Biggr\|_2
\\
&:=
\|{\mathbf A}_1+{\mathbf A}_2+{\mathbf A}_3+{\mathbf A}_4\|_2
\\
&\le
\|{\mathbf A}_1\|_2+\|{\mathbf A}_2\|_2
+\|{\mathbf A}_3\|_2+\|{\mathbf A}_4\|_2,
\end{align*}
}
\endgroup
where
\(
\boldsymbol\mu_{i}^{(k)}=[\mathcal M(Y_i)]_{(k)},
\boldsymbol\mu_{sj}^{(k)}=[\mathcal M(Y_{sj})]_{(k)},
\overline{\boldsymbol\mu}_{s}^{(k)}=\frac1l\sum_{j=1}^l\boldsymbol\mu_{sj}^{(k)},
\)
\(
\boldsymbol\mu_{s}^{(k)}
=
\mathbb E(\mathcal X_{(k)}\mid Y\in J_s),
\widehat{\boldsymbol\mu}_{s}^{(k)}
=
\frac1l\sum_{j=1}^l\mathcal X_{sj,(k)}.
\)
Here,
$\mathbf K_k^{\mathrm{or}}$ and
$\mathbf K_k^{\mathrm S}$ denote the sliced Kendall's tau
matrices constructed from $\overline{\boldsymbol\mu}_{s}^{(k)}$ and
$\boldsymbol\mu_{s}^{(k)}$, respectively.

For $\mathbf A_1$, Lemma~\ref{thm:tensor_kendall_concentration}
and Assumption~\ref{ass:tensor_bounded_inverse_kendall} give
\[
\|\mathbf A_1\|_2
=O_p\!\left\{
\|\mathbf K_k^{\mathrm{IR}}\|_2
\sqrt{\frac{r^*(\mathbf K_k^{\mathrm{IR}})\log p_k}{n}}
\right\}
=O_p\!\left(\sqrt{\frac{p_k\log p_k}{n}}\right),
\qquad
r^*(\mathbf K_k^{\mathrm{IR}})
=\frac{\operatorname{tr}(\mathbf K_k^{\mathrm{IR}})}
{\|\mathbf K_k^{\mathrm{IR}}\|_2}\le p_k.
\]

For the second part ${{\mathbf A}_2}$,
{
\begin{align*}
\widetilde{\mathbf K}_{k}^{\mathrm{IR}}/p_k
&=
\frac{2}{n(n-1)}
\left\{
\sum_{s'<s}
\sum_{j=1}^l\sum_{b=1}^l
\frac{
(\boldsymbol\mu_{sj}^{(k)}-\boldsymbol\mu_{s'b}^{(k)})
(\boldsymbol\mu_{sj}^{(k)}-\boldsymbol\mu_{s'b}^{(k)})^\top
}{
\|\boldsymbol\mu_{sj}^{(k)}-\boldsymbol\mu_{s'b}^{(k)}\|_F^2
}
\right.
\\
&\qquad\qquad\left.
+
\sum_{s=1}^H\sum_{j<b}
\frac{
(\boldsymbol\mu_{sj}^{(k)}-\boldsymbol\mu_{sb}^{(k)})
(\boldsymbol\mu_{sj}^{(k)}-\boldsymbol\mu_{sb}^{(k)})^\top
}{
\|\boldsymbol\mu_{sj}^{(k)}-\boldsymbol\mu_{sb}^{(k)}\|_F^2
}
\right\}
\\
&=
\frac{
l^2\binom{H}{2}
}{
\binom{n}{2}
}
\frac{1}{
l^2\binom{H}{2}
}
\sum_{s'<s}
\sum_{j=1}^l\sum_{b=1}^l
\frac{
(\boldsymbol\mu_{sj}^{(k)}-\boldsymbol\mu_{s'b}^{(k)})
(\boldsymbol\mu_{sj}^{(k)}-\boldsymbol\mu_{s'b}^{(k)})^\top
}{
\|\boldsymbol\mu_{sj}^{(k)}-\boldsymbol\mu_{s'b}^{(k)}\|_F^2
}
\\
&\quad+
\frac{
H\binom{l}{2}
}{
\binom{n}{2}
}
\frac{1}{
H\binom{l}{2}
}
\sum_{s=1}^H\sum_{j<b}
\frac{
(\boldsymbol\mu_{sj}^{(k)}-\boldsymbol\mu_{sb}^{(k)})
(\boldsymbol\mu_{sj}^{(k)}-\boldsymbol\mu_{sb}^{(k)})^\top
}{
\|\boldsymbol\mu_{sj}^{(k)}-\boldsymbol\mu_{sb}^{(k)}\|_F^2
}
\\
&=: {\mathbf B}_1+{\mathbf B}_2 .
\end{align*}
}
{
\begin{align*}
{\mathbf A}_2/p_k
&=
\frac{2}{H(H-1)}
\sum_{s'<s}
\frac{
(\overline{\boldsymbol\mu}_{s}^{(k)}-\overline{\boldsymbol\mu}_{s'}^{(k)})
(\overline{\boldsymbol\mu}_{s}^{(k)}-\overline{\boldsymbol\mu}_{s'}^{(k)})^\top
}{
\|\overline{\boldsymbol\mu}_{s}^{(k)}-\overline{\boldsymbol\mu}_{s'}^{(k)}\|_F^2
}
-
{\mathbf B}_1-{\mathbf B}_2                                                      \\
&=: {\mathbf B}_3-{\mathbf B}_1-{\mathbf B}_2 ,
\end{align*}
}
thus 
{
\[
\|{\mathbf A}_2\|_2/p_k
\le
\|{\mathbf B}_3-{\mathbf B}_1\|_2+\|{\mathbf B}_2\|_2 .
\]
}
Since
\(
\left\|
\frac{\mathbf U\mathbf U^\top}{\|\mathbf U\|_F^2}
\right\|_2
=
\frac{\|\mathbf U\|_2^2}{\|\mathbf U\|_F^2}
\le1,
\)
we have
\({
\|{\mathbf B}_2\|_2
\le
\frac{
H\binom{l}{2}
}{
\binom{n}{2}
}
=
O(H^{-1}).
}\)
Furthermore,
{
\begin{align*}
\|{\mathbf B}_3-{\mathbf B}_1\|_2
&=
\Biggl\|
\frac{1}{
l^2\binom{H}{2}
}
\sum_{s'<s}\sum_{j=1}^l\sum_{b=1}^l
\left\{
\frac{
(\boldsymbol\mu_{sj}^{(k)}-\boldsymbol\mu_{s'j}^{(k)})
(\boldsymbol\mu_{sb}^{(k)}-\boldsymbol\mu_{s'b}^{(k)})^\top
}{
\|\overline{\boldsymbol\mu}_{s}^{(k)}-\overline{\boldsymbol\mu}_{s'}^{(k)}\|_F^2
}
\right.
\\
&\qquad\qquad\left.
-
\frac{
(\boldsymbol\mu_{sj}^{(k)}-\boldsymbol\mu_{s'b}^{(k)})
(\boldsymbol\mu_{sj}^{(k)}-\boldsymbol\mu_{s'b}^{(k)})^\top
}{
\|\boldsymbol\mu_{sj}^{(k)}-\boldsymbol\mu_{s'b}^{(k)}\|_F^2
}
\right\}
\\
&\qquad+
\left(
1-
\frac{
l^2\binom{H}{2}
}{
\binom{n}{2}
}
\right)
\frac{1}{
l^2\binom{H}{2}
}
\sum_{s'<s}\sum_{j=1}^l\sum_{b=1}^l
\frac{
(\boldsymbol\mu_{sj}^{(k)}-\boldsymbol\mu_{s'b}^{(k)})
(\boldsymbol\mu_{sj}^{(k)}-\boldsymbol\mu_{s'b}^{(k)})^\top
}{
\|\boldsymbol\mu_{sj}^{(k)}-\boldsymbol\mu_{s'b}^{(k)}\|_F^2
}
\Biggr\|_2
\\
&=: \|{\mathbf D}_1+{\mathbf D}_2\|_2
\\
&\le
\|{\mathbf D}_1\|_2+\|{\mathbf D}_2\|_2 .
\end{align*}
}

Write
\(
\mathbf u_{sj,s'b}
=
\boldsymbol\mu_{sj}^{(k)}-\boldsymbol\mu_{s'b}^{(k)},
\mathbf v_{ss'}
=
\overline{\boldsymbol\mu}_{s}^{(k)}-\overline{\boldsymbol\mu}_{s'}^{(k)},
\mathbf V_{ss'}
=
\boldsymbol\mu_{s}^{(k)}-\boldsymbol\mu_{s'}^{(k)}.
\)
Since
\(
\mathbf v_{ss'}
=
\frac{1}{l^2}
\sum_{j=1}^l\sum_{b=1}^l
\mathbf u_{sj,s'b},
\)
we can write
{
\begin{align*}
{\mathbf D}_1
&=
\frac{1}{l^2\binom{H}{2}}
\sum_{s'<s}\sum_{j=1}^l\sum_{b=1}^l
\left\{
\frac{
\mathbf v_{ss'}\mathbf v_{ss'}^\top
}{
\|\mathbf v_{ss'}\|_F^2
}
-
\frac{
\mathbf u_{sj,s'b}\mathbf u_{sj,s'b}^\top
}{
\|\mathbf u_{sj,s'b}\|_F^2
}
\right\}.
\end{align*}
}

By adding and subtracting
$\mathbf V_{ss'}\mathbf V_{ss'}^\top/
\|\mathbf V_{ss'}\|_F^2$, and with Assumption~\ref{ass:tensor_slice_regular}, we obtain
\begin{align*}
\|{\mathbf D}_1\|_2
&\le
\frac{1}{l^2\binom{H}{2}}
\sum_{s'<s}\sum_{j=1}^l\sum_{b=1}^l
\Biggl\{
\left\|
\frac{
\mathbf v_{ss'}\mathbf v_{ss'}^\top
}{
\|\mathbf v_{ss'}\|_F^2
}
-
\frac{
\mathbf V_{ss'}\mathbf V_{ss'}^\top
}{
\|\mathbf V_{ss'}\|_F^2
}
\right\|_2
\\
&\qquad\qquad+
\left\|
\frac{
\mathbf u_{sj,s'b}\mathbf u_{sj,s'b}^\top
}{
\|\mathbf u_{sj,s'b}\|_F^2
}
-
\frac{
\mathbf V_{ss'}\mathbf V_{ss'}^\top
}{
\|\mathbf V_{ss'}\|_F^2
}
\right\|_2
\Biggr\} \\
&\le
\frac{4}{l^2\binom{H}{2}}
\sum_{s'<s}\sum_{j=1}^l\sum_{b=1}^l
\frac{
\|\mathbf v_{ss'}-\mathbf V_{ss'}\|_F
+
\|\mathbf u_{sj,s'b}-\mathbf V_{ss'}\|_F
}{
\|\mathbf V_{ss'}\|_F
}
\\
&\le
\frac{4C}{l^2\binom{H}{2}}
\sum_{s'<s}\sum_{j=1}^l\sum_{b=1}^l
\Bigl\{
\|\mathbf v_{ss'}-\mathbf V_{ss'}\|_F
+
\|\mathbf u_{sj,s'b}-\mathbf V_{ss'}\|_F
\Bigr\}.
\end{align*}

Let
\(
R_H
=
\max_s\max_{1\le j\le l}
\|\boldsymbol\mu_{sj}^{(k)}-\boldsymbol\mu_{s}^{(k)}\|_F.
\)
Since
\(
\overline{\boldsymbol\mu}_{s}^{(k)}
=
\frac1l\sum_{j=1}^l\boldsymbol\mu_{sj}^{(k)},
\)
we have
\(
\max_s\|\overline{\boldsymbol\mu}_{s}^{(k)}-\boldsymbol\mu_{s}^{(k)}\|_F
\le R_H.
\)
Consequently,
\(
\|\mathbf v_{ss'}-\mathbf V_{ss'}\|_F
\le 2R_H,
\|\mathbf u_{sj,s'b}-\mathbf V_{ss'}\|_F
\le 2R_H.
\)
It follows that
\({
\|{\mathbf D}_1\|_2
\le 16CR_H.
}\)

Therefore, using Assumption~\ref{ass:tensor_smoothness}
and the slice-width condition in
Assumption~\ref{ass:tensor_slice_regular}, we obtain
\begin{align*}
\|\boldsymbol\mu_{sj}^{(k)}-\boldsymbol\mu_{s}^{(k)}\|_F
&=
\left\|
\mathbb E\left[
[\mathcal M(Y_{sj})]_{(k)}-[\mathcal M(Y)]_{(k)}
\mid Y\in J_s
\right]
\right\|_F
\\
&\le
\mathbb E\left[
\|[\mathcal M(Y_{sj})]_{(k)}-[\mathcal M(Y)]_{(k)}\|_F
\mid Y\in J_s
\right]
\\
&\le
C\sqrt{p_kp_{-k}}\,
\max_{1\le j,b\le l}|Y_{sj}-Y_{sb}|.
\end{align*}
Thus,
\(
R_H
=
O_p(\sqrt{p_kp_{-k}}H^{-1}),
\)
which gives
\({
\|{\mathbf D}_1\|_2
=
O_p(\sqrt{p_kp_{-k}}H^{-1}).
}\)

Moreover,
{
\[
\|{\mathbf D}_2\|_2
\le
\left|
1-
\frac{
l^2\binom{H}{2}
}{
\binom{n}{2}
}
\right|
=
\frac{
H\binom{l}{2}
}{
\binom{n}{2}
}
=
O(H^{-1}).
\]
}
Combining the above bounds, we obtain
{
\[
\|{\mathbf A}_2\|_2
\le
p_k\{\|{\mathbf D}_1\|_2+\|{\mathbf D}_2\|_2+\|{\mathbf B}_2\|_2\}
=
O_p(p_k^{3/2}p_{-k}^{1/2}H^{-1}).
\]
}

For $\mathbf A_3$, let
$\boldsymbol\Delta_s
=\overline{\boldsymbol\mu}_s^{(k)}-\boldsymbol\mu_s^{(k)}$.
The preceding bound for $R_H$ gives
$\max_s\|\boldsymbol\Delta_s\|_F\le R_H
=O_p(\sqrt{p_kp_{-k}}H^{-1})$.
Since $\mathbf v_{ss'}-\mathbf V_{ss'}
=\boldsymbol\Delta_s-\boldsymbol\Delta_{s'}$, the normalized
outer-product inequality and Assumption~\ref{ass:tensor_slice_regular} yield
\begin{align*}
\frac{\|\mathbf A_3\|_2}{p_k}
&\le\frac{2}{H(H-1)}\sum_{s<s'}
\left\|
\frac{\mathbf V_{ss'}\mathbf V_{ss'}^\top}{\|\mathbf V_{ss'}\|_F^2}
-\frac{\mathbf v_{ss'}\mathbf v_{ss'}^\top}{\|\mathbf v_{ss'}\|_F^2}
\right\|_2\\
&\le\frac{8C}{H(H-1)}\sum_{s<s'}
\|\boldsymbol\Delta_s-\boldsymbol\Delta_{s'}\|_F\\
&\le\frac{8C}{H}\sum_{s=1}^H\|\boldsymbol\Delta_s\|_F
\le8CR_H=O_p(\sqrt{p_kp_{-k}}H^{-1}).
\end{align*}
Hence $\|\mathbf A_3\|_2=O_p(p_k^{3/2}p_{-k}^{1/2}H^{-1})$.

For ${{\mathbf A}_4}$, write
\(
\mathbf e_s=\widehat{\boldsymbol\mu}_{s}^{(k)}-\overline{\boldsymbol\mu}_{s}^{(k)}
=\frac1l\sum_{j=1}^l(\mathcal X_{sj,(k)}-\boldsymbol\mu_{sj}^{(k)})
\)
and
\(
\widehat{\mathbf V}_{ss'}
=\widehat{\boldsymbol\mu}_{s}^{(k)}-\widehat{\boldsymbol\mu}_{s'}^{(k)}.
\)
Then
\(
\widehat{\mathbf V}_{ss'}-\mathbf V_{ss'}
=(\mathbf e_s-\mathbf e_{s'})
+(\boldsymbol\Delta_s-\boldsymbol\Delta_{s'}).
\)
Consequently,
\[
\begin{aligned}
\frac{2}{H(H-1)}
\sum_{s<s'}
\|\widehat{\mathbf V}_{ss'}-\mathbf V_{ss'}\|_F
&\le
\frac2H\sum_{s=1}^H\|\mathbf e_s\|_F
+\frac2H\sum_{s=1}^H\|\boldsymbol\Delta_s\|_F\\
&\le
2\left\{\frac1H\sum_{s=1}^H\|\mathbf e_s\|_F^2\right\}^{1/2}
+2\max_s\|\boldsymbol\Delta_s\|_F\\
&=
O_p\!\left(\sqrt{\frac{p_kp_{-k}H}{n}}\right)
+O_p(\sqrt{p_kp_{-k}}H^{-1}).
\end{aligned}
\]

By Assumption~\ref{ass:tensor_moment}, \(
\|\widehat{\boldsymbol\mu}_{s}^{(k)}\|_F = O_p(\sqrt{p_kp_{-k}}).
\) Therefore,
\(
\|\widehat{\mathbf V}_{ss'}\|_F
=
\|\widehat{\boldsymbol\mu}_{s}^{(k)}-\widehat{\boldsymbol\mu}_{s'}^{(k)}\|_F
\le
\|\widehat{\boldsymbol\mu}_{s}^{(k)}\|_F+\|\widehat{\boldsymbol\mu}_{s'}^{(k)}\|_F
=
O_p(\sqrt{p_kp_{-k}}).
\) Moreover, using the bound for $\boldsymbol\Delta_s$ established in ${{\mathbf A}_3}$,
\[
\begin{aligned}
\|\mathbf V_{ss'}\|_F
&\le\|\mathbf V_{ss'}-\widehat{\mathbf V}_{ss'}\|_F
+\|\widehat{\mathbf V}_{ss'}\|_F\\
&\le\|\mathbf e_s\|_F+\|\mathbf e_{s'}\|_F
+\|\boldsymbol\Delta_s\|_F+\|\boldsymbol\Delta_{s'}\|_F
+\|\widehat{\mathbf V}_{ss'}\|_F\\
&=O_p\!\left(\sqrt{\frac{p_kp_{-k}H}{n}}\right)
+O_p(\sqrt{p_kp_{-k}}H^{-1})+O_p(\sqrt{p_kp_{-k}})\\
&=O_p(\sqrt{p_kp_{-k}}),
\end{aligned}
\]
Then, we have

\[
\begin{aligned}
\|{\mathbf A}_4\|_2/p_k
&=
\left\|
\frac{2}{H(H-1)}
\sum_{s<s'}
\left[
\frac{\widehat{\mathbf V}_{ss'}
      \widehat{\mathbf V}_{ss'}^\top}
     {\|\widehat{\mathbf V}_{ss'}\|_F^2}
-
\frac{\mathbf V_{ss'}\mathbf V_{ss'}^\top}
     {\|\mathbf V_{ss'}\|_F^2}
\right]
\right\|_2\\
&\le
\frac{8}{H(H-1)}
\sum_{s<s'}
\frac{\|\widehat{\mathbf V}_{ss'}-\mathbf V_{ss'}\|_F}
     {\|\mathbf V_{ss'}\|_F}\\
&\le
4\max_{s<s'}
\frac{\|\mathbf V_{ss'}\|_F+
      \|\widehat{\mathbf V}_{ss'}\|_F}
     {\|\mathbf V_{ss'}\|_F^2}
\frac{2}{H(H-1)}
\sum_{s<s'}
\|\widehat{\mathbf V}_{ss'}-\mathbf V_{ss'}\|_F.
\end{aligned}
\]

Therefore
\[
\begin{aligned}
\|{\mathbf A}_4\|_2
&=
p_k\,O_p((p_kp_{-k})^{-1/2})
\left\{
O_p\!\left(\sqrt{\frac{p_kp_{-k}H}{n}}\right)
+O_p(\sqrt{p_kp_{-k}}H^{-1})
\right\}\\
&=
O_p(p_kH^{1/2}n^{-1/2})
+O_p(p_kH^{-1}).
\end{aligned}
\]

Combining ${{\mathbf A}_1,{\mathbf A}_2,{\mathbf A}_3,{\mathbf A}_4}$,
\[
\left\|
\widehat{\mathbf K}_{k}^{\mathrm{IR}}
-
\mathbf K^{\mathrm{IR}}_{k}
\right\|_2
=
O_p(p_k^{3/2}p_{-k}^{1/2}H^{-1})
+
O_p(pH^{1/2}n^{-1/2}).
\]
\end{proof}

\paragraph{Proof of Proposition~\ref{ck:thm:Ck}}

\begin{proof}
For mode $k$. \(
\mathcal X_{(k)}\in\mathbb R^{p_k\times p_{-k}}
\sim\mathcal{MED}_{p_k\times p_{-k}}
\left(
\mathcal M_{(k)},
\boldsymbol\Sigma_k\otimes\boldsymbol\Sigma_{-k},
\varphi
\right),
\boldsymbol\Sigma_{-k}
=\bigotimes_{\ell\ne k}\boldsymbol\Sigma_\ell.
\) And write
\(
\mathbf D_{ij}^{(k)}
=\mathcal X_{i,(k)}-\mathcal X_{j,(k)}, N=\binom n2.
\) By Lemma~\ref{lem:tensor_eigenvalue}, $\mathbf D_{ij}^{(k)}/\|\mathbf D_{ij}^{(k)}\|_F\overset d=\boldsymbol\Sigma_k^{1/2}\mathbf Z\boldsymbol\Sigma_{-k}^{1/2}/\|\boldsymbol\Sigma_k^{1/2}\mathbf Z\boldsymbol\Sigma_{-k}^{1/2}\|_F$, where $Z_{ab}\overset{\mathrm{iid}}{\sim}N(0,1)$.
Let $\widehat{\mathbf U}_k
=(\widehat{\mathbf u}_{k,1},\ldots,\widehat{\mathbf u}_{k,p_k})$
be the orthonormal eigenbasis used in the calibration.
Denote
\(
\lambda_{k,a}^{\circ}
=\widehat{\mathbf u}_{k,a}^{\top}
\boldsymbol\Sigma_k\widehat{\mathbf u}_{k,a},
\boldsymbol\Phi_k^{\circ}
=\operatorname{diag}(\lambda_{k,1}^{\circ},\ldots,
\lambda_{k,p_k}^{\circ}),
\mathbf S_k^{\circ}
=\widehat{\mathbf U}_k\boldsymbol\Phi_k^{\circ}
\widehat{\mathbf U}_k^\top.
\)
We use the decomposition
\begin{equation}\label{eq:tensor_cal_decomp}
\begin{aligned}
\|\widehat{\mathbf K}_k^{\mathrm{cal}}
-\boldsymbol\Sigma_k\|_F
&\le
\|\widehat{\mathbf K}_k^{\mathrm{cal}}-\mathbf S_k^{\circ}\|_F
+\|\mathbf S_k^{\circ}-\boldsymbol\Sigma_k\|_F.
\end{aligned}
\end{equation}

\medskip\noindent
\textbf{Part I: Kendall eigenbasis error.}
Let $\kappa_{k,a}$ and $\lambda_{k,a}$ be the corresponding
eigenvalues of $\mathbf K_k$ and $\boldsymbol\Sigma_k$.
By the angular representation,
\[
\kappa_{k,a}
=p_k\,\mathbb E
\frac{\lambda_{k,a}\tau_a}
{\sum_b\lambda_{k,b}\tau_b},
\qquad
\tau_a=\sum_{j=1}^{p_{-k}}\omega_j Z_{aj}^2,
\]
where $\omega_j$ are the eigenvalues of
$\boldsymbol\Sigma_{-k}$.
Under Assumption~\ref{ass:tensor_kendall_scaling},
\[
\frac{|\kappa_{k,a}-\kappa_{k,b}|}
{|\lambda_{k,a}-\lambda_{k,b}|}
\ge \frac{c_{\Sigma_k}}{2C_{\Sigma_k}^2},
\qquad \lambda_{k,a}\ne\lambda_{k,b}.
\]
When $\lambda_{k,a}=\lambda_{k,b}$, exchangeability gives
$\kappa_{k,a}=\kappa_{k,b}$.
Hence there exists a scalar Lipschitz function $f_k$ satisfying
\(
f_k(\kappa_{k,a})=\lambda_{k,a},
f_k(\mathbf K_k)=\boldsymbol\Sigma_k,
|f_k(x)-f_k(y)|\le \frac{2C_{\Sigma_k}^2}{c_{\Sigma_k}}|x-y|.
\)
Then we have
\[
\begin{aligned}
\|\mathbf S_k^\circ-\boldsymbol\Sigma_k\|_F
&=\|\mathbf S_k^\circ-f_k(\mathbf K_k)\|_F\\
&\le
\|\mathbf S_k^\circ-f_k(\widehat{\mathbf K}_k)\|_F
+\|f_k(\widehat{\mathbf K}_k)-f_k(\mathbf K_k)\|_F\\
&=
\left\{
\sum_a
\left[\lambda_{k,a}-f_k(\widehat\kappa_{k,a})\right]^2
\right\}^{1/2}
+\|f_k(\widehat{\mathbf K}_k)-f_k(\mathbf K_k)\|_F\\
&\le
2\|f_k(\widehat{\mathbf K}_k)-f_k(\mathbf K_k)\|_F\\
&\le
\frac{4C_{\Sigma_k}^{\,2}}{c_{\Sigma_k}}
\|\widehat{\mathbf K}_k-\mathbf K_k\|_F.
\end{aligned}
\]
The U-statistic second-moment bound gives $\mathbb E\|\widehat{\mathbf K}_k-\mathbf K_k\|_F^2\le Cp_k^2/n$.
Hence
\begin{equation}\label{eq:tensor_cal_basis}
\|\mathbf S_k^{\circ}-\boldsymbol\Sigma_k\|_F
=O_p(p_k/\sqrt n).
\end{equation}

\medskip\noindent
\textbf{Part II: spectral-repair eigenvalue error.}
Let \(\mathcal H_{p_k}
=\{\boldsymbol\theta\in\mathbb R^{p_k}:
\mathbf1_{p_k}^\top\boldsymbol\theta=0\}\) and
\(\boldsymbol\theta_k^\circ
=\log\boldsymbol\lambda_k^\circ
-p_k^{-1}(\mathbf1_{p_k}^\top
\log\boldsymbol\lambda_k^\circ)\mathbf1_{p_k}\).
For \(g_a(\boldsymbol\theta)
=p_ke^{\theta_a}/\sum_b e^{\theta_b}\),
we have \(g(\boldsymbol\theta_k^\circ)
=\boldsymbol\lambda_k^\circ\).
Define \(s_{ij,a}^{(k)}
=\|\widehat{\mathbf u}_{k,a}^\top\mathbf D_{ij}^{(k)}\|_2^2\),
the empirical calibration loss is
\[L_{n,k}(\boldsymbol\theta)
=\mathbf1_{p_k}^\top\boldsymbol\theta
+\frac{p_k}{N}\sum_{i<j}
\log\left\{
\frac{\sum_{a=1}^{p_k}s_{ij,a}^{(k)}e^{-\theta_a}}
{\sum_{a=1}^{p_k}s_{ij,a}^{(k)}}
\right\}.\] And write
\(w_{ij,a}(\boldsymbol\theta)
=s_{ij,a}^{(k)}e^{-\theta_a}/
(\sum_b s_{ij,b}^{(k)}e^{-\theta_b})\), and
\(\mathbf w_{ij}(\boldsymbol\theta)
=(w_{ij,1}(\boldsymbol\theta),\ldots,
w_{ij,p_k}(\boldsymbol\theta))^\top\).
Direct differentiation gives
\[
\begin{aligned}
\nabla L_{n,k}(\boldsymbol\theta)
&=\mathbf1_{p_k}
-\frac{p_k}{N}\sum_{i<j}\mathbf w_{ij}(\boldsymbol\theta),\\
\nabla^2L_{n,k}(\boldsymbol\theta)
&=\frac{p_k}{N}\sum_{i<j}
\left\{\operatorname{diag}(\mathbf w_{ij}(\boldsymbol\theta))
-\mathbf w_{ij}(\boldsymbol\theta)
\mathbf w_{ij}(\boldsymbol\theta)^\top\right\}.
\end{aligned}
\]
Since \(\nabla^2L_{n,k}(\boldsymbol\theta)\succeq0\) and
\(\mathbf1_{p_k}^\top\nabla L_{n,k}(\boldsymbol\theta)=0\),
\(L_{n,k}\) is convex and its gradient lies in
\(\mathcal H_{p_k}\). Define
\(\boldsymbol\lambda_k
=(\lambda_{k,1},\ldots,\lambda_{k,p_k})^\top\) and
\(g(\boldsymbol\theta)
=(g_1(\boldsymbol\theta),\ldots,
g_{p_k}(\boldsymbol\theta))^\top\).
Under the parameterization
\(\boldsymbol\lambda_k=g(\boldsymbol\theta)\), we have
\(w_{ij,a}(\boldsymbol\theta)
=(s_{ij,a}^{(k)}/\lambda_{k,a})/
(\sum_b s_{ij,b}^{(k)}/\lambda_{k,b})\).
Therefore, \(\nabla L_{n,k}(\boldsymbol\theta)=0\)
is equivalent to the trace-normalized calibration equations 
\[
\lambda_{k,a}
=\frac{p_k}{N}\sum_{i<j}
\frac{s_{ij,a}^{(k)}}
{\sum_b s_{ij,b}^{(k)}/\lambda_{k,b}},
\qquad a=1,\ldots,p_k,
\qquad \sum_a\lambda_{k,a}=p_k.
\]
Define
\[
\mathbf R
=(\boldsymbol\Phi_k^{\circ})^{-1/2}
\widehat{\mathbf U}_k^\top\boldsymbol\Sigma_k
\widehat{\mathbf U}_k(\boldsymbol\Phi_k^{\circ})^{-1/2}.
\]
Assumption~\ref{ass:tensor_kendall_scaling} gives
$c_{\Sigma_k}\le\lambda_{k,a}^{\circ}\le C_{\Sigma_k}$
and $R_{aa}=1$. Together with
\eqref{eq:tensor_cal_basis}, this yields
\[
\|\mathbf R-\mathbf I_{p_k}\|_F
\le c_{\Sigma_k}^{-1}
\|\mathbf S_k^{\circ}-\boldsymbol\Sigma_k\|_F
=O_p(p_k/\sqrt n).
\]
Define
\[
\mathbf G_{ij}
=
\frac{\boldsymbol\Sigma_k^{-1/2}\mathbf D_{ij}^{(k)}
(\mathbf D_{ij}^{(k)})^\top\boldsymbol\Sigma_k^{-1/2}}
{\operatorname{tr}\{
\boldsymbol\Sigma_k^{-1}\mathbf D_{ij}^{(k)}
(\mathbf D_{ij}^{(k)})^\top\}},
\qquad
\mathbf M_n=\frac{p_k}{N}\sum_{i<j}\mathbf G_{ij},
\]
with \(\mathbf G_{ij}\succeq0\),
\(\operatorname{tr}(\mathbf G_{ij})=1\),
\(\|\mathbf G_{ij}\|_F\le1\), and
\(\mathbb E\mathbf G_{ij}=\mathbf I_{p_k}/p_k\).
The U-statistic second-moment bound gives
\[
\begin{aligned}
\mathbb E\|\mathbf M_n-\mathbf I_{p_k}\|_F^2
&\le \frac1{b_n}
\mathbb E\|p_k\mathbf G_{12}-\mathbf I_{p_k}\|_F^2\\
&=\frac1{b_n}
\left\{p_k^2\mathbb E\|\mathbf G_{12}\|_F^2
-\|\mathbf I_{p_k}\|_F^2\right\}\\
&\le \frac{p_k^2}{b_n}\mathbb E\|\mathbf G_{12}\|_F^2
\le \frac{p_k^2}{b_n}
\le \frac{Cp_k^2}{n},
\qquad b_n=\lfloor n/2\rfloor.
\end{aligned}
\]

Markov's inequality and $p_k/\sqrt n\to0$ imply
\[
\eta_{n,k}:=\|\mathbf M_n-\mathbf I_{p_k}\|_F
=O_p(p_k/\sqrt n)=o_p(1),
\qquad
\|\mathbf M_n\|_2
\le1+\|\mathbf M_n-\mathbf I_{p_k}\|_F
=1+\eta_{n,k}.
\]

Let \(\mathbf B=(\boldsymbol\Phi_k^\circ)^{-1/2}
\widehat{\mathbf U}_k^\top\boldsymbol\Sigma_k^{1/2}\) and
\(\mathbf O=\mathbf R^{-1/2}\mathbf B\).
Since \(\mathbf B\mathbf B^\top=\mathbf R\),
\(\mathbf O\) is orthogonal and
\(\mathbf B=\mathbf R^{1/2}\mathbf O\).
Define \(\overline{\mathbf G}_{ij}
=\mathbf O\mathbf G_{ij}\mathbf O^\top\) and
\(\overline{\mathbf M}_n=\mathbf O\mathbf M_n\mathbf O^\top\).
Then \(\operatorname{tr}(\overline{\mathbf G}_{ij})=1\),
\(\|\overline{\mathbf M}_n-\mathbf I_{p_k}\|_F=\eta_{n,k}\),
and \(\|\overline{\mathbf M}_n\|_2=\|\mathbf M_n\|_2\).
Set
{
\[
{\mathbf S}_{n,k}(\mathbf R)
=\frac{p_k}{N}\sum_{i<j}
\frac{\overline{\mathbf G}_{ij}}
{\operatorname{tr}(\mathbf R\overline{\mathbf G}_{ij})},
\qquad
{\mathbf F}_{n,k}(\mathbf R)
=\mathbf R^{1/2}{\mathbf S}_{n,k}(\mathbf R)\mathbf R^{1/2}.
\]
}
By the definitions,
\[
\frac{\mathbf B\mathbf G_{ij}\mathbf B^\top}
{\operatorname{tr}(\mathbf B\mathbf G_{ij}\mathbf B^\top)}
=
\frac{
(\boldsymbol\Phi_k^\circ)^{-1/2}
\widehat{\mathbf U}_k^\top
\mathbf D_{ij}^{(k)}(\mathbf D_{ij}^{(k)})^\top
\widehat{\mathbf U}_k
(\boldsymbol\Phi_k^\circ)^{-1/2}
}{
\operatorname{tr}\!\left\{
(\boldsymbol\Phi_k^\circ)^{-1/2}
\widehat{\mathbf U}_k^\top
\mathbf D_{ij}^{(k)}(\mathbf D_{ij}^{(k)})^\top
\widehat{\mathbf U}_k
(\boldsymbol\Phi_k^\circ)^{-1/2}
\right\}
}.
\]
Its \(a\)th diagonal entry is
\((s_{ij,a}^{(k)}/\lambda_{k,a}^\circ)/
(\sum_b s_{ij,b}^{(k)}/\lambda_{k,b}^\circ)
=w_{ij,a}(\boldsymbol\theta_k^\circ)\).
Together with \(\mathbf B=\mathbf R^{1/2}\mathbf O\),
this gives
\({\operatorname{diagvec}\{{\mathbf F}_{n,k}(\mathbf R)\}
=(p_k/N)\sum_{i<j}\mathbf w_{ij}(\boldsymbol\theta_k^\circ)}\).
Hence
\begin{equation}\label{eq:tensor_cal_score_identity}
\nabla L_{n,k}(\boldsymbol\theta_k^\circ)
=\mathbf1_{p_k}
-\operatorname{diagvec}\{{\mathbf F}_{n,k}(\mathbf R)\}.
\end{equation}

Let $c_*=c_{\Sigma_k}/C_{\Sigma_k}$ and
$C_*=C_{\Sigma_k}/c_{\Sigma_k}$, so that
$c_*\mathbf I_{p_k}\preceq\mathbf R\preceq C_*\mathbf I_{p_k}$.
Since $\operatorname{tr}(\mathbf R\overline{\mathbf G}_{ij})\ge c_*$,
Cauchy--Schwarz and
$\{\operatorname{tr}(\mathbf H\mathbf G)\}^2
\le\operatorname{tr}(\mathbf H^2\mathbf G)$
for $\mathbf G\succeq0$ with $\operatorname{tr}(\mathbf G)=1$
give, for symmetric $\mathbf H$,
\[
\begin{aligned}
&\left|
\operatorname{tr}\left\{
\mathbf H\bigl(\mathbf S_{n,k}(\mathbf R)
-\overline{\mathbf M}_n\bigr)
\right\}
\right|\\
&\quad=
\left|
\frac{p_k}{N}\sum_{i<j}
\frac{
\operatorname{tr}(\mathbf H\overline{\mathbf G}_{ij})
\operatorname{tr}\bigl((\mathbf R-\mathbf I_{p_k})
\overline{\mathbf G}_{ij}\bigr)
}{
\operatorname{tr}(\mathbf R\overline{\mathbf G}_{ij})
}
\right|\\
&\quad\le c_*^{-1}
\left[
\frac{p_k}{N}\sum_{i<j}
\left\{\operatorname{tr}(\mathbf H\overline{\mathbf G}_{ij})\right\}^2
\right]^{1/2}
\left[
\frac{p_k}{N}\sum_{i<j}
\left\{
\operatorname{tr}\bigl((\mathbf R-\mathbf I_{p_k})
\overline{\mathbf G}_{ij}\bigr)
\right\}^2
\right]^{1/2}\\
&\quad\le c_*^{-1}
\left\{\operatorname{tr}(\mathbf H^2\overline{\mathbf M}_n)\right\}^{1/2}
\left\{
\operatorname{tr}\bigl((\mathbf R-\mathbf I_{p_k})^2
\overline{\mathbf M}_n\bigr)
\right\}^{1/2}\\
&\quad\le c_*^{-1}
\|\overline{\mathbf M}_n\|_2
\|\mathbf H\|_F
\|\mathbf R-\mathbf I_{p_k}\|_F.
\end{aligned}
\]
If $\mathbf S_{n,k}(\mathbf R)=\overline{\mathbf M}_n$,
the bound below is immediate. Otherwise, choose
$\mathbf H=\{\mathbf S_{n,k}(\mathbf R)-\overline{\mathbf M}_n\}/
\|\mathbf S_{n,k}(\mathbf R)-\overline{\mathbf M}_n\|_F$.
The preceding inequality gives
\({\|{\mathbf S}_{n,k}(\mathbf R)-\overline{\mathbf M}_n\|_F
\le c_*^{-1}\|\mathbf M_n\|_2
\|\mathbf R-\mathbf I_{p_k}\|_F}\).
Together with
\[
\begin{aligned}
{\mathbf F}_{n,k}(\mathbf R)-\mathbf I_{p_k}
={}&\mathbf R^{1/2}
\bigl({\mathbf S}_{n,k}(\mathbf R)-\overline{\mathbf M}_n\bigr)
\mathbf R^{1/2}\\
&+\mathbf R^{1/2}
(\overline{\mathbf M}_n-\mathbf I_{p_k})\mathbf R^{1/2}
+(\mathbf R-\mathbf I_{p_k}),
\end{aligned}
\]
$\|\mathbf R\|_2\le C_*$ and
\eqref{eq:tensor_cal_score_identity} give
\begin{equation}\label{eq:tensor_cal_score}
\begin{aligned}
\|\nabla L_{n,k}(\boldsymbol\theta_k^\circ)\|_2
&=\|\operatorname{diagvec}\{\mathbf I_{p_k}-\mathbf F_{n,k}(\mathbf R)\}\|_2
\le\|\mathbf F_{n,k}(\mathbf R)-\mathbf I_{p_k}\|_F\\
&\le C_*\|\mathbf S_{n,k}(\mathbf R)-\overline{\mathbf M}_n\|_F
+C_*\eta_{n,k}+\|\mathbf R-\mathbf I_{p_k}\|_F\\
&\le\{1+C_*c_*^{-1}\|\mathbf M_n\|_2\}
\|\mathbf R-\mathbf I_{p_k}\|_F+C_*\eta_{n,k}
=O_p(p_k/\sqrt n).
\end{aligned}
\end{equation}

Since \(
\mathcal X_{(k)}\in\mathbb R^{p_k\times p_{-k}}
\sim\mathcal{MED}_{p_k\times p_{-k}}
\left(
\mathcal M_{(k)},
\boldsymbol\Sigma_k\otimes\boldsymbol\Sigma_{-k},
\varphi
\right),
\) the bound in \eqref{cr:eq:T} can also be applied to the mode-$k$ unfolding:
\[
\frac{p_k}{N}\sum_{i<j}
\{\operatorname{tr}(\mathbf A\mathbf G_{ij})\}^2
\le
\left(\frac{2}{p_k+2}+\zeta_{n,k}\right)\|\mathbf A\|_F^2,
\qquad
\zeta_{n,k}=O_p(p_k/\sqrt n),
\]
for $\forall\,\mathbf A\in\mathbb R^{p_k\times p_k}
\text{ with }\mathbf A^\top=\mathbf A
\text{ and }\operatorname{tr}(\mathbf A)=0$.

Let \({\rho_{n,k}
=\|{\mathbf F}_{n,k}(\mathbf R)-\mathbf I_{p_k}\|_2
=O_p(p_k/\sqrt n)}\).
By the Hessian estimate leading to \eqref{cr:eq:H0}, we have
\[
\begin{aligned}
\mathbf v^\top\nabla^2L_{n,k}
(\boldsymbol\theta_k^\circ)\mathbf v
&\ge
\left[
1-\rho_{n,k}
-\left\{\frac{\lambda_{\max}(\mathbf R)}
{\lambda_{\min}(\mathbf R)}\right\}^{2}
\left\{\frac{2}{p_k+2}+\zeta_{n,k}\right\}
\right]\|\mathbf v\|_2^2\\
&\ge \frac14\|\mathbf v\|_2^2,
\qquad \mathbf v\in\mathcal H_{p_k},
\end{aligned}
\]
with probability tending to one.

For $\|\mathbf h\|_\infty\le\delta$,
\[
w_{ij,a}(\boldsymbol\theta_k^{\circ}+\mathbf h)
\ge e^{-2\delta}w_{ij,a}(\boldsymbol\theta_k^{\circ}).
\]
Hence,
\begin{equation}\label{eq:tensor_cal_local}
\mathbf v^\top\nabla^2L_{n,k}
(\boldsymbol\theta_k^{\circ}+\mathbf h)\mathbf v
\ge c_\delta\|\mathbf v\|_2^2,
\qquad
c_\delta=\tfrac14e^{-4\delta},
\quad
\mathbf v,\mathbf h\in\mathcal H_{p_k},\quad
\|\mathbf h\|_\infty\le\delta.
\end{equation}

By Taylor's formula and \eqref{eq:tensor_cal_local},
\[
L_{n,k}(\boldsymbol\theta_k^{\circ}+\mathbf h)
-L_{n,k}(\boldsymbol\theta_k^{\circ})
\ge
-\|\nabla L_{n,k}(\boldsymbol\theta_k^{\circ})\|_2\|\mathbf h\|_2
+\frac{c_\delta}{2}\|\mathbf h\|_2^2.
\]

For fixed $\delta>0$,
$\|\nabla L_{n,k}(\boldsymbol\theta_k^\circ)\|_2=o_p(1)$
implies that the preceding lower bound is strictly positive
uniformly over
$\{\mathbf h\in\mathcal H_{p_k}:\|\mathbf h\|_2=\delta\}$
with probability tending to one.
Hence, convexity and local strong convexity yield a unique global
minimizer $\widehat{\boldsymbol\theta}_k$ on $\mathcal H_{p_k}$,
lying in the interior of the corresponding ball.
Using $L_{n,k}(\widehat{\boldsymbol\theta}_k)
\le L_{n,k}(\boldsymbol\theta_k^\circ)$ gives
\[
\|\widehat{\boldsymbol\theta}_k-\boldsymbol\theta_k^\circ\|_2
\le \frac{2}{c_\delta}
\|\nabla L_{n,k}(\boldsymbol\theta_k^\circ)\|_2
=O_p(p_k/\sqrt n).
\]

Finally,
\[
Dg(\boldsymbol\theta)
=\operatorname{diag}(g(\boldsymbol\theta))
-\frac1{p_k}g(\boldsymbol\theta)g(\boldsymbol\theta)^\top,
\qquad
\|Dg(\boldsymbol\theta)\|_2
\le\max_a g_a(\boldsymbol\theta).
\]
Since \(g_a(\boldsymbol\theta)
\le e^{2\delta}\lambda_{k,a}^\circ
\le e^{2\delta}C_{\Sigma_k}\) whenever
\(\|\boldsymbol\theta-\boldsymbol\theta_k^\circ\|_\infty
\le\delta\), \(g\) is Lipschitz on this neighborhood.

Using $g(\boldsymbol\theta_k^\circ)=\boldsymbol\lambda_k^\circ$, we obtain
\[
\begin{aligned}
\|\widehat{\mathbf K}_k^{\mathrm{cal}}-\mathbf S_k^\circ\|_F
&=\|\widehat{\mathbf U}_k\operatorname{diag}
(\widehat{\boldsymbol\lambda}_k-\boldsymbol\lambda_k^\circ)
\widehat{\mathbf U}_k^\top\|_F
=\|\widehat{\boldsymbol\lambda}_k-\boldsymbol\lambda_k^\circ\|_2\\
&\le e^{2\delta}C_{\Sigma_k}
\|\widehat{\boldsymbol\theta}_k-\boldsymbol\theta_k^\circ\|_2
=O_p(p_k/\sqrt n).
\end{aligned}
\]

Combining this bound with \eqref{eq:tensor_cal_basis} gives
\[
\|\widehat{\mathbf K}_k^{\mathrm{cal}}
-\boldsymbol\Sigma_k\|_2
\le
\|\widehat{\mathbf K}_k^{\mathrm{cal}}
-\boldsymbol\Sigma_k\|_F
\le
\|\widehat{\mathbf K}_k^{\mathrm{cal}}-\mathbf S_k^{\circ}\|_F
+\|\mathbf S_k^{\circ}-\boldsymbol\Sigma_k\|_F
=O_p(p_k/\sqrt n).
\]
\end{proof}

\paragraph{Proof of Theorem~\ref{thm:tensor_mesir_operator_consistency}}
\begin{proof}
By Assumption~\ref{ass:tensor_kendall_scaling}
and Proposition~\ref{ck:thm:Ck},
$\|\widehat{\mathbf K}_k^{\mathrm{cal}}-\boldsymbol\Sigma_k\|_2
=O_p(p_k/\sqrt n)=o_p(1)$.
Weyl's inequality gives
\[
\lambda_{\min}(\widehat{\mathbf K}_k^{\mathrm{cal}})
\ge c_{\Sigma_k}
-\|\widehat{\mathbf K}_k^{\mathrm{cal}}-\boldsymbol\Sigma_k\|_2
=c_{\Sigma_k}-o_p(1),
\]
so $\|(\widehat{\mathbf K}_k^{\mathrm{cal}})^{-1}\|_2=O_p(1)$.
Then we have
\[
\begin{aligned}
\|(\widehat{\mathbf K}_k^{\mathrm{cal}})^{-1}-\boldsymbol\Sigma_k^{-1}\|_2
&=\|\boldsymbol\Sigma_k^{-1}
(\boldsymbol\Sigma_k-\widehat{\mathbf K}_k^{\mathrm{cal}})
(\widehat{\mathbf K}_k^{\mathrm{cal}})^{-1}\|_2\\
&\le\|\boldsymbol\Sigma_k^{-1}\|_2
\|\widehat{\mathbf K}_k^{\mathrm{cal}}-\boldsymbol\Sigma_k\|_2
\|(\widehat{\mathbf K}_k^{\mathrm{cal}})^{-1}\|_2
=O_p(p_k/\sqrt n).
\end{aligned}
\]
Combining these bounds with
Assumption~\ref{ass:tensor_bounded_inverse_kendall}
and Proposition~\ref{prop:tensor_slice_kendall_error}, we obtain
\begin{equation}\label{eq:tensor_operator_bound}
\begin{aligned}
&\|(\widehat{\mathbf K}_k^{\mathrm{cal}})^{-1}
\widehat{\mathbf K}_k^{\mathrm{IR}}
-\boldsymbol\Sigma_k^{-1}\mathbf K_k^{\mathrm{IR}}\|_2\\
&\quad\le
\|(\widehat{\mathbf K}_k^{\mathrm{cal}})^{-1}-\boldsymbol\Sigma_k^{-1}\|_2
\|\mathbf K_k^{\mathrm{IR}}\|_2
+\|(\widehat{\mathbf K}_k^{\mathrm{cal}})^{-1}\|_2
\|\widehat{\mathbf K}_k^{\mathrm{IR}}-\mathbf K_k^{\mathrm{IR}}\|_2\\
&\quad=O_p(p_kn^{-1/2})
+O_p(p_k^{3/2}p_{-k}^{1/2}H^{-1})
+O_p(p_kH^{1/2}n^{-1/2})\\
&\quad=O_p(p_k^{3/2}p_{-k}^{1/2}H^{-1})
+O_p(p_kH^{1/2}n^{-1/2}),
\end{aligned}
\end{equation}
where the calibration term is absorbed since $H\ge1$.
\end{proof}

\paragraph{Proof of Corollary~\ref{cor:tensor_mode_subspace_consistency}}

\begin{proof}
Fix $k\in\{1,\ldots,K\}$, and let
$a_{n,k}=p_k^{3/2}p_{-k}^{1/2}H^{-1}
+p_kH^{1/2}n^{-1/2}$.
Let $\boldsymbol\Gamma_k$ and $\widehat{\boldsymbol\Gamma}_k$
contain the leading $d_k$
orthonormal eigenvectors of $\mathbf K_k^{\mathrm{IR}}$ and
$\widehat{\mathbf K}_k^{\mathrm{IR}}$,
respectively. By Assumption~\ref{ass:tensor_bounded_inverse_kendall},
\(
 \delta_k:=\lambda_{d_k}(\mathbf K_k^{\mathrm{IR}})
 -\lambda_{d_k+1}(\mathbf K_k^{\mathrm{IR}})\ge c_k>0.
\)
Applying the Davis--Kahan theorem in \citep{Han02012018}
to these symmetric matrices gives
\[
 \|\mathbf P_{\widehat{\boldsymbol\Gamma}_k}
 -\mathbf P_{\boldsymbol\Gamma_k}\|_2
 \le \frac{C}{\delta_k}
 \|\widehat{\mathbf K}_k^{\mathrm{IR}}
 -\mathbf K_k^{\mathrm{IR}}\|_2
 =O_p(a_{n,k}).
\]

Since
$\widehat{\mathbf B}_k=
(\widehat{\mathbf K}_k^{\mathrm{cal}})^{-1}
\widehat{\boldsymbol\Gamma}_k$
and
$\operatorname{Span}(\mathbf B_k)
=\operatorname{Span}(\boldsymbol{\Sigma}_k^{-1}\boldsymbol\Gamma_k)$,
then we have
\[
\begin{aligned}
 \|\mathbf P_{\widehat{\mathbf B}_k}
 -\mathbf P_{\mathbf B_k}\|_2
 &\le C\left\{
 \|\mathbf P_{\widehat{\boldsymbol\Gamma}_k}
 -\mathbf P_{\boldsymbol\Gamma_k}\|_2
 +\|(\widehat{\mathbf K}_k^{\mathrm{cal}})^{-1}
 -{\boldsymbol\Sigma}_k^{-1}\|_2
 \right\}\\
 &\le C\left\{
 \delta_k^{-1}
 \|\widehat{\mathbf K}_k^{\mathrm{IR}}
 -\mathbf K_k^{\mathrm{IR}}\|_2
 +\|\widehat{\mathbf K}_k^{\mathrm{cal}}
 -\boldsymbol{\Sigma}_k\|_2
 \right\}\\
 &=O_p(a_{n,k})+O_p(p_k/\sqrt n)\\
 &=O_p(p_k^{3/2}p_{-k}^{1/2}H^{-1}
 +p_kH^{1/2}n^{-1/2}).
\end{aligned}
\]
\end{proof}

\paragraph{Proof of Theorem~\ref{theo:TESIR-Auto}}

\begin{proof}
We prove the result for $\widehat d_k$. Let
\[
\Delta_{n,k}
=
\left\|
\widehat{\mathbf K}_{k}^{\mathrm{IR}}
-
\mathbf K^{\mathrm{IR}}_{k}
\right\|_2 .
\]
By Proposition~\ref{prop:tensor_slice_kendall_error} and Weyl's inequality,
\begin{align*}
\Delta_{n,k}
&=
O_p(a_{n,k}),\qquad
\max_{1\le j\le p_k}
\left|
\widehat\lambda_j^{(k)}
-
\lambda_j^{(k)}
\right|
\le
\Delta_{n,k},\qquad
\tau_{n,k}
=
a_{n,k}^{\gamma}\to0,
\qquad
\frac{\Delta_{n,k}}{\tau_{n,k}}
=
O_p(a_{n,k}^{1-\gamma})
=
o_p(1).
\end{align*}
Hence, for
\(\mathcal E_n=\{\Delta_{n,k}\le\tau_{n,k}\}\),
we have \(P(\mathcal E_n)\to1\). Recall that
\[
R_{n,j}^{(k)}
=
\frac{
\widehat\lambda_j^{(k)}+\tau_{n,k}
}{
\widehat\lambda_{j+1}^{(k)}+\tau_{n,k}
},
\qquad
1\le j\le d_{k,\max}.
\]

On $\mathcal E_n$, Assumption~\ref{ass:tensor_bounded_inverse_kendall} and Weyl's inequality imply
\(c_k-\Delta_{n,k}\le\widehat\lambda_j^{(k)}
\le C_k+\Delta_{n,k}\) for \(1\le j\le d_1\),
whereas
\(0\le\widehat\lambda_j^{(k)}\le\Delta_{n,k}\) for \(j>d_k\).
Therefore,
\begin{align*}
\max_{1\le j<d_k}R_{n,j}^{(k)}
&\le
\frac{
C_k+\Delta_{n,k}+\tau_{n,k}
}{
c_k-\Delta_{n,k}+\tau_{n,k}
}
\le
\frac{C_k+2\tau_{n,k}}{c_k}
=
O(1),\\
R_{n,d_k}^{(k)}
&\ge
\frac{
c_k-\Delta_{n,k}+\tau_{n,k}
}{
\Delta_{n,k}+\tau_{n,k}
}
\ge
\frac{c_k}{2\tau_{n,k}}
\longrightarrow\infty,\\
\max_{d_k<j\le d_{k,\max}}R_{n,j}^{(k)}
&\le
\frac{
\Delta_{n,k}+\tau_{n,k}
}{
\tau_{n,k}
}
=
1+\frac{\Delta_{n,k}}{\tau_{n,k}}
=
O_p(1).
\end{align*}

Thus,
\[
R_{n,d_k}^{(k)}
\xrightarrow{p}\infty,
\qquad
\max_{\substack{1\le j\le d_{k,\max}\\ j\ne d_k}}
R_{n,j}^{(k)}
=
O_p(1),
\]
which gives
\[
P\left\{
R_{n,d_k}^{(k)}
>
\max_{\substack{1\le j\le d_{k,\max}\\ j\ne d_k}}
R_{n,j}^{(k)}
\right\}
\to1.
\]
\end{proof}
\end{document}

%% file: manuscript-macros.tex
\providecommand{\bi}{\bmf{i}}

\providecommand{\RR}{\mathbb{R}}

\providecommand*{\one}{{\bm 1}}